\documentclass[aip,graphicx]{revtex4-1}

\usepackage[english]{babel}
\usepackage{graphicx}
\usepackage{epsf}
\usepackage{amstext}
\usepackage{amsmath}
\usepackage{amssymb}
\usepackage{flafter}
\usepackage{appendix}
\usepackage{float}
\restylefloat{figure}

\draft 

\begin{document}


\title{Effect of density changes on tokamak plasma confinement} 



\author{F. Spineanu}
\email[]{florin.spineanu@euratom.ro}
\author{M. Vlad}
\affiliation{National Institute of Laser, Plasma and Radiation Physics Bucharest, Romania}


\date{\today}

\begin{abstract}
A change of the particle density (by gas puff, pellets or impurity seeding)
during the plasma discharge in tokamak produces a radial current and implicitly
a torque and rotation that can modify the state of confinement. After
ionization the newly born ions will evolve toward the periodic neoclassical
orbits (trapped or circulating) but the first part of their excursion, which
precedes the periodicity, is an effective radial current. It is short,
spatially finite and unique for each new ion, but multiplied by the rate of
ionization and it can produce a substantial total radial current. The
associated torque induces rotation which modify the transport processes. We
derive the magnitude of the radial current induced by ionization by three
methods: the analysis of a simple physical picture, a numerical model and
the neoclassical drift-kinetic treatment. The results of the three
approaches are in agreement and show that the current can indeed be
substantial. Many well known experimental observations can be reconsidered
under this perspective. In reactor-grade plasma the confinement can be
strongly influenced by adequate particle fuelling.
\end{abstract}


\maketitle 


\section{Introduction} \label{introduction}

Every event of ionization of a neutral particle in tokamak plasma it
followed by the displacement of the newly born charges (electron and ion)
towards the equilibrium orbits. They leave the magnetic surface where they
have been created, due to the neoclassical drift, and evolve towards
stationary trajectories. Since the electrons drift much less than the ions,
the main effect is associated with the newly born ions. Neglecting
collisions, the ions will settle on circulating or trapped orbits and during
the periodic motions they depart radially, relative to a certain magnetic
surface, alternatively to larger and respectively to smaller radius. Since
these positive and negative radial deviations relative to the magnetic
surface compensate, the time average shows no effective radial displacement:
the orbit has an effective \textquotedblleft center\textquotedblright\ which
corresponds to the spatial average of the successive positions of the ion
(for example: the \textquotedblleft center\textquotedblright\ of a banana,
for the trapped ion; we neglect smaller neoclassical motions of this
\textquotedblleft center\textquotedblright , like the toroidal drift).
However there is a part in the radial excursion of the new ion which remains
uncompensated. This is precisely the first interval, just after the
ionization, when the ion evolves to take the periodic trajectory, and its
successive positions do not yet average to the \textquotedblleft
center\textquotedblright . This displacement, from the position where
ionization takes place, towards the \textquotedblleft
center\textquotedblright\ of the periodic trajectory, is an effective radial
current. At the end of this finite, transitory part, the motion becomes
periodic and there is no radial current. The radial current of the first
part is a source of torque, implicitly rotation and in this way has an
impact on the quality of the confinement. It effectively makes a connection
between the density change (via pellets, gas puff or impurity seeding) and
the change of the confinement. We note that there is a considerable
experimental evidence that the variation of density (during the discharge)
produces a change in the quality of the confinement.

On a fast time scale a radial electric field is generated by the charge
separation: the electrons are almost tied to the magnetic surface, while the
ions will travel with the neoclassical drift velocity to their
\textquotedblleft center\textquotedblright , on a distance about half of a
banana width. We estimate the radial current and obtain an order of
magnitude of the rate of the torque. Compared with the damping rate of a
poloidal rotation by transit time magnetic pumping, the ionization-induced
rate can be substantially higher. Some improved regimes in JET, as
\textquotedblleft pellet enhanced performance\textquotedblright\ (PEP) \cite%
{pepjetreview}, \cite{Hugon}, DIII-D \cite{impurityseedingd3d}, \cite%
{improvedcorefuelling} \ and confinement changes observed in many devices 
\cite{itbalcatorfiore}, \cite{valovicmast}, \cite{globaltoresupra}, \cite%
{startrotationtournianski} appear to be connected with this effect of
density variation.

The ionization-induced torque has a direction which is fixed by neoclassical
orbit's geometry and it interacts with any pre-existing rotation which may
have been induced by Reynolds stress, Stringer mechanism or by external
factors (NBI, ICRH). The new torque can enhance the pre-existing rotation or
can act against it, which makes difficult to predict its consequences in all
situations.

\bigskip

We suggest that this process may be a unifying connection between a wide
class of regimes where it has been noted a correlation between a dynamic
change of density (within a discharge) and the change of the confinement.

According to the preceding explanation, there are two mechanisms that are
responsible for this connection

(1) the change of density via ionization (of a pellet, gas puff, impurity
seeding or influx of neutral atoms from the edge) means that $\sqrt{%
\varepsilon }$ ($\varepsilon =r/R$) fraction of the newly created ions are
trapped and \ are moving radially to occupy the positions (the
\textquotedblleft centers\textquotedblright ) which are the averages of
positions on the trapped (banana) orbits. While after arriving there the
bounce averaged radial displacement is zero for the trapped ion, the first
step, when the ion moves from the place where it has been created to the
\textquotedblleft center\textquotedblright\ of the banana is a net radial
current, a single and unrepeatable event for every ionization event. The
ensemble of such events is a radial current that produces a torque which
generates poloidal rotation and sustains it against magnetic pumping \cite%
{florinmadieps13}; the sheared poloidal rotation is a barrier that reduce
the turbulence and enhances the confinement.

(2) every conversion of a trapped ion into a circulating one (and equally
the reversed process), is accompanied by a substantial radial drift. This is
because the \textquotedblleft centers\textquotedblright\ of the two kinds of
orbits are different and the change from one type of periodic motion (\emph{%
e.g.} trapped) to the other type of periodic motion (circulating) goes
through an intermediate regime, unique and transitory. It consists of the
last part of the motion on banana, when the periodicity is lost, followed by
the first part of the motion, until the new periodicity is established. Both
these parts are unique and transitory and are manifested as a radial current
which produces a torque. Then any dynamic process which implies a (slow or
fast) change in the velocity space in the region: trapped/circulating ions,
will produce a torque. It is interesting to note that this is a mechanism of 
\emph{direct} coupling between the toroidal and poloidal rotation. If for
some reason a toroidal flow occurs in plasma, a number of trapped ions will
have increased their parallel velocity and will change from trapped to
circulating \cite{nycanderyankovtrapped}. The associated radial current
produces a poloidal torque. Conversely, stopping the toroidal rotation leads
a subset of the ion population to convert from circulating to trapped, which
again produces a transient radial current and further poloidal rotation.

\bigskip

In the present work we concentrate on only the first of the two processes.

\bigskip

Several effects can be connected with this ionization-induced torque.
Regimes that are confirmed by experiments, like the mentioned PEP, density
peaking and/or anomalous density pinch may have a connection with this
torque. The dynamic charge separation that occurs when the ions move from
the place of ionization to the \textquotedblleft center\textquotedblright\
of the neoclassical periodic orbit induces a return current of the
background ions. Since the number of new ions generated by ionization of a
pellet is episodically comparable with the local background ion density, the
response motion of this latter population takes the aspect of a massive,
even if short, radial displacement. On a relatively large space interval, on
which the radial derivative of the rate of ionization [$\partial S$/$%
\partial x$ in Eq.(\ref{jtapp}) below] keeps the same sign, the displacement
of the background ions has a unique direction and is sustained all along the
total time of ionization. This appears as a density pinch, eventually
contributing to the density peaking. Since this equally involves the
impurity ions, it can provide a new mechanism for the impurity accumulation.

We note that the ionization of impurity atoms leads to much larger radial
drifts and in consequence larger radial currents and torque. This must be
examined in relation with impurity (argon) seeding at the edge and with 
\emph{Li} pellets in the core. In general any influx of neutral atoms in the
plasma will be a source of rotation which affects the local conditions,
including the possibility of being a trigger for the \emph{L} to \emph{H}
mode transition.

We give in the next Sections a simple description of the statistical
build-up of a radial current associated with the ionization. At this level
of description the collisions are neglected, as are the dispersions of the
absolute magnitude of velocities, and of the parallel velocities. The
intention is to draw attention to the high amplitude of this current.
Further, our result is confirmed by the drift-kinetic neoclassical approach,
developed in parallel to the classical treatments of Rosenbluth and Hinton
for the similar cases of the torque induced by \emph{alpha} particles \cite%
{rosenbluthhintonalpha} and by \emph{neutral beam injection} (NBI) \cite%
{hintonrosenbluthnbi}.

\section{Estimation of the radial current generated at ionization} \label{section1}

\subsection{The contributions to the current}

For an easier discussion we adopt a simple picture, mainly having in mind
the pellet injection. The source is considered to be limited to a finite
segment $\left[ r_{1},r_{2}\right] $ on the radius $r$, placed somewhere
between $r=0$ (magnetic axis) and $r=a$ (edge). Due to the symmetry we take
the segment as lying in the equatorial plane. The words \textquotedblleft
left\textquotedblright\ or \textquotedblleft right\textquotedblright\ refer
to this segment, with left being closer to the magnetic axis. The newly born
ions will move to place themselves on the periodic neoclassical
trajectories, banana or circulating. A radial current is produced during
only the first, unique and transitory, part of the trajectory, which is
about half of the width of the banana.

We start with a purely geometric example. Consider the motion of a point on
a circle $C$ with radius $R$ and center at $\left( x=R,y=0\right) $ on the $%
Ox$ axis. We introduce the angle $\theta \left( t\right) $ between the $Ox$
axis and the radius from the center of the circle $\left( R,0\right) $ to
the position of the point on the circle $\left( x,y\right) $. $\theta $
increases clockwise. The point starts from the origin $\left( x=0,y=0\right) 
$, $\theta \left( t=0\right) =0$. The equations are: $x\left( t\right)
=R-R\cos \left[ \theta \left( t\right) \right] $ and $y\left( t\right)
=R\sin \left[ \theta \left( t\right) \right] $. The motion on $C$ is assumed
uniform $\theta \left( t\right) =\omega t$. The average of the positions of
the point $\left( x\left( t\right) ,y\left( t\right) \right) $ up to the
current time $t$ are $\overline{x}\left( t\right) =\frac{1}{t}%
\int_{0}^{t}dt^{\prime }x\left( t^{\prime }\right) =R-\frac{1}{t}R\frac{1}{%
\omega }\sin \left( \omega t\right) $ with $\overline{x}\left( t=0\right) =0$
and $\overline{y}\left( t\right) =\frac{1}{t}\int_{0}^{t}dt^{\prime }R\sin
\left( \omega t^{\prime }\right) =\frac{R}{\omega t}\left[ 1-\cos \left(
\omega t\right) \right] $, with $\overline{y}\left( t=0\right) =0$. Clearly
the asymptotic $t\rightarrow \infty $ average position will be $\left(
R,0\right) $, which is the \textquotedblleft center\textquotedblright\ and,
after a transient phase, the motion is periodic. The speed with which the
average position moves is $d\overline{x}/dt=\left( R/\omega \right)
t^{-2}\sin \left( \omega t\right) -Rt^{-1}\cos \left( \omega t\right) $. We
note that the $y$-projection of the average, $\overline{y}\left( t\right) $,
is always positive and that for small time, $t\ll \omega ^{-1}$, the $x$%
-projected average is linear in time $\overline{x}\left( t\right) \approx
\left( R\omega ^{2}/3\right) t$. Both properties will also be found for
banana orbits.

\bigskip

The speed of the ions on this transient part of its motion is the
neoclassical drift velocity $\mathbf{v}_{Di}=\left( 1/\Omega _{ci}\right) 
\widehat{\mathbf{n}}\times \left( \mu \mathbf{\nabla }B+v_{\parallel
}^{2}\left( \widehat{\mathbf{n}}\cdot \mathbf{\nabla }\right) \widehat{%
\mathbf{n}}\right) \approx \Omega _{ci}^{-1}\left( v_{\perp
}^{2}/2+v_{\parallel }^{2}\right) /R$. Here $\Omega _{ci}$ is the ion
cyclotron frequency, $v_{\perp ,\parallel }$ are respectively the
perpendicular and the parallel velocity of the ion, $\mu =v_{\perp
}^{2}/\left( 2B\right) $ is the magnetic moment, $\widehat{\mathbf{n}}=%
\mathbf{B}/\left\vert \mathbf{B}\right\vert $ is the versor of the magnetic
field $\mathbf{B}$ and $R$ is the radius of curvature of the magnetic line.
We introduce the notation $v_{Di}$ , constant and positive.  The sign \ of
the velocity will be given according to the particular type of ion's motion
and according to the helical orientation of the magnetic field line. The
latter is given by the direction of the plasma current, which we take in the
following anti-parallel to the main magnetic field.

We consider first the trapped $\left( t\right) $ ions that have, at the
moment of ionization, parallel velocity in the same direction with the
magnetic field $\left( +\parallel \right) $ and we note that their banana is
entirely outside the magnetic surface on which the ionization has taken
place. The ion radial displacement, on the length $d=\Delta ^{t+}\approx $
half of the ion banana and only on the time interval of this displacement $%
\delta t=\Delta ^{t+}/v_{Di}$, is toward the edge.

%
\begin{figure}[h]
\includegraphics[height=6cm]{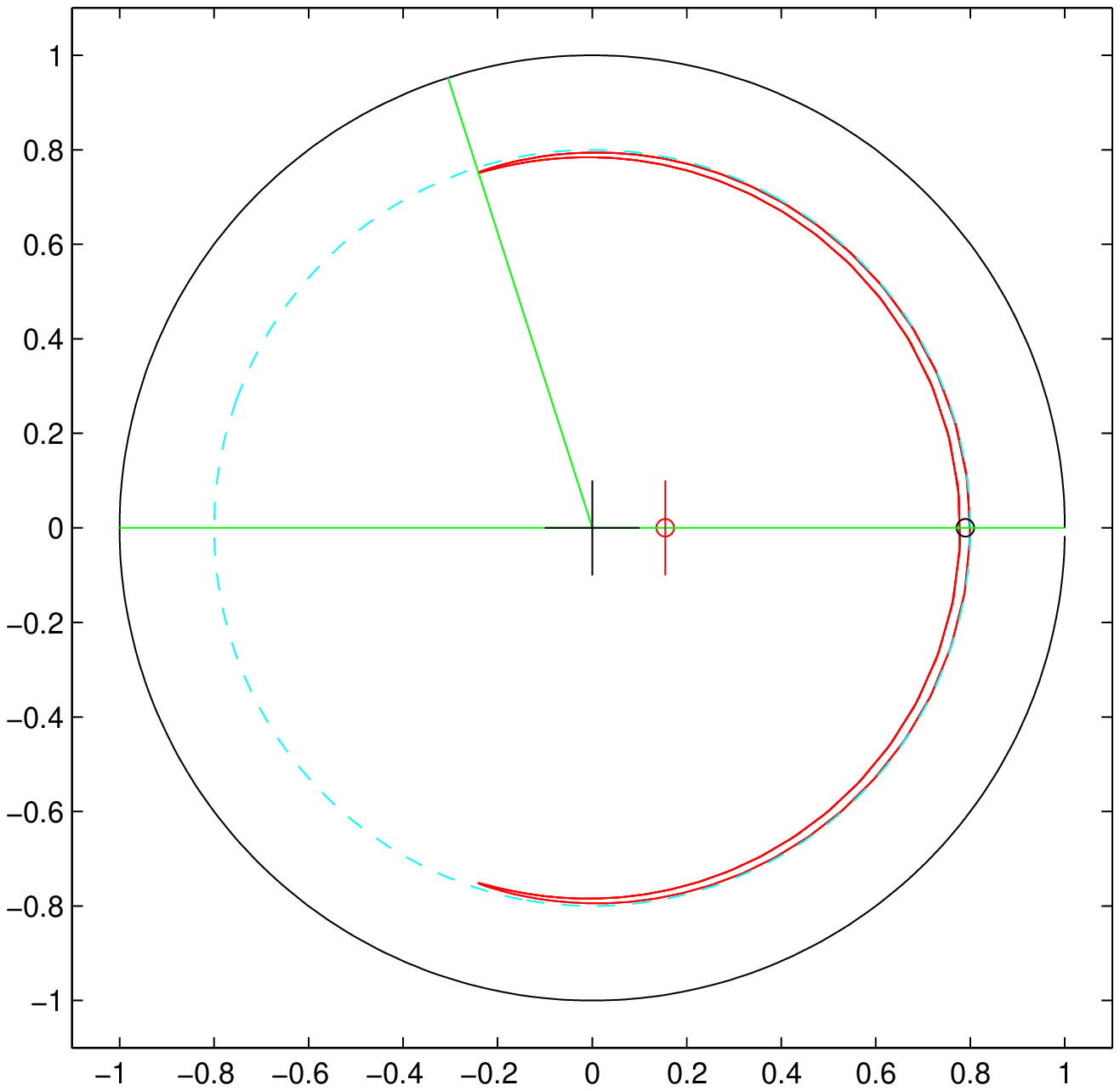}
\includegraphics[height=6cm]{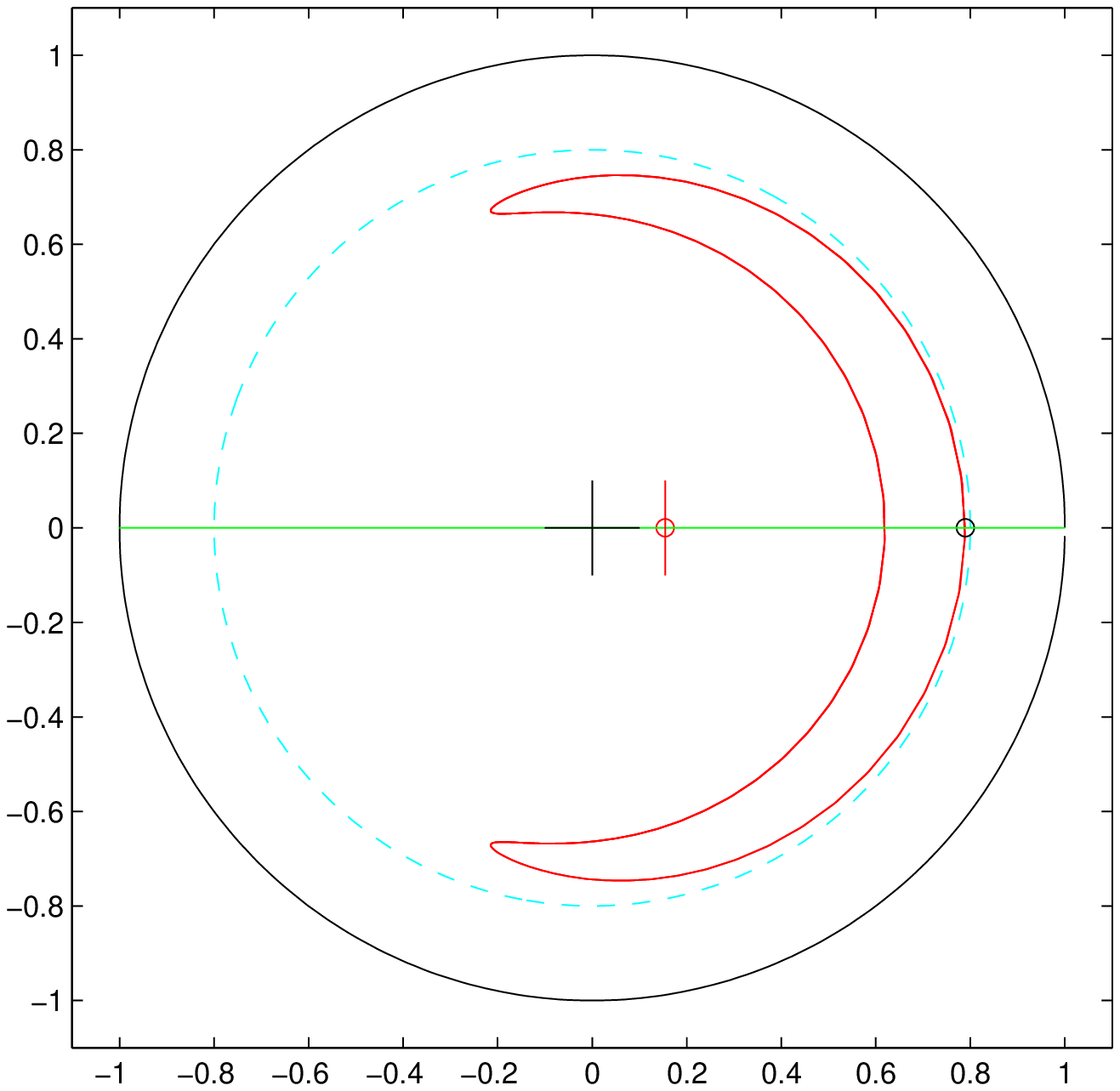}
\caption{\small{The banana orbit of an ion, which is fully inside the magnetic surface where it has
been created.  The magnetic surface is represented by the \emph{cyan} dashed
circle. The open black dot corresponds to the \emph{center} of the positions 
$r\left( t\right) $. The open red dot (and the small red vertical line)
indicate the \emph{center} of the positions on the coordinate $x\left(
t\right) $. Just for better visibility we show the orbit dilated along the
radius by an arbitrary factor $\left( \times 10\right) $.}}
\label{figure1a2}
\end{figure}
%

%
\begin{figure}[h]
\includegraphics[height=6cm]{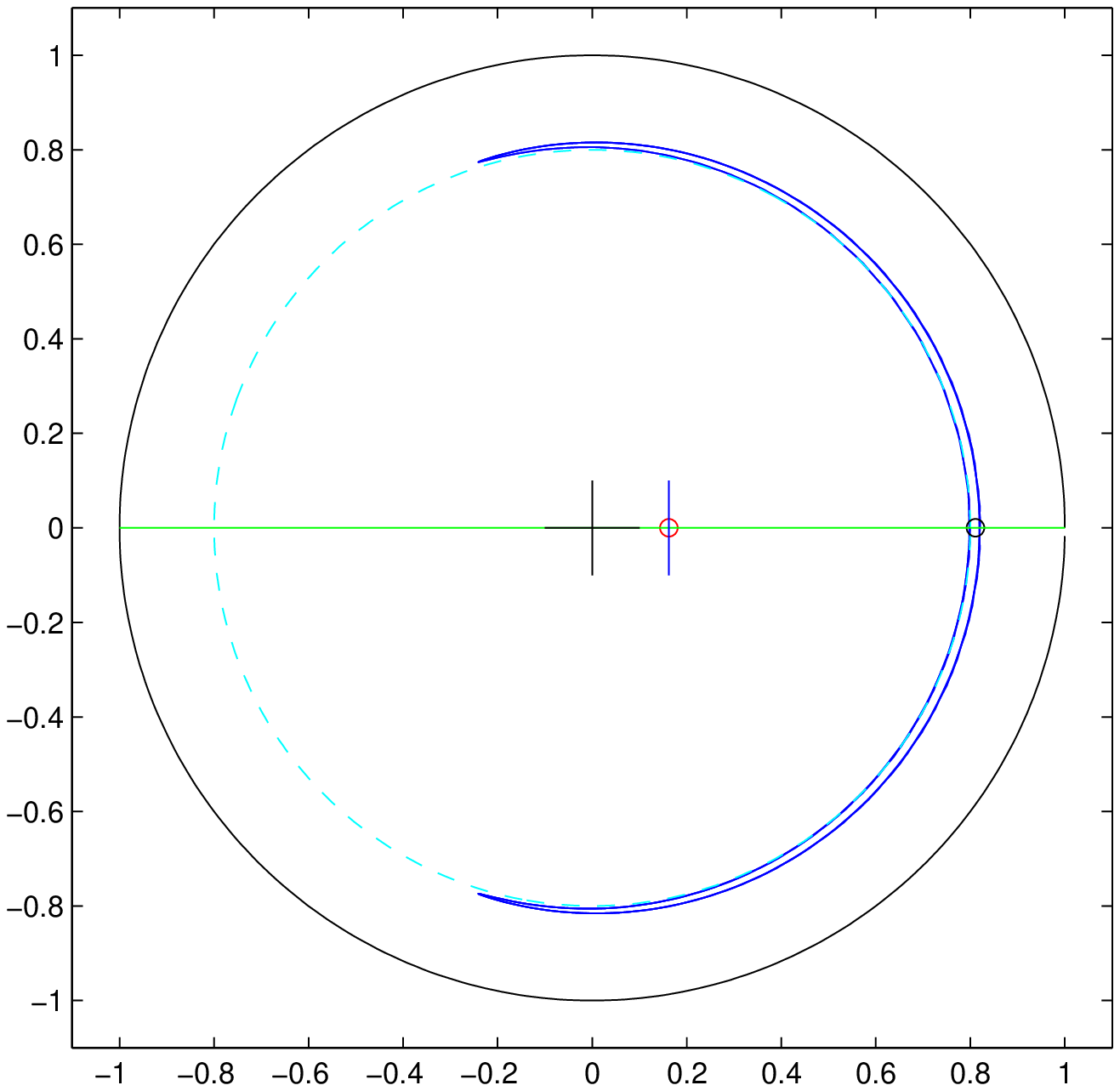}
\includegraphics[height=6cm]{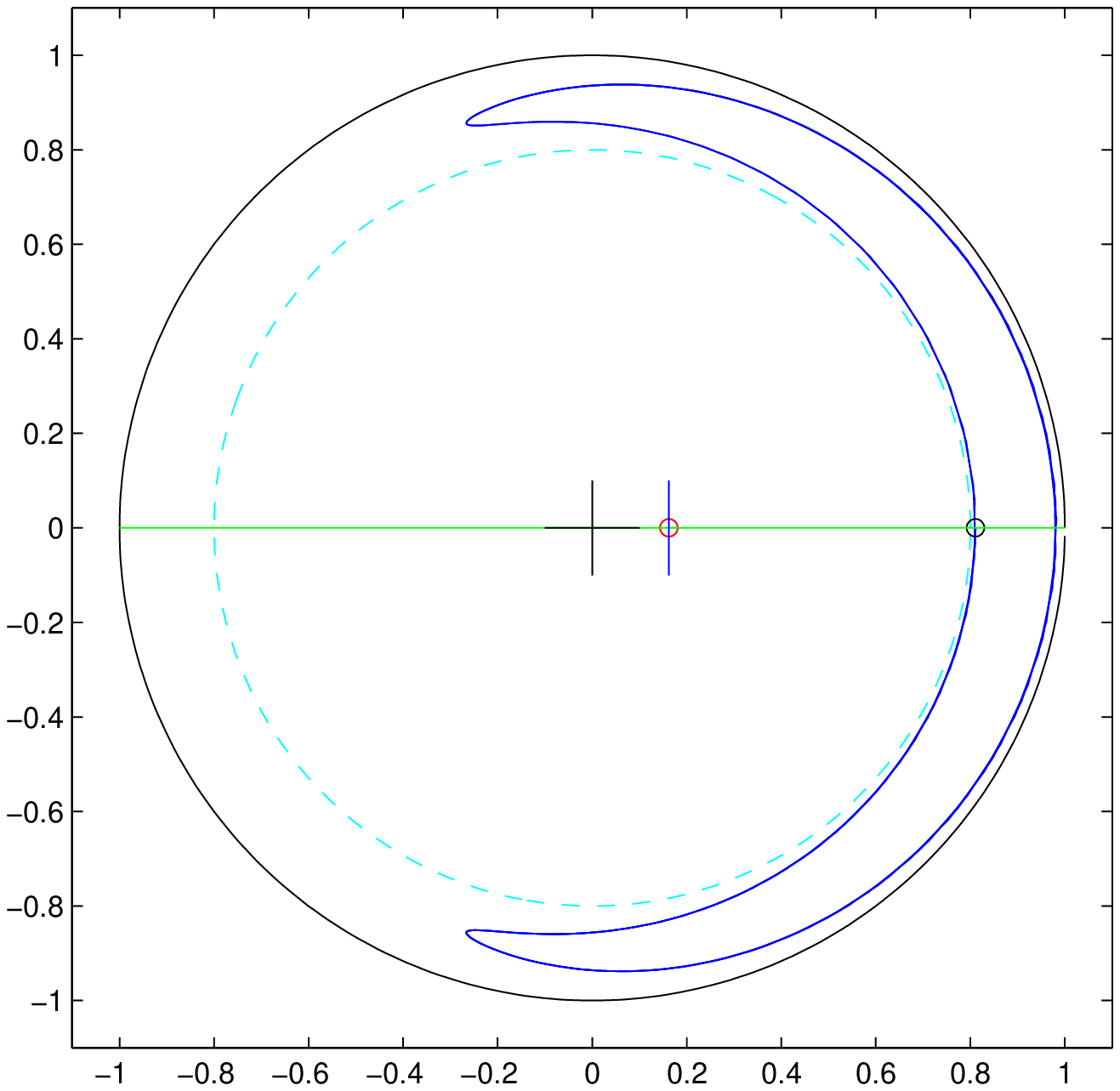}
\caption{\small{The banana orbit of an ion, which is outside the magnetic surface.  The
magnetic surface is represented by the \emph{cyan} dashed circle. The open
black dot corresponds to the \emph{center} of the positions $r\left(
t\right) $. The open red dot (and the small blue vertical line) indicate the 
\emph{center} of the positions on the coordinate $x\left( t\right) $. Just
for better visibility we show the orbit dilated along the radius by an
arbitrary factor $\left( \times 10\right) $.}}
\label{figure3a4}
\end{figure}
%

%
\begin{figure}[h]
\includegraphics[height=10cm]{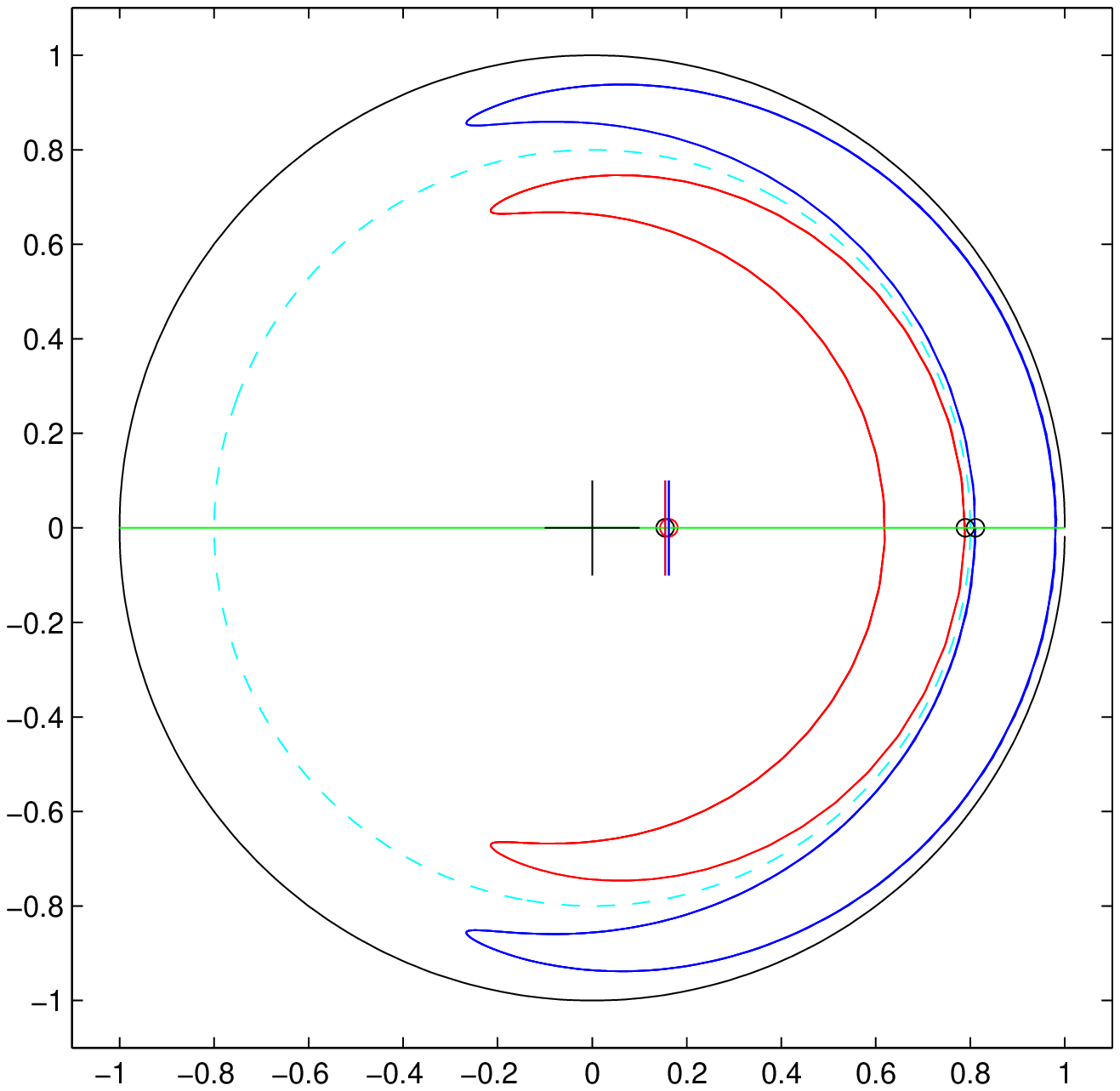}
\caption{\small{The banana orbits of two ions. One is inside (red) the other is outside
(blue) the magnetic surface. The magnetic surface is represented by the 
\emph{cyan} dashed circle. The two open black dots correspond to the \emph{%
centers} of the positions $r\left( t\right) $, for each orbit. The two open
red and black dots (and the small red and blue vertical lines) indicate the 
\emph{centers} of the positions on the coordinates $x\left( t\right) $ for
both orbits. For better visibility we multiply the radius coordinate of each
orbit by an arbitrary factor $\left( \times 10\right) $.}}
\label{figure5}
\end{figure}
%

The trapped ions that have in the point of ionization a velocity
anti-parallel to the magnetic field vector, $\left( -\parallel \right) $
have banana orbit entirely inside the magnetic surface and the transitory
displacement of the new ion, until the \textquotedblleft
center\textquotedblright\ of this banana, is towards smaller $r$ radius, $%
d=-\Delta ^{t-}<0$ (where $\Delta ^{t-}$ is defined positive).

%
\begin{figure}[h]
\includegraphics[height=6cm]{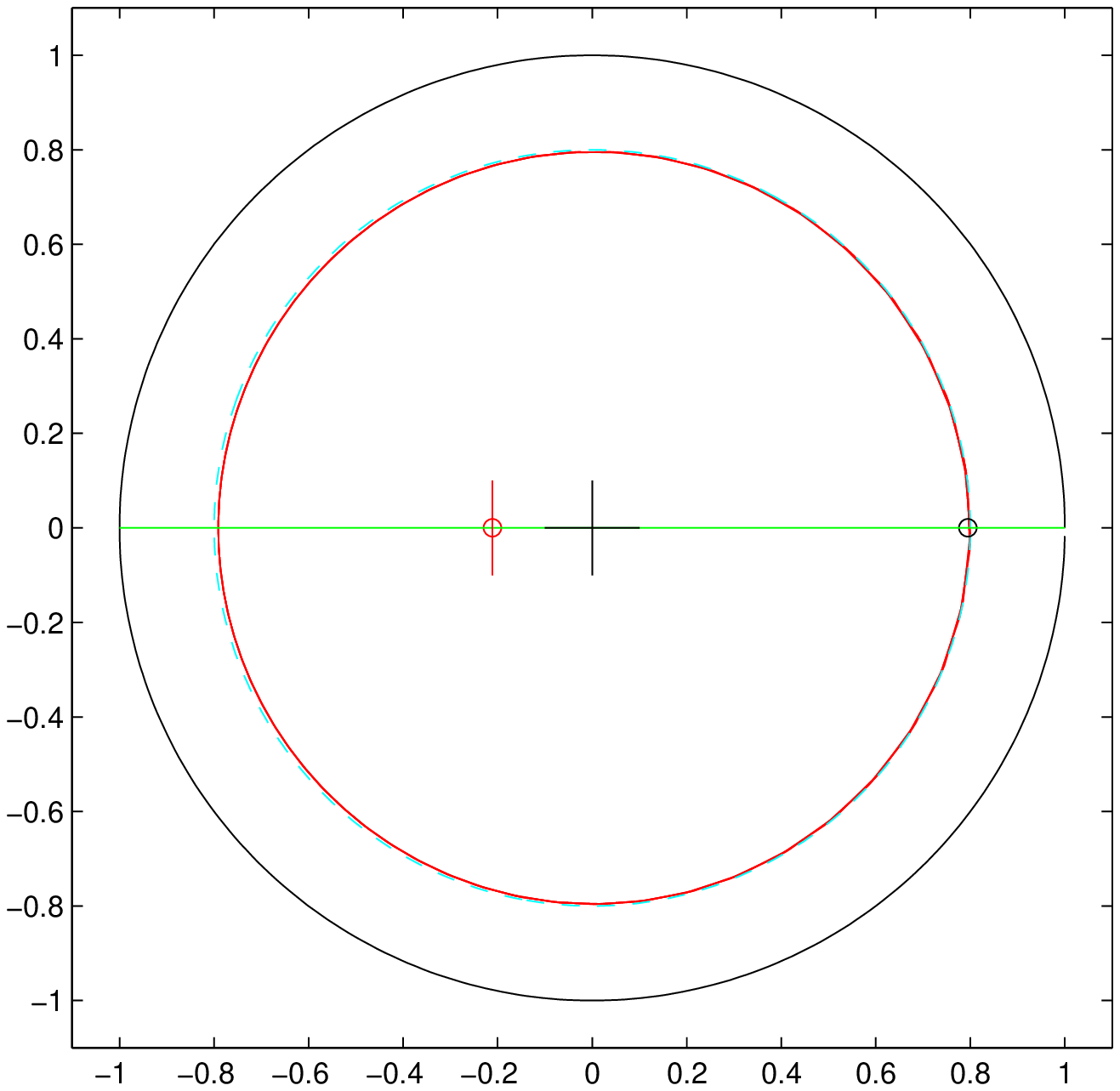}
\includegraphics[height=6cm]{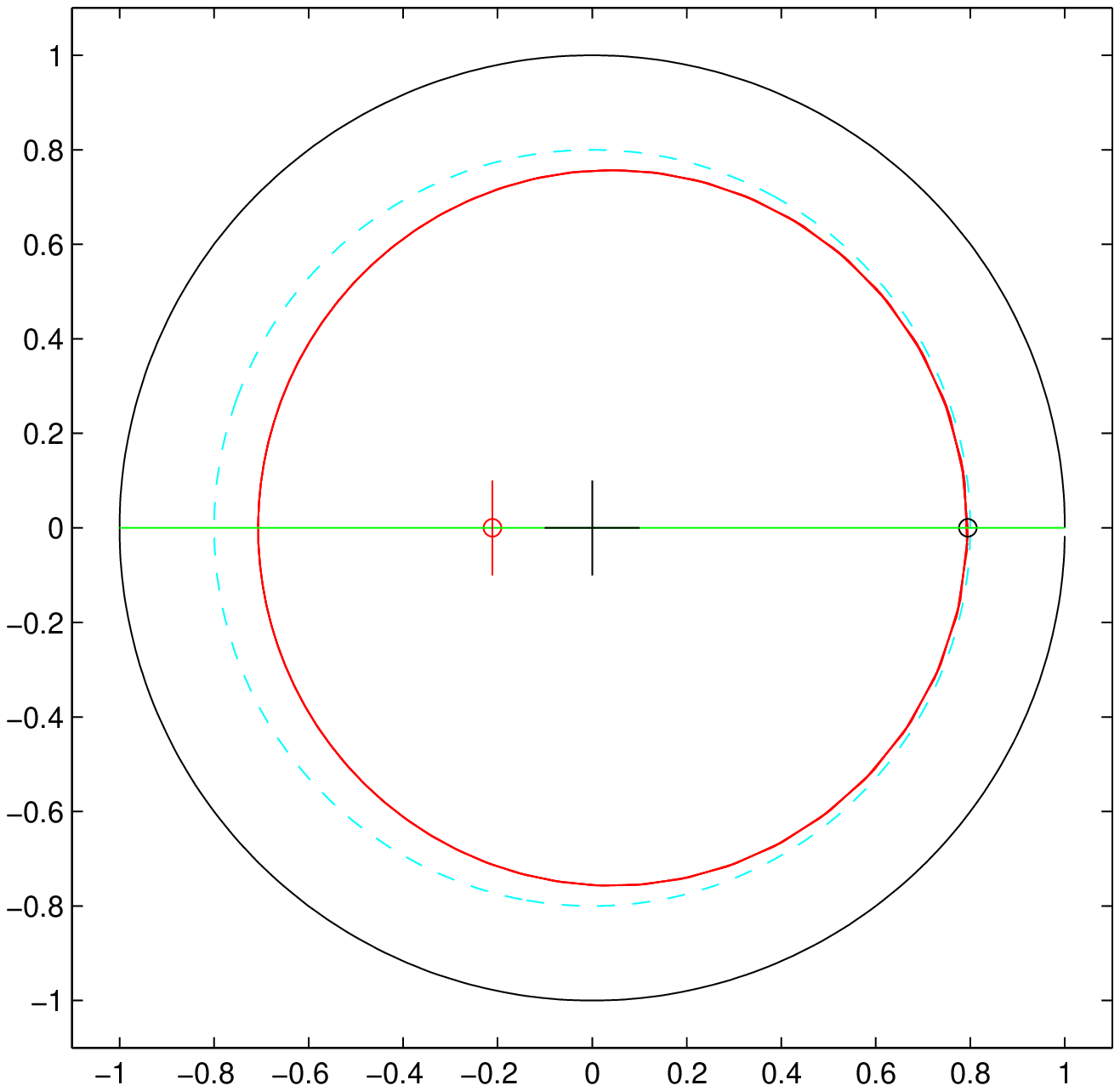}
\caption{\small{The untrapped (circulating) orbit of an ion, fuly inside the magnetic
surface. The magnetic surface is represented by the \emph{cyan} dashed
circle. The open black dot corresponds to the \emph{center} of the positions 
$r\left( t\right) $. The open red dot (and the small red vertical line)
indicate the \emph{center} of the positions on the coordinate $x\left(
t\right) $. Just for better visibility we also show the orbit dilated along
the radius by an arbitrary factor $\left( \times 10\right) $.}}
\label{figure6a7}
\end{figure}
%

In addition there are circulating $\left( c\right) $ new ions, \emph{i.e.}
on untrapped orbits. Those that have at the initial point a momentum
directed parallel to the magnetic field $\left( +\parallel \right) $ have a
circular orbit which is entirely inside the magnetic surface on which the
ionization has taken place. The effective displacement to the virtual center
is toward larger radius, $d=\Delta ^{c+}>0$ ($\Delta ^{c+}$ is defined
positive).

%
\begin{figure}[h]
\includegraphics[height=6cm]{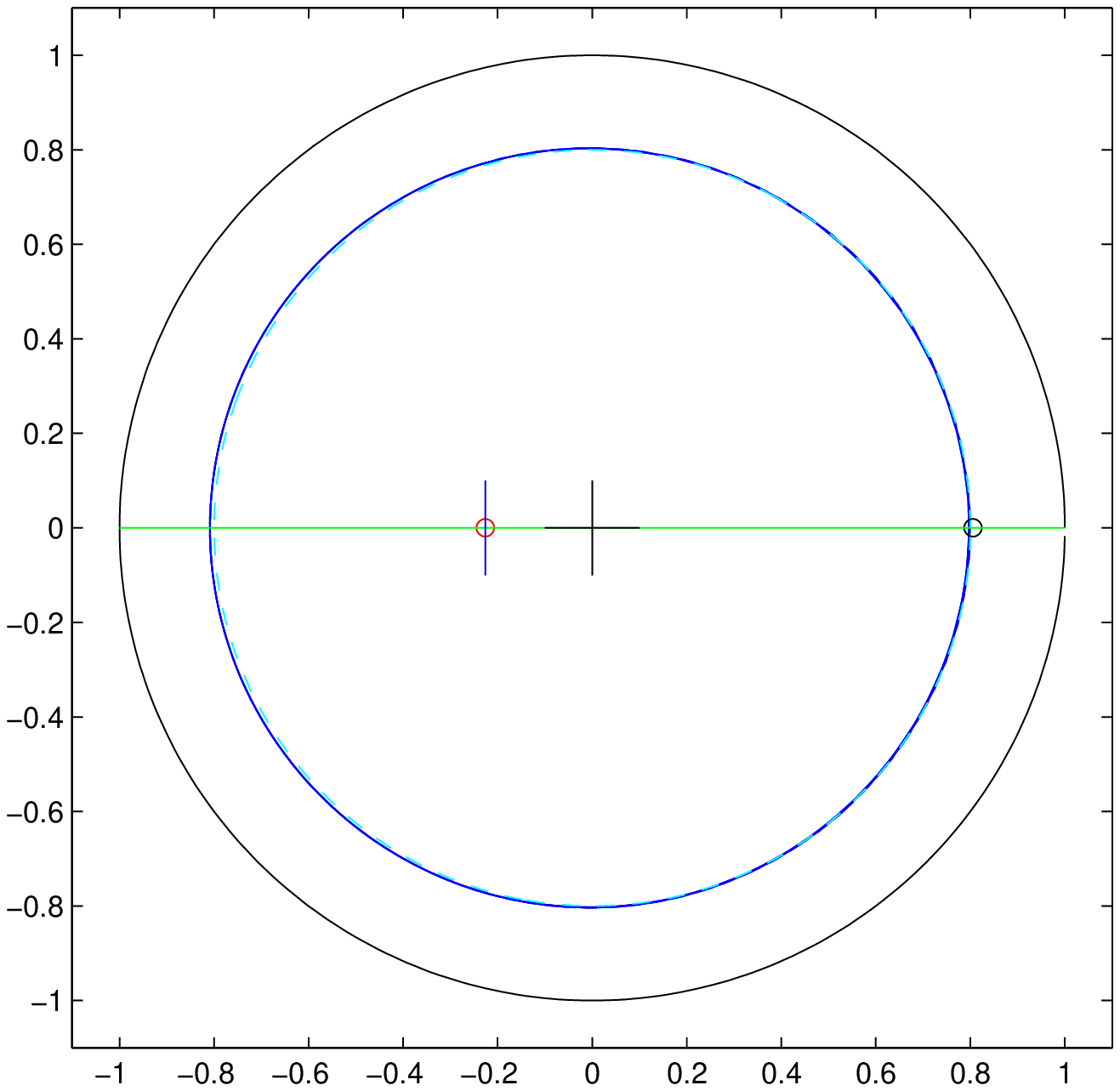}
\includegraphics[height=6cm]{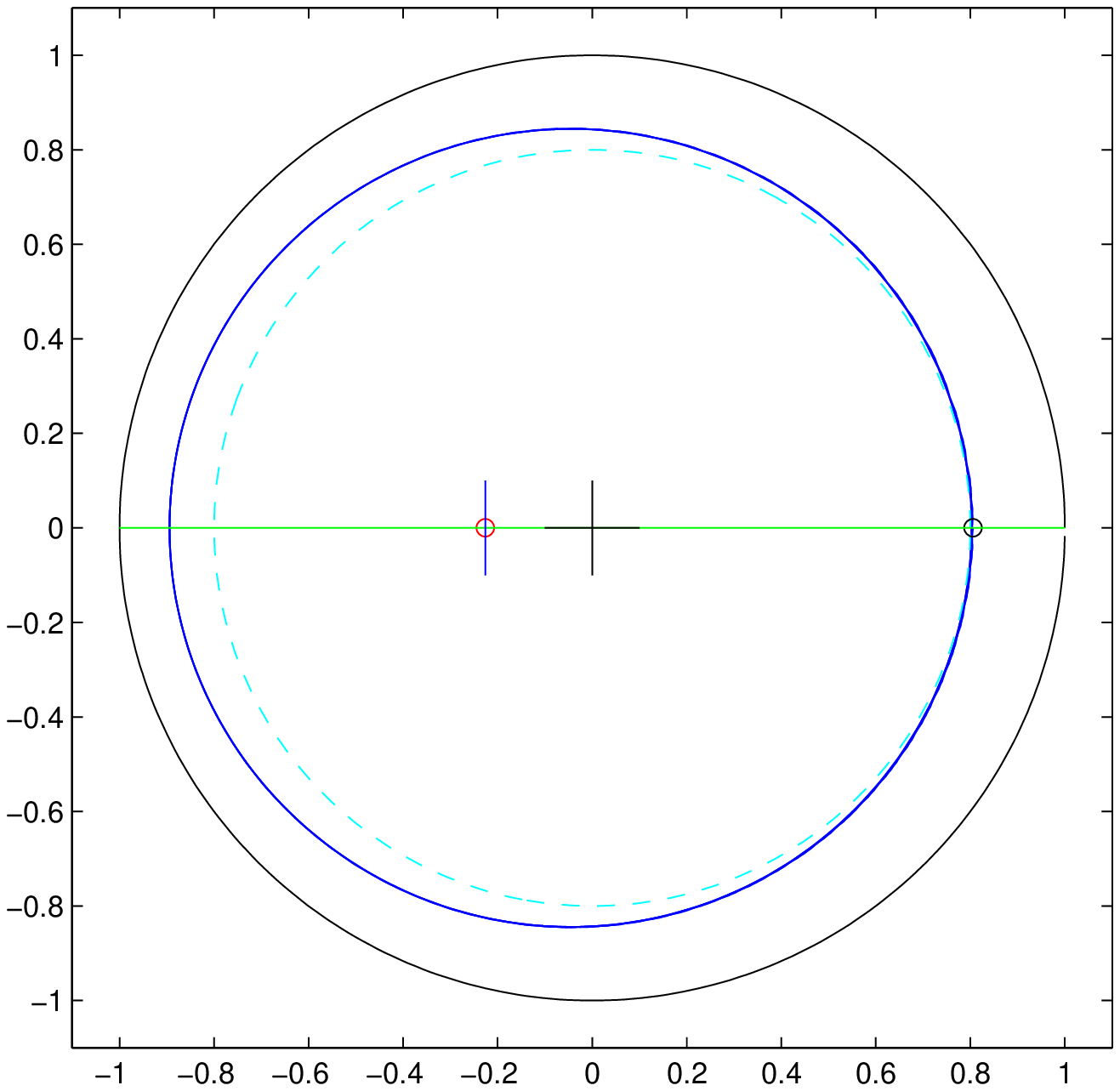}
\caption{\small{The untrapped (circulating) orbit of an ion, fully outside the magnetic
surface. The magnetic surface is represented by the \emph{cyan} dashed
circle. The open black dot corresponds to the \emph{center} of the positions 
$r\left( t\right) $. The open red dot (and the small red vertical line)
indicate the \emph{center} of the positions on the coordinate $x\left(
t\right) $. Just for better visibility we show the orbit dilated along the
radius by an arbitrary factor $\left( \times 10\right) $.}}
\label{figure8a9}
\end{figure}
%

%
\begin{figure}[h]
\includegraphics[height=10cm]{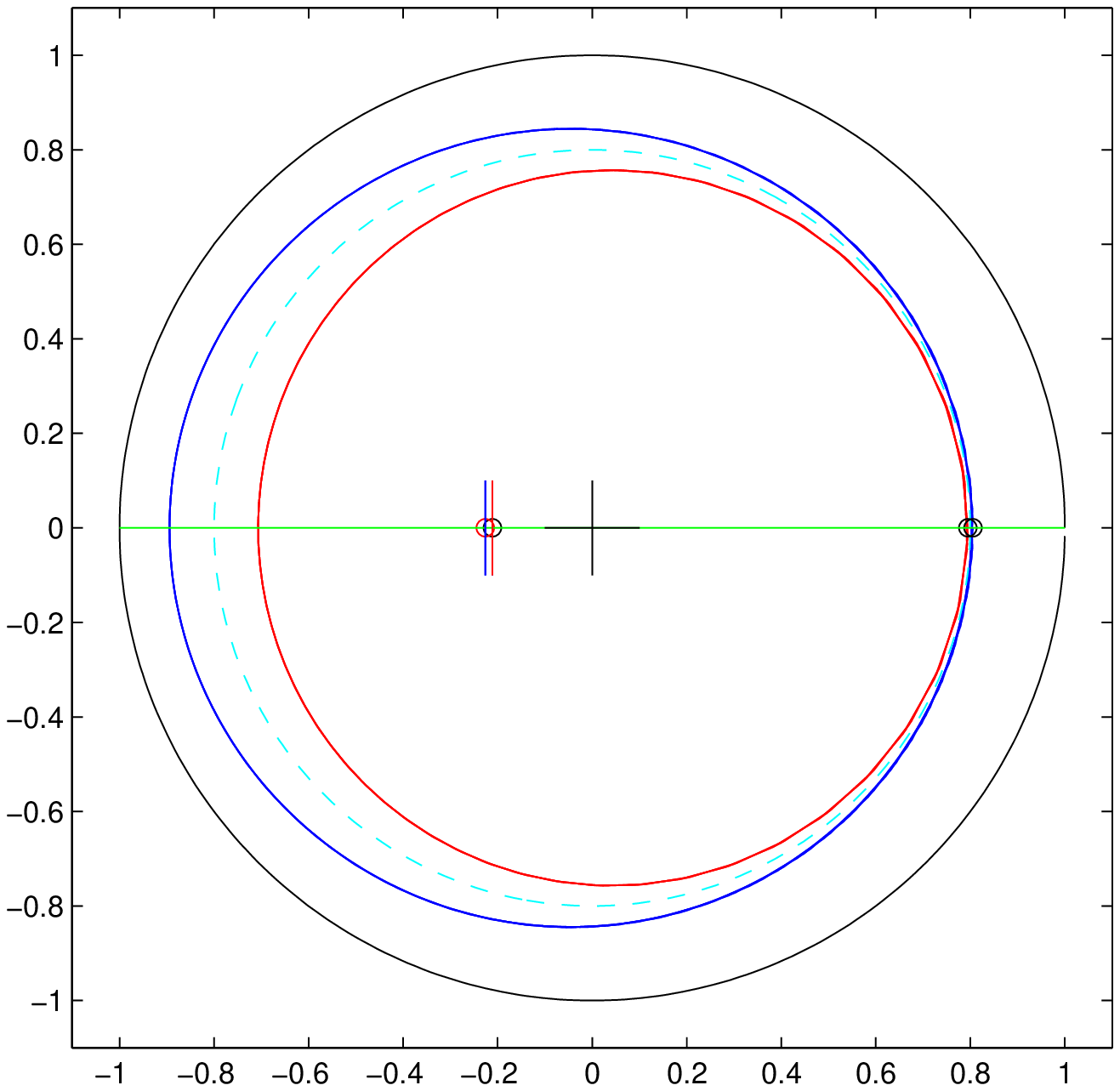}
\caption{\small{The untrapped (circulating) orbits of two ions fully enclosing (blue)
respectively fully inside (red) the magnetic surface. The magnetic surface
is represented by the \emph{cyan} dashed circle. The two open black dots
correspond to the \emph{centers} of the positions $r\left( t\right) $, for
each orbit. The two open red and black dots (and the small red and blue
vertical lines) indicate the \emph{centers} of the positions on the
coordinates $x\left( t\right) $ for both orbits. For better visibility we
multiply the radius coordinate of each orbit by an arbitrary factor $\left(
\times 10\right) $.}}
\label{figure10}
\end{figure}
%

The last type consists of ions that are circulating and with their velocity
at the initial point anti-parallel to the magnetic field $\left( -\parallel
\right) $. For them the closed orbit fully includes the magnetic surface
and\ it is equivalent to the displacement of the average position of the new
ion to smaller $r$, toward the magnetic axis, $d=-\Delta ^{c-}<0$ ($\Delta
^{c-}$ is defined positive).

%
\begin{figure}[h]
\includegraphics[height=8cm]{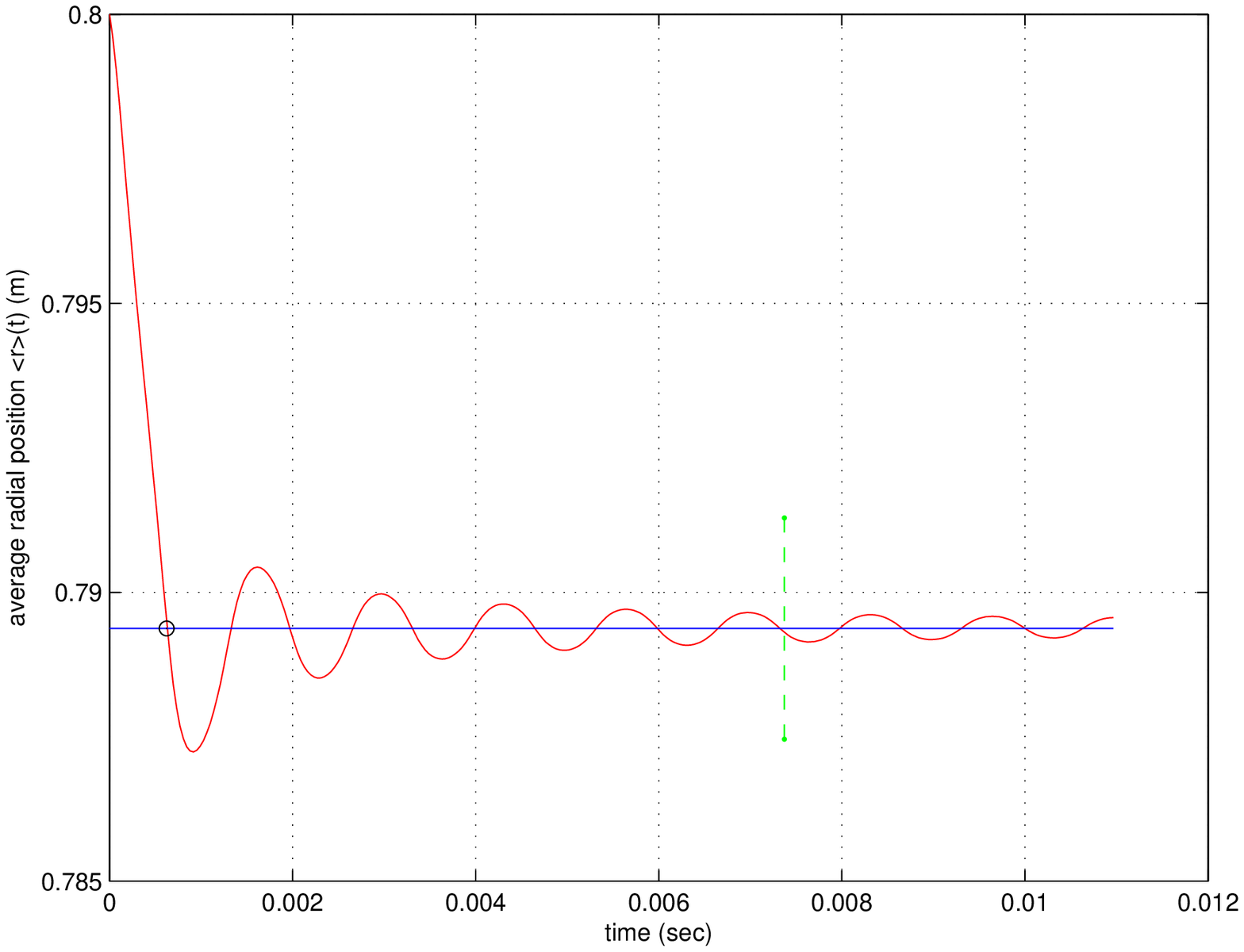}
\caption{\small{The time evolution of the average radial position $\overline{r}\left(
t\right) $ for an ions whose orbit is fully inside the magnetic surface. The
straight horizontal line is the asymptotic average position $r_{asymp}$ (the
\textquotedblleft center\textquotedblright\ of the periodic orbit). The
dashed vertical line in the asymptotic range is used to select the part of
the trajectory to determine $r_{asymp}$, far from the transient. The open
dot at the intersection of $r=r_{asymp}$ and $r=\overline{r}\left( t\right) $
determines the approximative end of the transitory regime, $\tau _{asymp}$ .
The drift velocity is estimated as $v_{Di}=\frac{r_{asymp}-r_{ini}}{\tau
_{asymp}}$, where $r_{ini}$ is the radial position at ionization.
}}
\label{figure11}
\end{figure}
%

%
\begin{figure}[h]
\includegraphics[height=8cm]{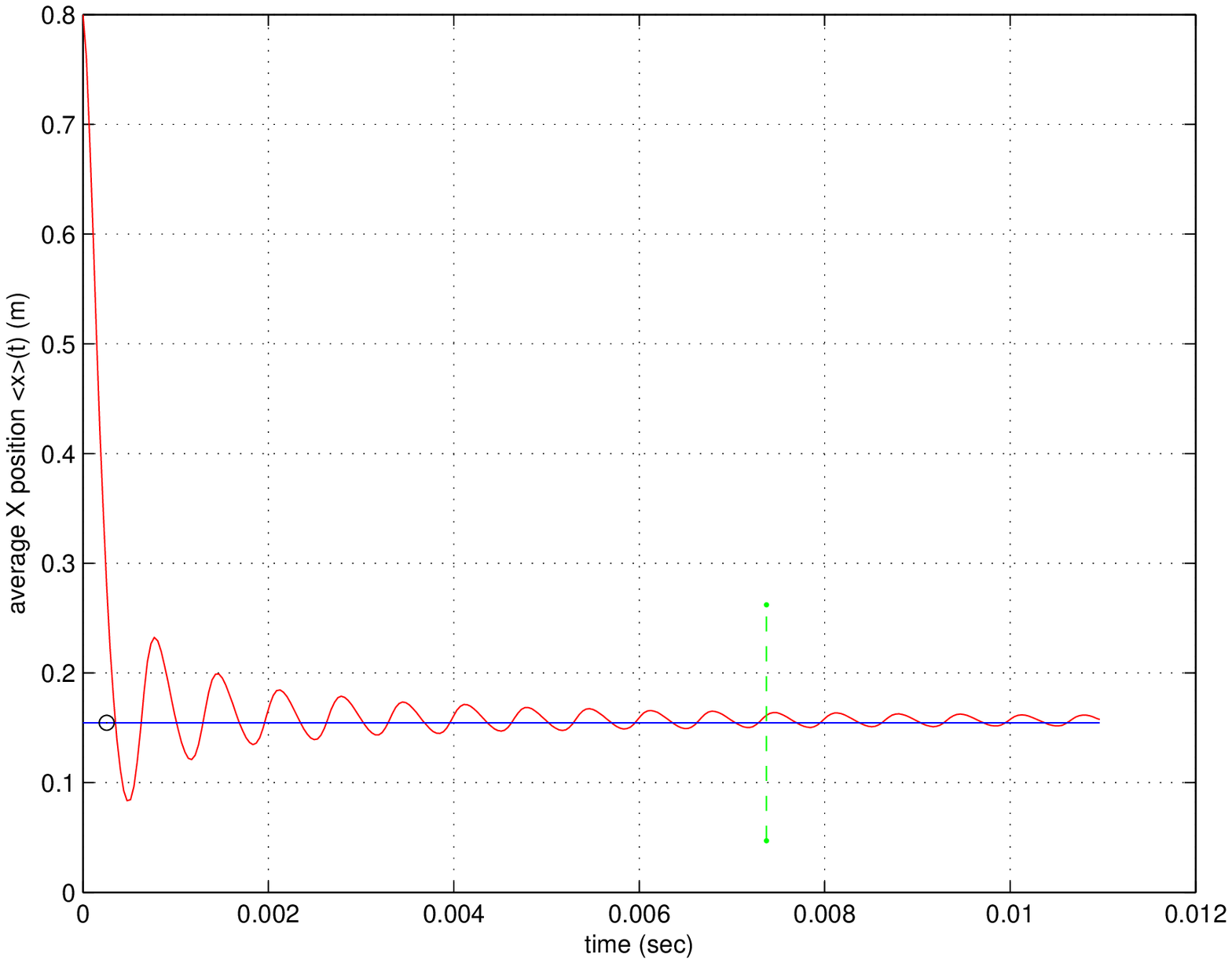}
\caption{\small{The time evolution of the average $x$-projection of the position, \emph{i.e.}
$\overline{x}\left( t\right) $,  for an ions whose orbit is fully inside the
magnetic surface. The details are the same as for Figure \ref{figure11}.
}}
\label{figure12}
\end{figure}
%

%
\begin{figure}[h]
\includegraphics[height=8cm]{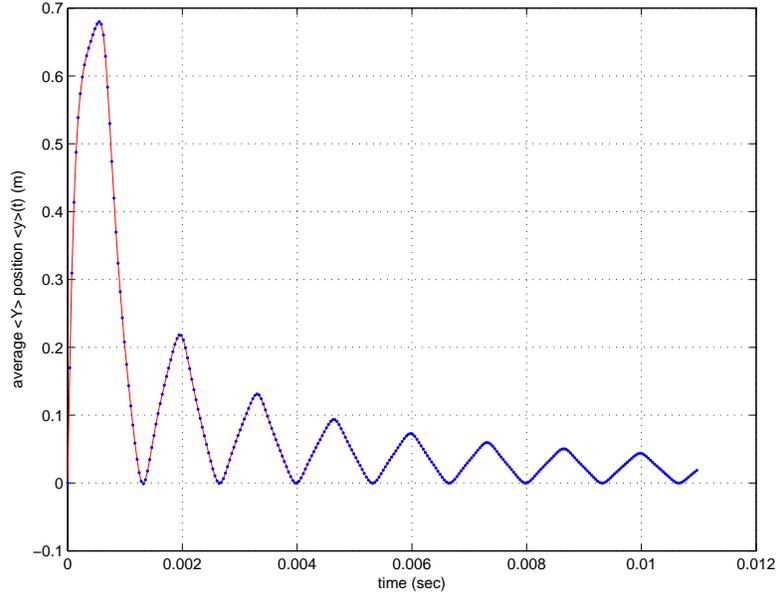}
\caption{\small{The time evolution of the average $y$-projection of the position, \emph{i.e.}
$\overline{y}\left( t\right) $, for an ions whose orbit is fully inside the
magnetic surface. We note that it remains positive at all times.
}}
\label{figure13}
\end{figure}
%

%
\begin{figure}[h]
\includegraphics[height=8cm]{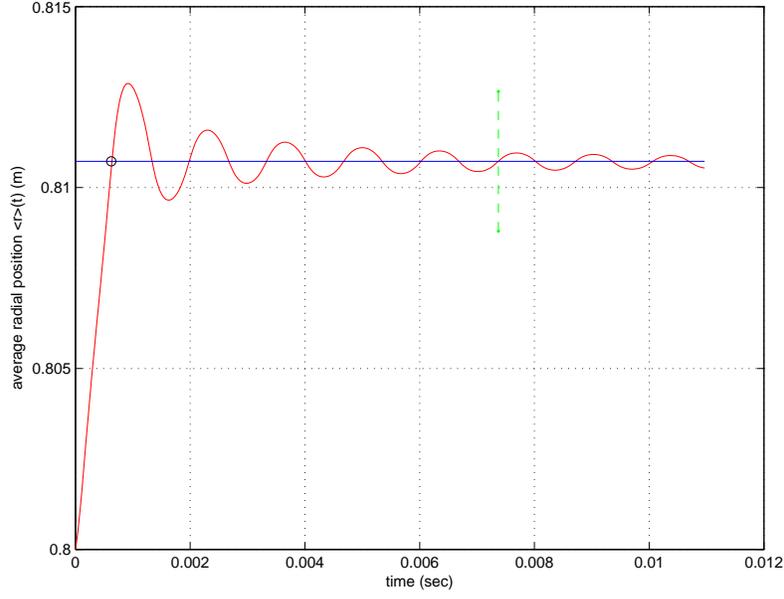}
\caption{\small{The time evolution of the average radial position $\overline{r}\left(
t\right) $ for an ions whose orbit is outside the magnetic surface. The
details are the same as for Figure \ref{figure11}.
}}
\label{figure14}
\end{figure}
%

%
\begin{figure}[h]
\includegraphics[height=8cm]{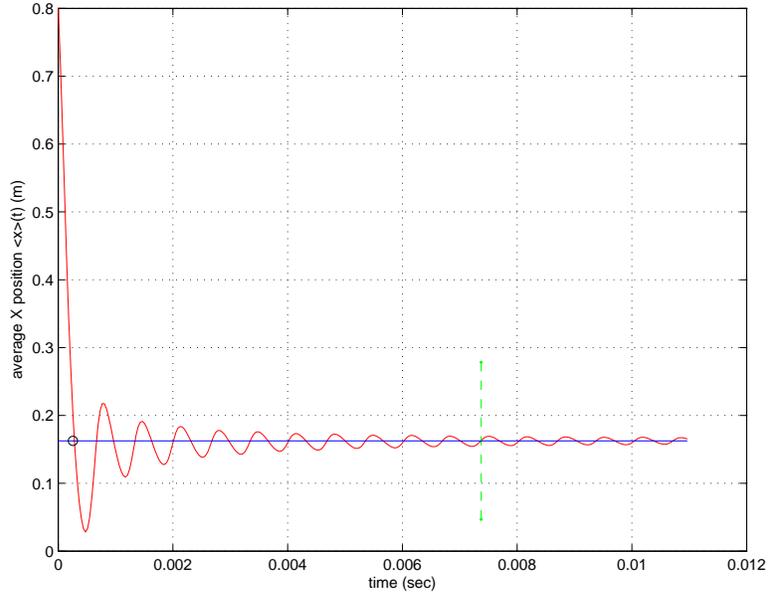}
\caption{\small{The time evolution of the average $x$-projection of the position, \emph{i.e.}
$\overline{x}\left( t\right) $, for an ions whose orbit is outside the
magnetic surface. The details of the Figure are the same as for Figure \ref{figure11}.
}}
\label{figure15}
\end{figure}
%

%
\begin{figure}[h]
\includegraphics[height=8cm]{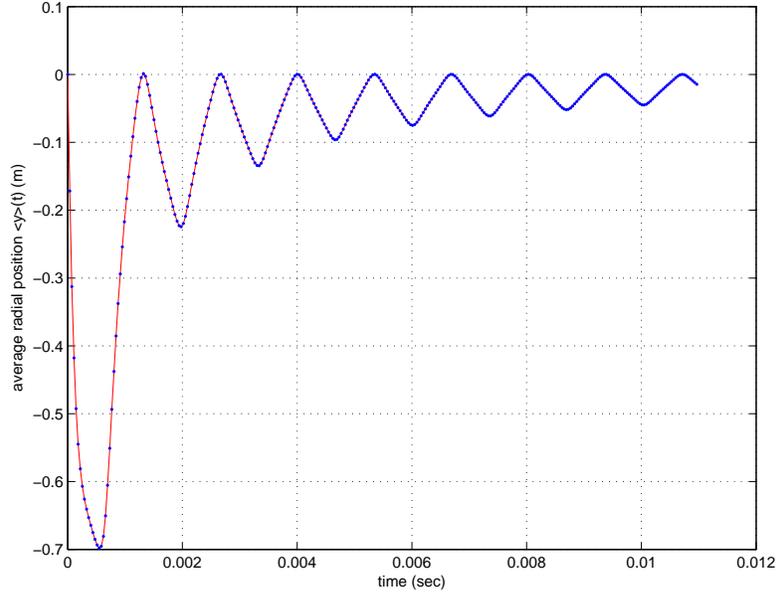}
\caption{\small{The time evolution of the average $y$-projection of the position, \emph{i.e.}
$\overline{y}\left( t\right) $, for an ions whose orbit is outside the
magnetic surface.
}}
\label{figure16}
\end{figure}
%

%
\begin{figure}[h]
\includegraphics[height=8cm]{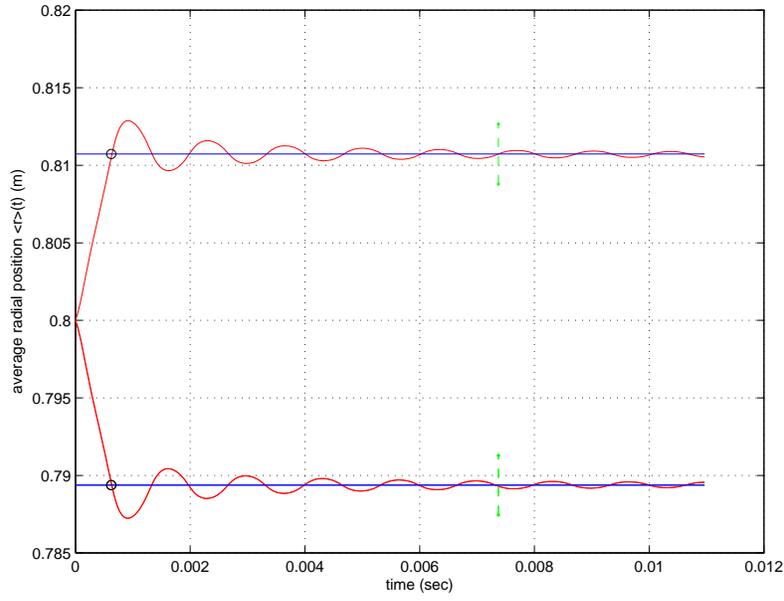}
\caption{\small{This plot represents the time evolutions of the average positions on $r$, 
\emph{i.e.} $\overline{r}\left( t\right) $, for the trapped particles whose
orbits are fully inside (lower curve) and respectively outside (the
upper curve) the magnetic surface. The details are the same as in Figure \ref{figure11}.
}}
\label{figure17}
\end{figure}
%

%
\begin{figure}[h]
\includegraphics[height=8cm]{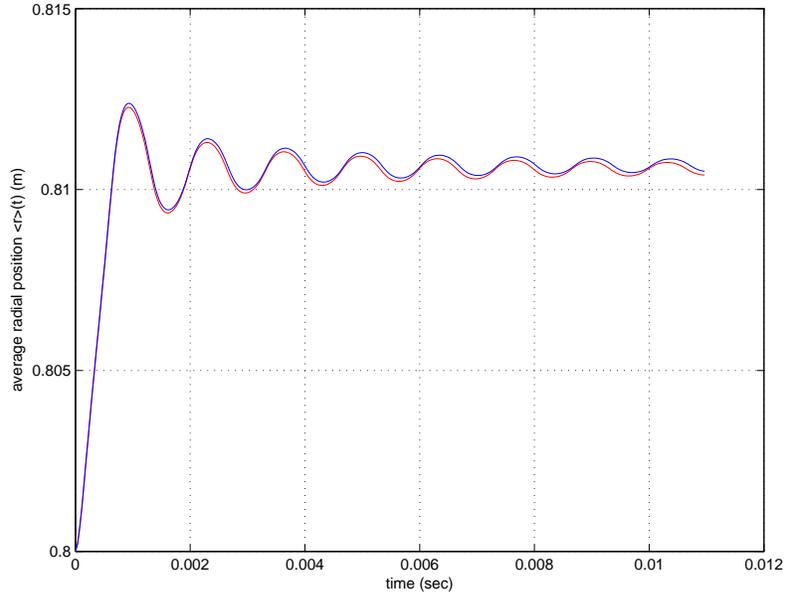}
\caption{\small{The time variation of the average positions on $r$, \emph{i.e.} $\overline{r}%
\left( t\right) $, for the two types of trapped orbits are represented here
with the purpose to give an idea of the smallness of the difference in the
average radial displacements. One of the curves (the lower one in Figure \ref{figure17})
has been reversed in order to make easier the comparison.
}}
\label{figure18}
\end{figure}
%

%
\begin{figure}[h]
\includegraphics[height=8cm]{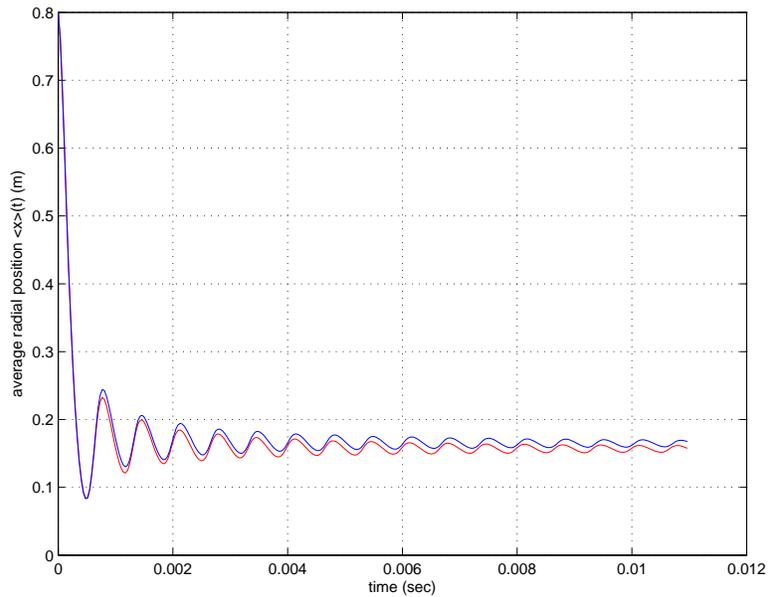}
\caption{\small{The same as Figure \ref{figure18} but for the $x$-projection.
}}
\label{figure19}
\end{figure}
%

%
\begin{figure}[h]
\includegraphics[height=8cm]{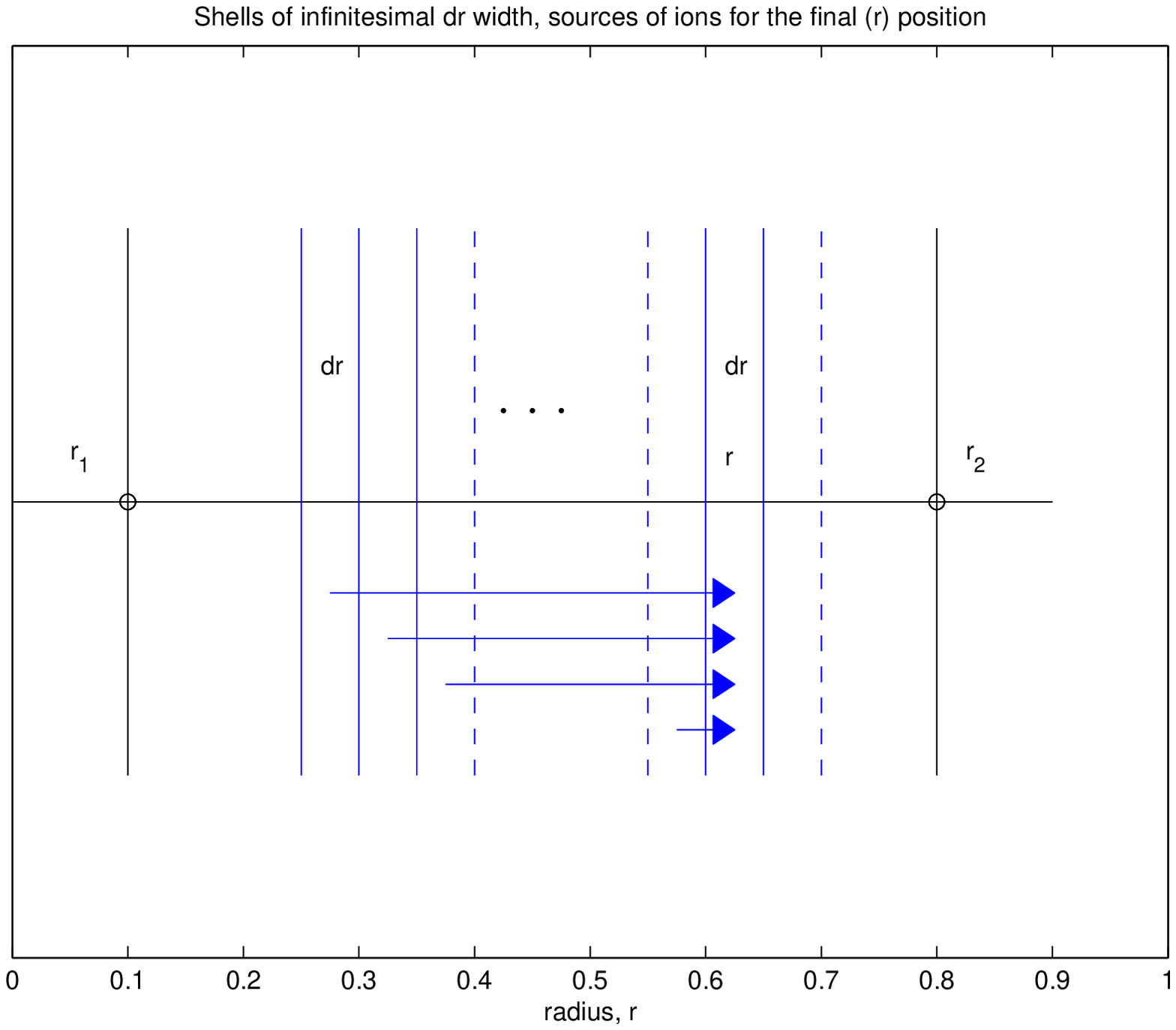}
\caption{\small{A simple representation of the fluxes that are traversing the surface placed
at the radius $r$, coming from various infinitesimal regions $\delta r$. The
sum over these contributions is the integral that defines the current
density in $\left( r,t\right) $ acoording to the text. The positions $r_{1}$
and $r_{2}$ are the limits of the domain of ionization.
}}
\label{figure20}
\end{figure}
%

The four contributions to the current through $\left( r,t\right) $, from the
two pairs (\emph{i.e.}$\left( \pm \parallel \right) $-trapped, respectively $%
\left( \pm \parallel \right) $-circulating) appear to be small and they
partly compensate, having opposite signs. However, even the small remaining
(effective) radial displacement, multiplied with the rate of generation of
new ions, leads to a significant radial current. There is another aspect:
the centers of the two bananas discussed above [$\left( +\parallel \right) $
and $\left( -\parallel \right) $] are spatially separated and we can
associate to them the full population of new ions generated in those two
positions. The rate of ionization, expressed in number of ions per $m^{3}$
per second has a significant radial variation, both for pellets and for
gas-puff: the rate of generation of new ions may differ substantially
between two radial positions, even if they have small separation, of the
order of centimeters. The two banana centers are separated by a distance $%
\Delta ^{t+}+\Delta ^{t-}$ which is in this range and suggests that the
radial variation of the rate of ionization is an important factor.

For any point $r$, the four contributions combine into a single,
short-lived, finite-spatial size - event of ion charge displacement, \emph{%
i.e.} a current. These events occur everywhere within the radial segment of
ionization and for all the time when there are still neutral atoms to be
ionized.

\bigskip

\subsection{Calculation of the flux of electric charge of the new ions into
a point $\left( r,t\right) $}

The pellet contains a total number of particles (neutral atoms) $N_{t}\ $\
and the ionization takes place in the toroidal volume between the surfaces $%
r_{1}$ and $r_{2}$, $V_{t}=2\pi ^{2}R\left( r_{2}^{2}-r_{1}^{2}\right) $ .
The \emph{total} time for ionization of the particles of the pellet is $\tau
^{ioniz}$. The rate of generation of ions per unit volume and per second has
the average magnitude%
\begin{equation}
\overset{\cdot }{n}^{ioniz}\left( r,t\right) \sim \frac{N_{t}}{\tau ^{ioniz}}%
\frac{1}{V_{t}}\   \label{npoint}
\end{equation}%
The rate of ionization has strong spatial and temporal variation and we find
more convenient to express it as%
\begin{equation}
\overset{\cdot }{n}^{ioniz}\left( r,t\right) =\overset{\cdot }{n}%
_{0}^{ioniz}S\left( r,t\right) \ \left( \frac{ions}{m^{3}s}\right)
\label{sxt}
\end{equation}%
where $S\left( r,t\right) \geq 0$ is by definition a nondimensional function
representing the space-time variation of the ionization rate. The maximum of 
$S$ is $1$ and its variable is normalized, $x\equiv r/a$. The constant
factor $\overset{\cdot }{n}_{0}^{ioniz}$(of order $\sim 10^{23}\
ions/m^{3}/s $) is the physical quantity that carries information on the
average rate of ionization and is taken from experimental observation.\ The
factors $\overset{\cdot }{n}_{0}^{ioniz}$ and $S$ are constraint by the
condition%
\begin{equation}
\int_{0}^{\tau ^{ioniz}}d\tau \int dV\overset{\cdot }{n}^{ioniz}\left(
r,t\right) =N_{t}  \label{nt}
\end{equation}%
For example, taking for simplicity a source that is constant in time for the
entire duration of the ionization $\left[ 0,\tau ^{ioniz}\right] $, we have%
\begin{equation}
\overset{\cdot }{n}_{0}^{ioniz}\tau ^{ioniz}\int_{r_{1}}^{r_{2}}S\left(
r\right) 4\pi ^{2}Rrdr=N_{t}  \label{constr}
\end{equation}

The finite volume $V_{t}$ is divided into toroidal shells of infinitesimal
width $dr$ on the minor radius, placed at the position $r$ , with volume $%
V_{dr}=Adr=4\pi ^{2}Rr\left( dr\right) $ where $A$ is the surface area. We
fix a reference position $r\in \left[ r_{1},r_{2}\right] $, and calculate
the net flux of ions that traverses the surface $A=4\pi ^{2}Rr$ at $r$.
Consider the bunch of \ ions that are produced in a time $\delta \tau $,
filling the elementary shell (denoted $D$) situated at a distance $\rho $
from $r$. Their number is $\overset{\cdot }{n}^{ioniz}\left( r-\rho ,t-\frac{%
\rho }{v_{Di}}\right) \times V_{dr}\delta \tau $. The ions from $D$ that are
generated at time $t-\rho /v_{Di}$ travel with constant velocity $v_{Di}$
and arrive in $r$ at time $t$ , traversing the surface $A$ in a time $\delta
\tau $. The flux at $\left( r,t\right) $, in $\left( ions/m^{2}/s\right) $,
is%
\begin{equation}
\Gamma \left( r,t\right) =\overset{\cdot }{n}^{ioniz}\left( r-\rho ,t-\frac{%
\rho }{v_{Di}}\right) V_{dr}\delta \tau \frac{1}{A}\frac{1}{\delta \tau }=%
\overset{\cdot }{n}^{ioniz}\left( r-\rho ,t-\frac{\rho }{v_{Di}}\right) dr\ 
\label{Gamma}
\end{equation}%
The maximum distance of travel on which a new ion generates a current is
from the point of ionization until the \textquotedblleft
center\textquotedblright\ of the periodic orbit, $\rho _{\max }=\Delta $
(which is one of $\Delta ^{t\pm }$). Between $r-\Delta $ and $r$ there are
many infinitesimal shells, at distance $\rho ^{\prime }$ from $r$ ($\Delta
\geq \rho ^{\prime }\geq 0$). The ions created in these intermediate cells
arrive at time $t$ in $r$ if they are generated at $t-\rho ^{\prime }/v_{Di}$%
. Summing these partial contributions Eq.(\ref{Gamma}), the origins of which
are in the interval $\left[ r-\Delta ,r\right] $ it results $\Gamma ^{\Delta
}\left( r,t\right) $%
\begin{eqnarray}
\Gamma ^{\Delta }\left( r,t\right) &=&\int_{0}^{\Delta }\overset{\cdot }{n}%
^{ioniz}\left( r-\rho ^{\prime },t-\frac{\rho ^{\prime }}{v_{Di}}\right)
d\rho ^{\prime }  \label{GammaDelta} \\
&\approx &\overset{\cdot }{n}_{0}^{ioniz}\int_{0}^{\Delta }d\rho ^{\prime }%
\left[ S\left( x,t\right) -\left( \frac{\partial S}{\partial r}\right) \rho
^{\prime }-\left( \frac{\partial S}{\partial t}\right) \frac{\rho ^{\prime }%
}{v_{Di}}\right]  \notag \\
&=&\overset{\cdot }{n}_{0}^{ioniz}\left[ S\left( r,t\right) \Delta -\frac{1}{%
2}\left( \frac{\partial S}{\partial r}\right) \Delta ^{2}-\frac{1}{2}\left( 
\frac{\partial S}{\partial t}\right) \frac{\Delta ^{2}}{v_{Di}}\right] 
\notag
\end{eqnarray}%
where the derivatives of $S$ are calculated in $\left( r,t\right) $. This
flux must be multiplied with the fractional number representing how many of
the new ions will settle on trapped, respectively circulating orbits. We
take the approximative values $\sqrt{\varepsilon }$ and respectively $1-%
\sqrt{\varepsilon }$. In addition, we assume that a fraction of $1/2$ new
ions have parallel, respectively anti-parallel initial velocities.

\subsection{The current density in $\left( r,t\right) $}

Taking into account the four type of ion's orbits, we use Eq.(\ref%
{GammaDelta}) to estimate the flows of new ions coming in, or leaving, the
point $\left( r,t\right) $. The contributions from neighbor points are%
\begin{eqnarray}
&&\Gamma _{in}^{t\left( +\parallel \right) }\left( r,t\right) =\frac{1}{2}%
\sqrt{\varepsilon }\int_{0}^{\Delta ^{t+}}\overset{\cdot }{n}^{ioniz}\left(
r-\rho ^{\prime },t-\frac{\rho ^{\prime }}{v_{di}}\right) d\rho ^{\prime }
\label{tplus} \\
&=&\frac{1}{2}\sqrt{\varepsilon }\overset{\cdot }{n}_{0}^{ioniz}\left[
S\left( r,t\right) \Delta ^{t+}-\frac{1}{2}\left( \frac{\partial S}{\partial
r}\right) \left( \Delta ^{t+}\right) ^{2}-\frac{1}{2}\left( \frac{\partial S%
}{\partial t}\right) \frac{\left( \Delta ^{t+}\right) ^{2}}{v_{Di}}\right]  
\notag
\end{eqnarray}%
This is the number of ions that are trapped and had an initial velocity
parallel with $\mathbf{B}$. They are produced in $\left( r-\rho ^{\prime },t-%
\frac{\rho ^{\prime }}{v_{Di}}\right) $, summed over $\left[ r-\Delta ^{t+},r%
\right] $, and flow toward $\left( r,t\right) $ from the left (\emph{i.e.}
their current is positive).%
\begin{eqnarray}
&&\Gamma _{in}^{t\left( -\parallel \right) }\left( r,t\right) =\frac{1}{2}%
\sqrt{\varepsilon }\int_{0}^{\Delta ^{t-}}\overset{\cdot }{n}^{ioniz}\left(
r+\rho ^{\prime },t-\frac{\rho ^{\prime }}{v_{di}}\right) d\rho ^{\prime }
\label{tminus} \\
&=&\frac{1}{2}\sqrt{\varepsilon }\overset{\cdot }{n}_{0}^{ioniz}\left[
S\left( r,t\right) \Delta ^{t-}+\frac{1}{2}\left( \frac{\partial S}{\partial
r}\right) \left( \Delta ^{t-}\right) ^{2}-\frac{1}{2}\left( \frac{\partial S%
}{\partial t}\right) \frac{\left( \Delta ^{t-}\right) ^{2}}{v_{Di}}\right]  
\notag
\end{eqnarray}%
This is the number of ions that are trapped and had an initial velocity
anti-parallel to $\mathbf{B}$. They are produced in $\left( r+\rho ^{\prime
},t-\frac{\rho ^{\prime }}{v_{Di}}\right) $, summed over the interval $\left[
r,r+\Delta ^{t-}\right] $ and flow towards $\left( r,t\right) $ from the
right (\emph{i.e.} their current is negative).%
\begin{eqnarray}
&&\Gamma _{in}^{c\left( +\parallel \right) }\left( r,t\right) =\frac{1}{2}%
\left( 1-\sqrt{\varepsilon }\right) \int_{0}^{\Delta ^{c+}}\overset{\cdot }{n%
}^{ioniz}\left( r-\rho ^{\prime },t-\frac{\rho ^{\prime }}{v_{di}}\right)
d\rho ^{\prime }  \label{cplus} \\
&=&\frac{1}{2}\left( 1-\sqrt{\varepsilon }\right) \overset{\cdot }{n}%
_{0}^{ioniz}\left[ S\left( r,t\right) \Delta ^{c+}-\frac{1}{2}\left( \frac{%
\partial S}{\partial r}\right) \left( \Delta ^{c+}\right) ^{2}-\frac{1}{2}%
\left( \frac{\partial S}{\partial t}\right) \frac{\left( \Delta ^{c+}\right)
^{2}}{v_{Di}}\right]   \notag
\end{eqnarray}%
This is the number of ions that are circulating and had an initial velocity
parallel with $\mathbf{B}$. They are produced in $\left( r-\rho ^{\prime },t-%
\frac{\rho ^{\prime }}{v_{Di}}\right) $, summed over the interval $\left[
r-\Delta ^{c+},r\right] $ and flow towards $\left( r,t\right) $ from the
left (\emph{i.e.} their current is positive).%
\begin{eqnarray}
&&\Gamma _{in}^{c\left( -\parallel \right) }\left( r,t\right) =\frac{1}{2}%
\left( 1-\sqrt{\varepsilon }\right) \int_{0}^{\Delta ^{c-}}\overset{\cdot }{n%
}^{ioniz}\left( r+\rho ^{\prime },t-\frac{\rho ^{\prime }}{v_{di}}\right)
d\rho ^{\prime }  \label{cminus} \\
&=&\frac{1}{2}\left( 1-\sqrt{\varepsilon }\right) \overset{\cdot }{n}%
_{0}^{ioniz}\left[ S\left( r,t\right) \Delta ^{c-}+\frac{1}{2}\left( \frac{%
\partial S}{\partial r}\right) \left( \Delta ^{c-}\right) ^{2}-\frac{1}{2}%
\left( \frac{\partial S}{\partial t}\right) \frac{\left( \Delta ^{c-}\right)
^{2}}{v_{Di}}\right]   \notag
\end{eqnarray}%
This is the number of ions that are circulating and had an initial velocity
anti-parallel to $\mathbf{B}$. They are produced in $\left( x+\rho ^{\prime
},t-\frac{\rho ^{\prime }}{v_{Di}}\right) $, summed over $\left[ r,r+\Delta
^{c-}\right] $ and flow towards $\left( r,t\right) $ from the right (\emph{%
i.e. }their current is negative).

The current density resulting from the \emph{in} fluxes is 
\begin{equation}
J^{in}\left( x,t\right) =\left\vert e\right\vert \left( \Gamma
_{in}^{t\left( +\parallel \right) }-\Gamma _{in}^{t\left( -\parallel \right)
}+\Gamma _{in}^{c\left( +\parallel \right) }-\Gamma _{in}^{c\left(
-\parallel \right) }\right)   \label{jin}
\end{equation}%
or%
\begin{eqnarray}
&&J^{in}\left( x,t\right)   \label{jinf} \\
&=&\frac{1}{2}\left\vert e\right\vert \overset{\cdot }{n}_{0}^{ioniz}\left\{
S\left( x,t\right) \left[ \sqrt{\varepsilon }\Delta ^{t+}-\sqrt{\varepsilon }%
\Delta ^{t-}+\left( 1-\sqrt{\varepsilon }\right) \Delta ^{c+}-\left( 1-\sqrt{%
\varepsilon }\right) \Delta ^{c-}\right] \right.   \notag \\
&&\ \ \ \ \ \ \ +\frac{1}{2}\left( \frac{\partial S}{\partial r}\right) %
\left[ -\sqrt{\varepsilon }\left( \Delta ^{t+}\right) ^{2}-\sqrt{\varepsilon 
}\left( \Delta ^{t-}\right) ^{2}-\left( 1-\sqrt{\varepsilon }\right) \left(
\Delta ^{c+}\right) ^{2}-\left( 1-\sqrt{\varepsilon }\right) \left( \Delta
^{c-}\right) ^{2}\right]   \notag \\
&&\left. +\frac{1}{2}\frac{1}{v_{Di}}\left( \frac{\partial S}{\partial t}%
\right) \left[ -\sqrt{\varepsilon }\left( \Delta ^{t+}\right) ^{2}+\sqrt{%
\varepsilon }\left( \Delta ^{t-}\right) ^{2}-\left( 1-\sqrt{\varepsilon }%
\right) \left( \Delta ^{c+}\right) ^{2}+\left( 1-\sqrt{\varepsilon }\right)
\left( \Delta ^{c-}\right) ^{2}\right] \right\}   \notag
\end{eqnarray}

\bigskip

We now have to calculate the flows (\emph{out}) that leave the shell of the
point $\left( r,t\right) $.The groups that are leaving $\left( r,t\right) $
are:

Trapped with parallel (\emph{i.e.} positive) initial velocity $t\left(
+\parallel \right) $%
\begin{equation}
n_{out}^{t\left( +\parallel \right) }\left( r,t\right) =\frac{1}{2}\sqrt{%
\varepsilon }\overset{\cdot }{n}^{ioniz}\left( r,t\right) \ \ \text{(toward
large }r\text{)}  \label{outtplus}
\end{equation}

The second group consists of trapped with anti-parallel (\emph{i.e.}
negative) initial velocity%
\begin{equation}
n_{out}^{t\left( -\parallel \right) }\left( r,t\right) =\frac{1}{2}\sqrt{%
\varepsilon }\overset{\cdot }{n}^{ioniz}\left( r,t\right) \ \ \text{(toward
small }r\text{)}  \label{outtminus}
\end{equation}

The third group consists of circulating ions with parallel initial velocity%
\begin{equation}
n_{out}^{c\left( +\parallel \right) }\left( r,t\right) =\frac{1}{2}\left( 1-%
\sqrt{\varepsilon }\right) \overset{\cdot }{n}^{ioniz}\left( r,t\right) \ \ 
\text{(toward large }r\text{)}  \label{outcplus}
\end{equation}

The fourth group consists of circulating ions with anti-parallel initial
velocity%
\begin{equation}
n_{out}^{c\left( -\parallel \right) }\left( r,t\right) =\frac{1}{2}\left( 1-%
\sqrt{\varepsilon }\right) \overset{\cdot }{n}^{ioniz}\left( r,t\right) \ \ 
\text{(toward small }r\text{)}  \label{outcminus}
\end{equation}

Summing the \emph{out} components after taking into account the signs
according to the description%
\begin{equation}
J^{out}\left( r,t\right) =\left\vert e\right\vert \left( \Gamma
_{out}^{t\left( \parallel \right) }-\Gamma _{out}^{t\left( -\parallel
\right) }+\Gamma _{out}^{c\left( \parallel \right) }-\Gamma _{out}^{c\left(
-\parallel \right) }\right) =0  \label{jout}
\end{equation}

\bigskip

Adding the flows \emph{in} and \emph{out} $J\left( r,t\right) =J^{in}\left(
r,t\right) -J^{out}\left( r,t\right) $ we obtain%
\begin{eqnarray}
&&J\left( r,t\right)   \label{j} \\
&=&\frac{1}{2}\left\vert e\right\vert \overset{\cdot }{n}_{0}^{ioniz}\left\{
S\left( r,t\right) \left[ \sqrt{\varepsilon }\left( \Delta ^{t+}-\Delta
^{t-}\right) +\left( 1-\sqrt{\varepsilon }\right) \left( \Delta ^{c+}-\Delta
^{c-}\right) \right] \right.   \notag \\
&&\ \ \ \ \ \ \ +\frac{1}{2}\left( \frac{\partial S}{\partial r}\right) %
\left[ -\sqrt{\varepsilon }\left( \left( \Delta ^{t+}\right) ^{2}+\left(
\Delta ^{t-}\right) ^{2}\right) -\left( 1-\sqrt{\varepsilon }\right) \left(
\left( \Delta ^{c+}\right) ^{2}+\left( \Delta ^{c-}\right) ^{2}\right) %
\right]   \notag \\
&&\left. +\frac{1}{2}\frac{1}{v_{Di}}\left( \frac{\partial S}{\partial t}%
\right) \left[ \sqrt{\varepsilon }\left( -\left( \Delta ^{t+}\right)
^{2}+\left( \Delta ^{t-}\right) ^{2}\right) +\left( 1-\sqrt{\varepsilon }%
\right) \left( -\left( \Delta ^{c+}\right) ^{2}+\left( \Delta ^{c-}\right)
^{2}\right) \right] \right\}   \notag
\end{eqnarray}

\bigskip

We show in \textbf{Appendix A} that the contributions of the circulating
ions is much smaller than that of the trapped ions and for the present
estimation can be neglected%
\begin{eqnarray}
&&J\left( r,t\right) \approx \frac{1}{2}\left\vert e\right\vert \overset{%
\cdot }{n}_{0}^{ioniz}\left\{ S\left( x,t\right) \sqrt{\varepsilon }\left[
\Delta ^{t+}-\Delta ^{t-}\right] \right.  \label{jrttr} \\
&&\ \ \ \ \ \ \ -\frac{1}{2}\left( \frac{\partial S}{\partial r}\right) 
\sqrt{\varepsilon }\left[ \left( \Delta ^{t+}\right) ^{2}+\left( \Delta
^{t-}\right) ^{2}\right]  \notag \\
&&\left. +\frac{1}{2}\frac{1}{v_{Di}}\left( \frac{\partial S}{\partial t}%
\right) \sqrt{\varepsilon }\left[ -\left( \Delta ^{t+}\right) ^{2}+\left(
\Delta ^{t-}\right) ^{2}\right] \right\}  \notag
\end{eqnarray}

\bigskip

In the numerical model (next Section) we take a time-independent source, $%
\partial S/\partial t=0$, and the current density becomes%
\begin{equation}
J\left( r\right) \approx \frac{1}{2}\left\vert e\right\vert \overset{\cdot }{%
n}_{0}^{ioniz}\sqrt{\varepsilon }\left\{ S\left( r,t\right) \left( \Delta
^{t+}-\Delta ^{t-}\right) -\frac{1}{2}\left( \frac{\partial S}{\partial r}%
\right) \left[ \left( \Delta ^{t+}\right) ^{2}+\left( \Delta ^{t-}\right)
^{2}\right] \right\}  \label{jrtr}
\end{equation}

The distance $\Delta ^{t\pm }$ travelled by the new ion from ionization to
the \textquotedblleft center\textquotedblright\ of the periodic motion (%
\emph{i.e.} the distance on which there is effective current) will be
calculated in the next Section by solving the equations of motion of the
ion, as initial value problem. For the present estimation we adopt
neoclassical approximations \cite{fonghahm}\textbf{\ }replacing $\Delta
^{t\pm }$ with the \textquotedblleft radius\textquotedblright\ of the
banana, 
\begin{equation}
\Delta ^{\pm }\approx v_{Di}\tau _{bounce}=\rho _{i}q\varepsilon ^{-1/2}
\label{deltatpm}
\end{equation}%
where ions $v_{Di}\approx \frac{1}{\Omega _{ci}}\frac{v_{th,i}^{2}}{R}$ and $%
\tau _{bounce}\approx r/v_{\theta }=\frac{rB}{B_{\theta }}\frac{1}{v_{th,i}}%
\sqrt{\frac{R}{r}}$ \cite{rhh1972}. In regions where the ionization rate has
strong spatial variation, the second term in Eq.(\ref{jrtr}) is large and we
can simplify the result as%
\begin{equation}
J\approx -\frac{1}{2}\left\vert e\right\vert \overset{\cdot }{n}%
_{0}^{ioniz}\left( \frac{\partial S}{\partial r}\right) \rho
_{i}^{2}q^{2}\varepsilon ^{-1/2}  \label{jtapp}
\end{equation}%
For an estimation we take $a=1\ \left( m\right) $, $R=3.5\ \left( m\right) $%
, $B_{T}=3.5\left( T\right) $, $T_{i}=1.5\ \left( keV\right) $, $%
N_{t}=3\times 10^{21}\ $neutral atoms in the pellet, $\tau ^{ioniz}\approx
4\times 10^{-3}\ \left( s\right) $ duration of the complete ionization
process \cite{improvedcorefuelling}, \cite{jetrefuelling} and the radial
extension of the zone of ionization is between $r_{1}=0.4a$ and $r_{2}=0.7a$%
. The energy of the new ions is a fraction ($\eta =0.75$) of the background
ion energy and the trapping parameter $\lambda =hv_{\perp }^{2}/v^{2}$, $%
h=1+\varepsilon \cos \theta $ is taken $\lambda =0.92$. It results $%
V_{t}\approx 22.8\ \left( m^{3}\right) $, $A\left( r\right) \approx 138r\
\left( m^{2}\right) $ and the average rate of ionization $\overset{\cdot }{n}%
^{ioniz}\sim 3.3\times 10^{22}\ \left( ions/m^{3}/s\right) $. Adopting \ for 
$S\left( r\right) $ a simple spatial profile limited between $r_{a}=0.475a$, 
$r_{b}=0.625a$ , with $\max S=1$ we find from the constraint Eq.(\ref{constr}%
) $\overset{\cdot }{n}_{0}^{ioniz}\approx 1.2\times 10^{23}$ $\left(
ions/m^{3}/s\right) $. Using values suggested by experiments \cite%
{jetrefuelling}, \cite{pelletjet1}, \cite{baylorjettftr}, \cite{baylorjetd3d}%
, \cite{improvedcorefuelling}, $\Delta ^{t\pm }$ from exact integration, and 
$a^{-1}\partial S/\partial x\sim 20\ \left( m^{-1}\right) $, we obtain from
Eq.(\ref{jrtr}) $\left\vert J\right\vert \sim 11.52\ \left( A/m^{2}\right) $.

The result is indeed high. For comparison we consider $nm_{i}\partial
v_{\theta }/\partial t\sim JB_{T}$. This means $\partial v_{\theta
}/\partial t\sim 23\times 10^{8}\ \left( m/s^{2}\right) $, or, in every
microsecond the poloidal speed would increase with more than $4\ \left(
km/s\right) $. In less than one tenth of a milliseconds $v_{\theta }$ rises
to the range of the ion thermal speed $v_{th,i}\sim 0.38\times 10^{3}\
\left( km/s\right) $. For comparison the transit time magnetic pumping decay
of the poloidal velocity would contribute with 
\begin{equation}
-\gamma ^{MP}=\frac{\partial }{\partial t}\ln v_{\theta }=\frac{3}{4}\left(
1+\frac{1}{2q^{2}}\right) \left( \frac{l}{qR}\right) ^{2}\nu _{ii}\ 
\label{gammp}
\end{equation}%
where $l$ is the mean free path and $\nu _{ii}$ is the ion-ion collision
time \cite{hassamkulsrud}. We find $-\gamma ^{MP}\sim 1.2\times 10^{4}\
\left( s^{-1}\right) $, and taking a poloidal velocity (as observed in some
experiments, \emph{e.g.} \cite{corepoloidaltftr}) $v_{\theta }\sim 10^{4}\
\left( m/s\right) $ it is estimated $\left\vert \left( \partial v_{\theta
}/\partial t\right) ^{MP}\right\vert \sim 1.2\times 10^{8}\ \left(
m/s^{2}\right) $. It results%
\begin{equation}
\left( \frac{\partial v_{\theta }}{\partial t}\right) ^{ioniz}>\left\vert
\left( \frac{\partial v_{\theta }}{\partial t}\right) ^{MP}\right\vert 
\label{compax}
\end{equation}%
Of course, this ionization torque acts for short time (few milliseconds) and
Eq.(\ref{jtapp}) is an overestimation as long as the neutral atoms' dynamics
(\emph{e.g.} the pellet cloud) and the bulk ion's reaction (return current)
are not described in detail. But this result is a strong suggestion that the
ionization torque is important.

\bigskip

\section{Numerical implementation} \label{section3}

The numerical simulation of this process has been done on a discrete mesh $%
\left\{ r_{i}\right\} _{i=1,NX}\times \left\{ t_{k}\right\} _{k=1,NT}$ for $%
NR=NT=500$. The simple physical picture described above has been
implemented.. For any cell $\left( r_{i},t_{k}\right) $ we calculate the
number of ions that are generated, $\overset{\cdot }{n}_{0}^{ioniz}S\left(
r,t\right) $, using the expression for $S\left( r\right) =\alpha +\beta
\left( r/a\right) +\gamma \left( r/a\right) ^{2}$. Imposing $S\left(
r_{a}\right) =S\left( r_{b}\right) =0$ and $\max S=1$ at $\left(
r_{a}+r_{b}\right) /2$, the coefficients are determined and, as explained
above, Eq.(\ref{constr}) determines the constant $\overset{\cdot }{n}%
_{0}^{ioniz}=1.2\times 10^{23}\ \left( ions/m^{3}/s\right) $. We now assume
that the energy of the new ions is a fraction $\left( 0.75\right) $ of the
background ion thermal energy (at $T_{i}=1.5keV$) and that the probability
of being trapped is $\sqrt{\varepsilon }$. Finally we assume that the
probability that the ion has an initial velocity which is parallel to $%
\mathbf{B}$ is $1/2$, equal with the probability to be anti-parallel. Now we
look at the way the ions move. The total excursion on $r$ is $\Delta ^{t+}$
(to larger $r$) and $\Delta ^{t-}$ (to smaller $r$). The displacement is
represented on the mesh $\left\{ r_{i^{\prime }},t_{k^{\prime }}\right\} $
and every cell $\left( i^{\prime },k^{\prime }\right) $ which is traversed
by the flux of ions stores this contribution, adding it to a variable that
will finally be the current flowing through it. There are several other
cells whose new ions traverse this cell $\left( i^{\prime },k^{\prime
}\right) $ and all contributions are counted and summed. Due to the assumed
constancy of $v_{Di}$, there is no decay along the path that starts from $%
\left( r_{i},t_{k}\right) $ and ends in the cell $\left( r_{i}+\Delta
^{t+},t+\Delta ^{t+}/v_{Di}\right) $, [respectively $\left( r_{i}-\Delta
^{t-},t+\Delta ^{t-}/v_{Di}\right) $ for the anti-parallel initial
velocity]. All cells traversed along this path retain the contribution from $%
\left( r_{i},t_{k}\right) $.

%
\begin{figure}[h]
\includegraphics[height=8cm]{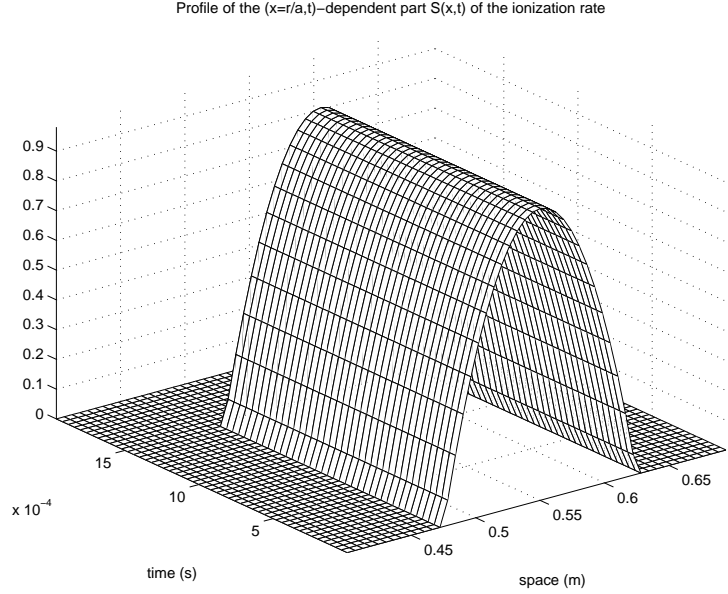}
\caption{\small{The space-time profile of the ionization source, the function $S\left(
r,t\right) $. The factor $\overset{\cdot }{n}_{0}\left( ions/m^{3}/s\right) $
multiplies this function to obtain the effective rate of production of new
ions.
}}
\label{figure21}
\end{figure}
%

%
\begin{figure}[h]
\includegraphics[height=8cm]{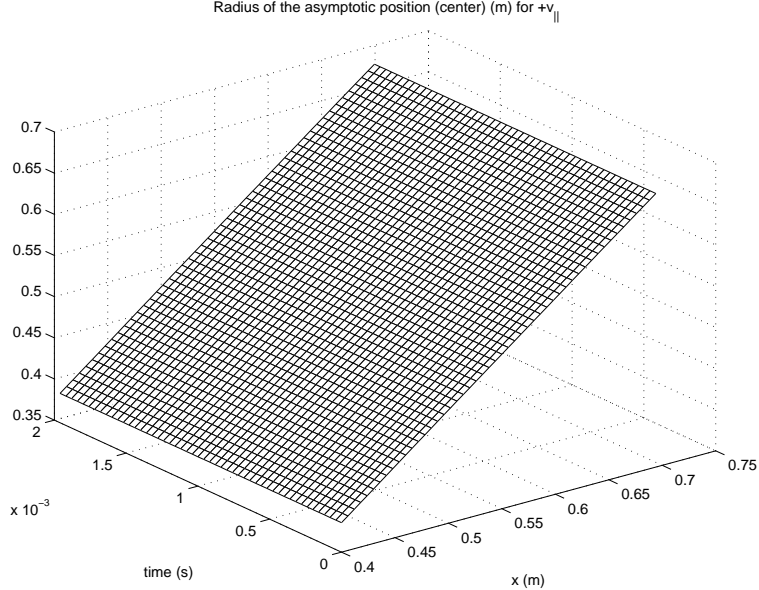}
\caption{\small{The $\left( r,t\right) $ profile of the final position (the
\textquotedblleft center\textquotedblright ) of the average $r\left(
t\right) $ for the ions whose orbit encloses the magnetic surface. This is
calculated by solving in every $\left( r_{i},t_{k}\right) $, for $i=1,NR$, $%
k=1,NTIME$ cell of the disctere mesh, the set of equations of motion for a
trapped ion.
}}
\label{figure22}
\end{figure}
%

%
\begin{figure}[h]
\includegraphics[height=8cm]{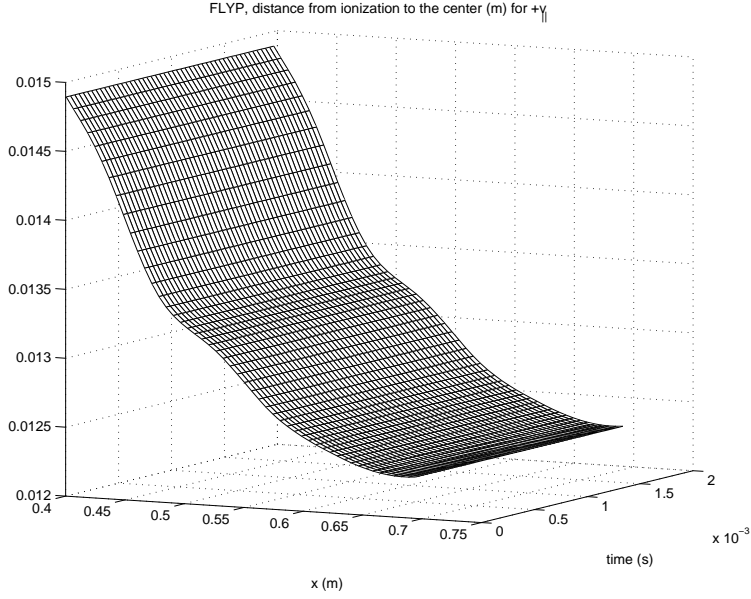}
\caption{\small{The variable $FLYP$, difference between the final position (the
\textquotedblleft center\textquotedblright\ of the banana) and the initial
radial position (where ionization occurs) for an ions whose orbit encloses
the magnetic surface. This is the amount of radial displacement of the ion
on which there is current.
}}
\label{figure23}
\end{figure}
%

%
\begin{figure}[h]
\includegraphics[height=8cm]{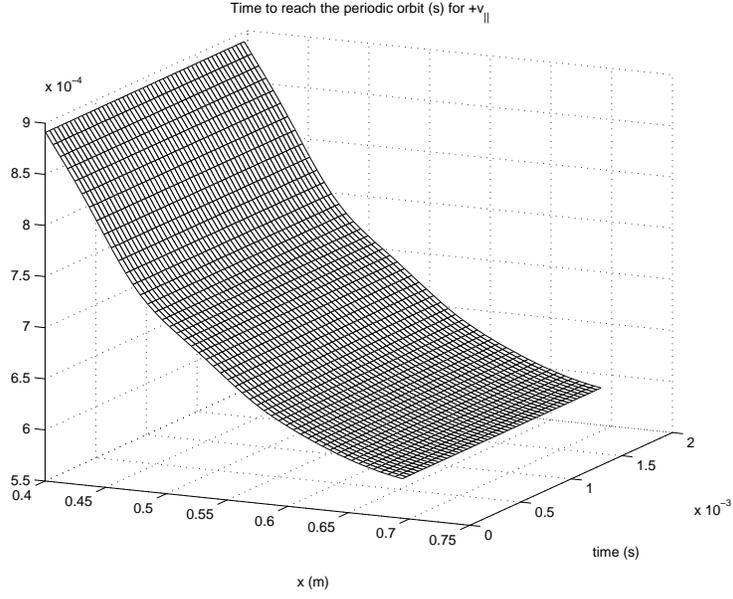}
\caption{\small{The $\left( r,t\right) $ profile of the calculated time for an ion to reach
the average position (the \textquotedblleft center\textquotedblright )
starting from the place of ionization.
}}
\label{figure24}
\end{figure}
%

%
\begin{figure}[h]
\includegraphics[height=8cm]{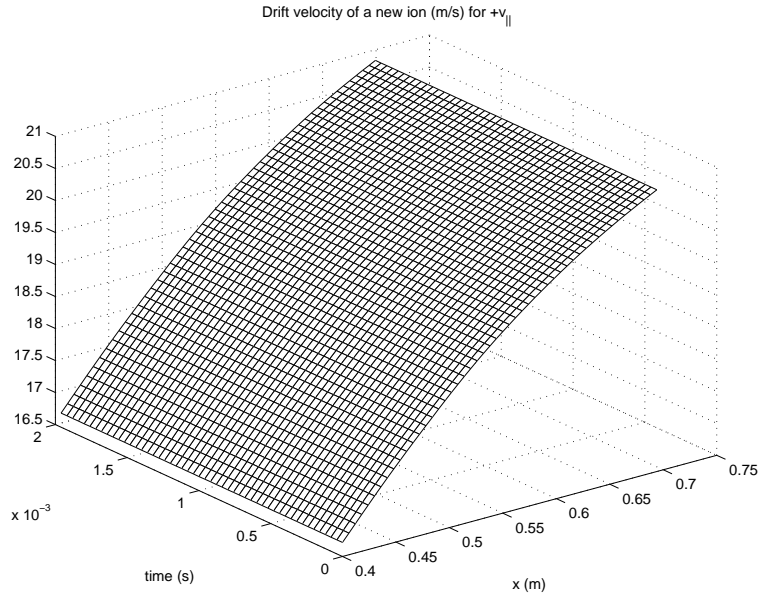}
\caption{\small{The $\left( r,t\right) $ profile of the calculated drift velocity $v_{Di}$,
as explained in Figure \ref{figure11}.
}}
\label{figure25}
\end{figure}
%

%
\begin{figure}[h]
\includegraphics[height=8cm]{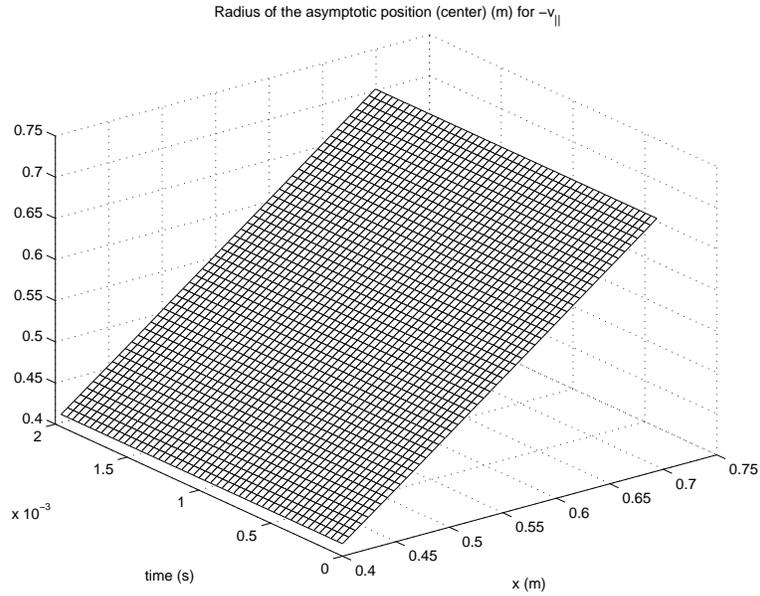}
\caption{\small{Same as Figure \ref{figure22}, but for ions whose orbits, calculated in every
mesh-cell, are inside the magnetic surface.
}}
\label{figure26}
\end{figure}
%

%
\begin{figure}[h]
\includegraphics[height=8cm]{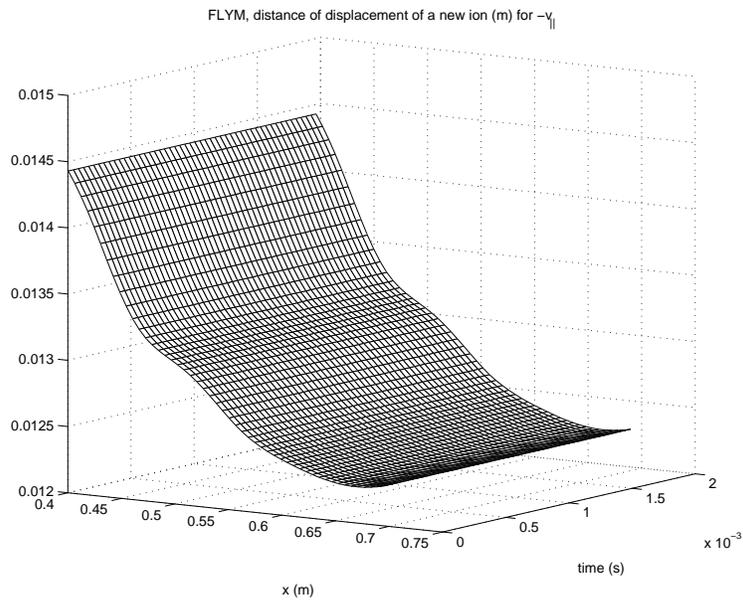}
\caption{\small{The variable $FLYM$, similar with $FLYP$ from Figure \ref{figure23}, but for the
\textquotedblleft smaller\textquotedblright\ banana.
}}
\label{figure27}
\end{figure}
%

%
\begin{figure}[h]
\includegraphics[height=8cm]{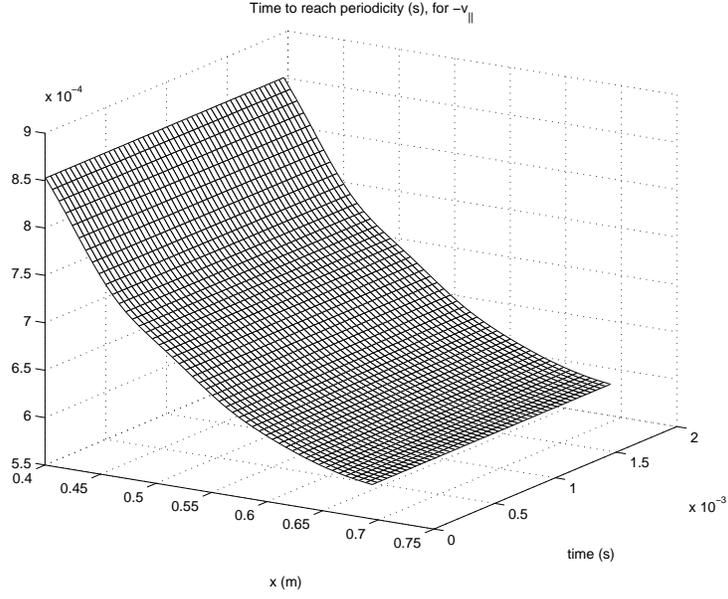}
\caption{\small{The time to reach the \textquotedblleft center\textquotedblright , for the
\textquotedblleft smaller\textquotedblright\ banana.
}}
\label{figure28}
\end{figure}
%

%
\begin{figure}[h]
\includegraphics[height=8cm]{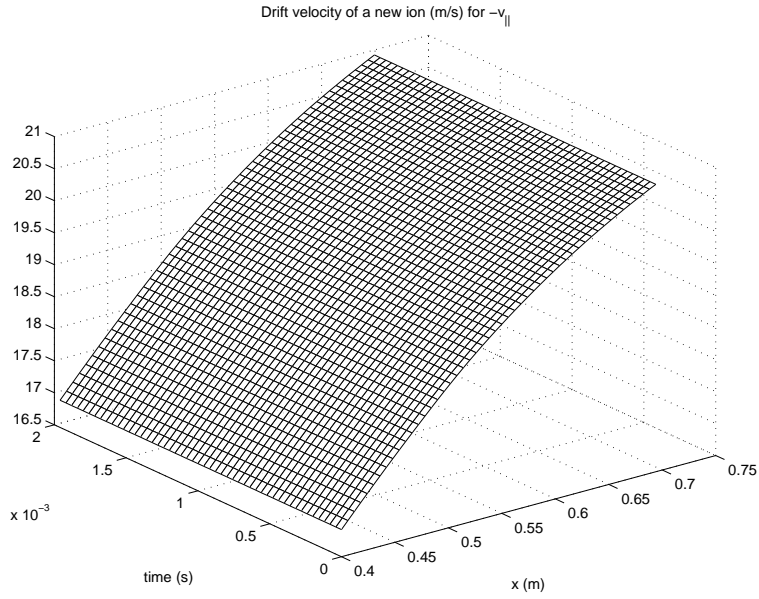}
\caption{\small{The drift velocity for the smaller banana.
}}
\label{figure29}
\end{figure}
%

%
\begin{figure}[h]
\includegraphics[height=6cm]{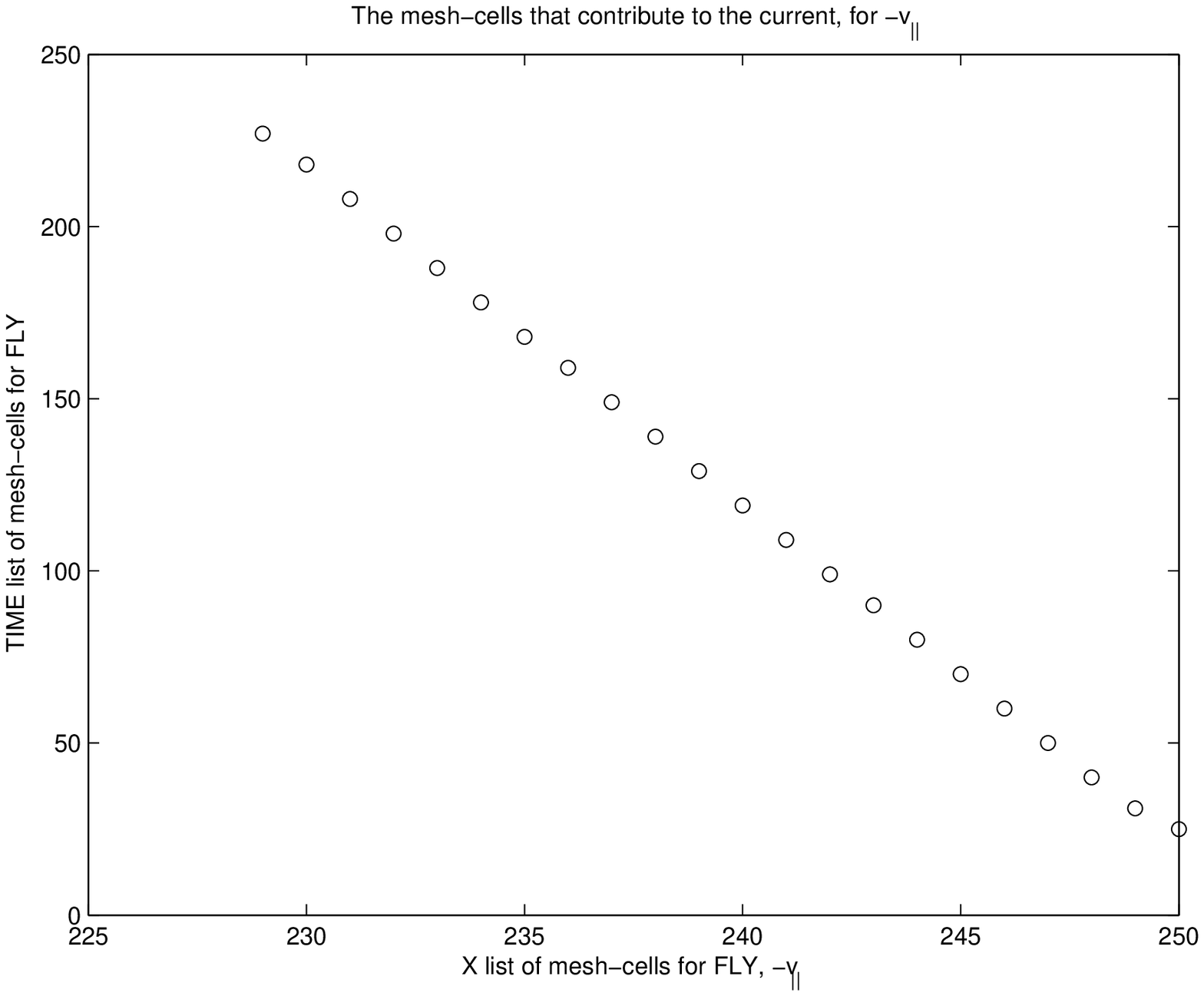}
\includegraphics[height=6cm]{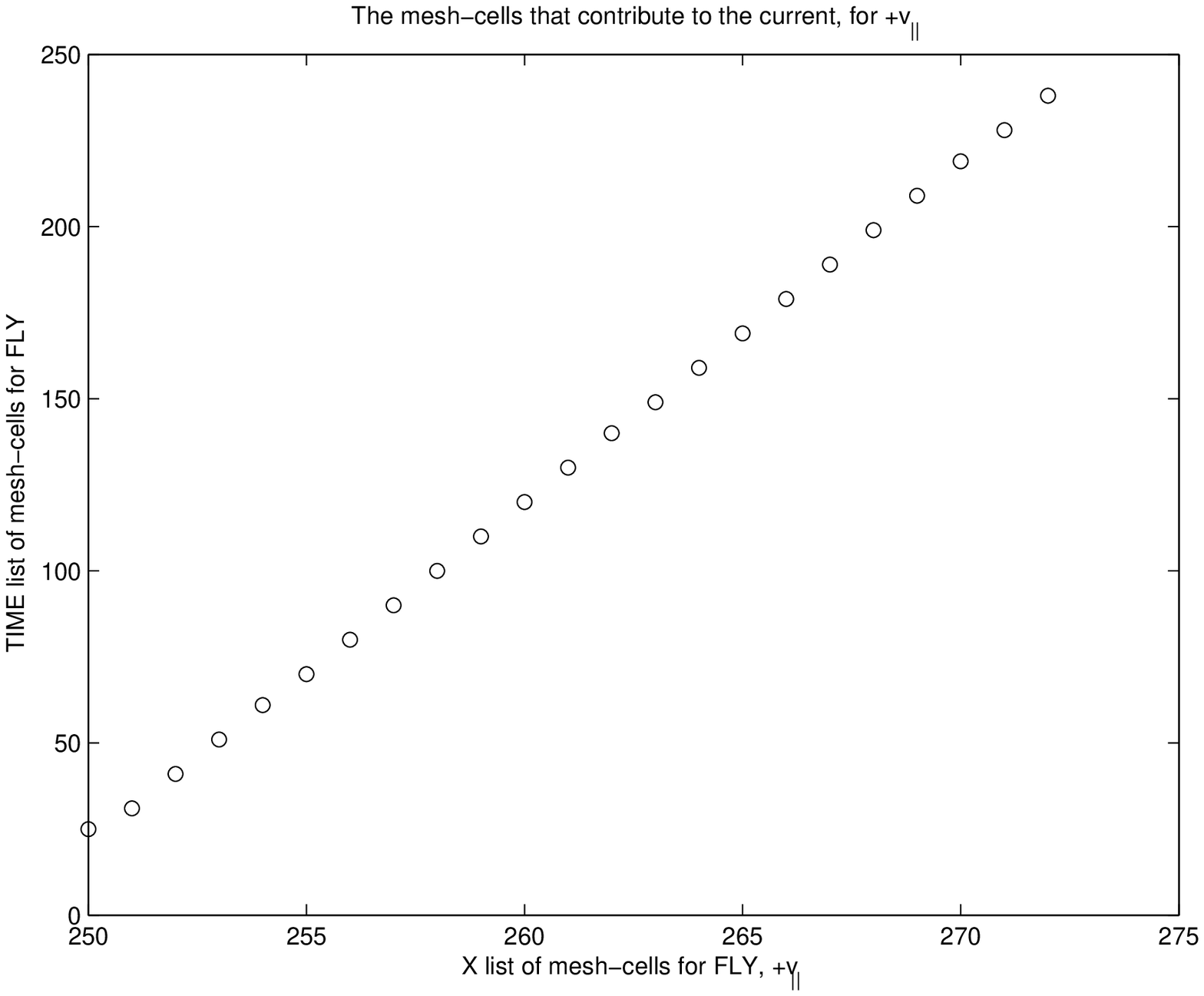}
\caption{\small{At left: the mesh-cells that are traversed by the ions generated in the point $\left(
ir=250,itime=25\right) $  that evolve towards smaller radii (their banana
orbits are fully inside the magnetic surface. At right: the mesh-cells that are traversed by the ions that evolve towards larger radii (their banana
orbit ecloses the magnetic surface).
}}
\label{figure3031}
\end{figure}
%

Instead of the neoclassical approximations for $\Delta ^{t\pm }$ and $v_{Di}$
\cite{rhh1972} we choose to solve the system of equations of motion \cite%
{morozovsolovev}, \cite{berkgaleev}, \cite{galeevsagdeev}, \cite{wongburrell}%
, \cite{fonghahm} in every cell $\left( r_{i},t_{k}\right) _{i=1,NX;k=1,NT}$.%
\begin{equation*}
\frac{dr}{dt}\approx -\frac{1}{\Omega }\left( \frac{v_{\perp }^{2}}{2}%
+v_{\parallel }^{2}\right) \frac{\sin \theta }{R_{0}}
\end{equation*}%
\begin{equation*}
\frac{d\theta }{dt}\approx \frac{v_{\parallel }}{qR_{0}}-\frac{1}{r}\frac{1}{%
\Omega }\left( \frac{v_{\perp }^{2}}{2}+v_{\parallel }^{2}\right) \frac{\cos
\theta }{R_{0}}
\end{equation*}%
\begin{equation*}
\frac{d\varphi }{dt}\approx \frac{v_{\parallel }}{R_{0}}
\end{equation*}%
\begin{eqnarray*}
\frac{d}{dt}\left( \frac{v_{\perp }^{2}}{2}\right) &=&\left( \frac{v_{\perp
}^{2}}{2}\right) v_{\parallel }\frac{B_{\theta }}{B_{T}}\frac{\sin \theta }{%
R_{0}} \\
\frac{dv_{\parallel }}{dt} &=&-\left( \frac{v_{\perp }^{2}}{2}\right) \frac{%
B_{\theta }}{B_{T}}\frac{\sin \theta }{R_{0}}
\end{eqnarray*}

From the solution we get the exact orbit of an ion born in the cell $\left(
r_{i},t_{k}\right) _{i=1,NR;k=1,NT}$ but we still have to operate the
separation of the transitory part, the part which represents the unique
manifestation of a current, from the periodic part of the trajectory, whose
average does not produce a current. We calculate for each trajectory the
time-dependent\ average position $\overline{r}\left( t\right) =\frac{1}{t}%
\int_{0}^{t}r\left( t^{\prime }\right) dt^{\prime }$ and leave the
integration sufficiently long such that the asymptotic quasi-static average
position $r_{asymp}=\overline{r}\left( t\rightarrow \infty \right) $ to be
clearly identified. This position $r_{asymp}$ (the \textquotedblleft
center\textquotedblright\ of the banana) is retained and the quantity $%
\Delta ^{t}$ is obtained as the difference between $r_{asymp}$ and the
initial position, which is the point of the ionization, $r_{ini}$. We still
need the estimation of the \emph{effective }time that is necessary for the
ion to reach this asymptotic position. The first intersection between the
asymptotic line $r=r_{asymp}$ and the evolution line $\overline{r}\left(
t\right) $ takes place at a moment $t_{asymp}$, which is retained as the
representative time of ion's travel to the center. This procedure is
admittedly approximative but we have tried several reasonably alternative
methods and the present one seems the best.

Once we know $r_{asymp}-r_{ini}\equiv \Delta $ and $t_{asymp}$ we find $%
v_{Di}=\Delta /t_{asymp}$ and all data for the displacements of the new ions
originating from $\left( r_{i},t_{k}\right) $ on the mesh are now available.
In every cell the system is solved for both parallel and anti-parallel
initial velocities and we calculate the distances $\Delta ^{t\pm }$, the
times $t_{asymp}^{\pm }$ of this excursion and the drift velocities $%
v_{Di}^{\pm }=\Delta ^{t\pm }/t_{asymp}^{\pm }$. This is shown in Figure 1
for ions with parallel, respectively anti-parallel initial velocity. The
asymptotic position $r_{asymp}$ is obtained by averaging a set of late
values of $\overline{r}\left( t\right) $, for $t$ beyond the vertical dashed
lines. This ensures a good precision of identification of the
\textquotedblleft center\textquotedblright . The straight line $r=r_{asymp}$
intersects the evolution line $\overline{r}\left( t\right) $ is a point
marked by a open dot, the time $t_{asymp}$.

%
\begin{figure}[h]
\includegraphics[height=8cm]{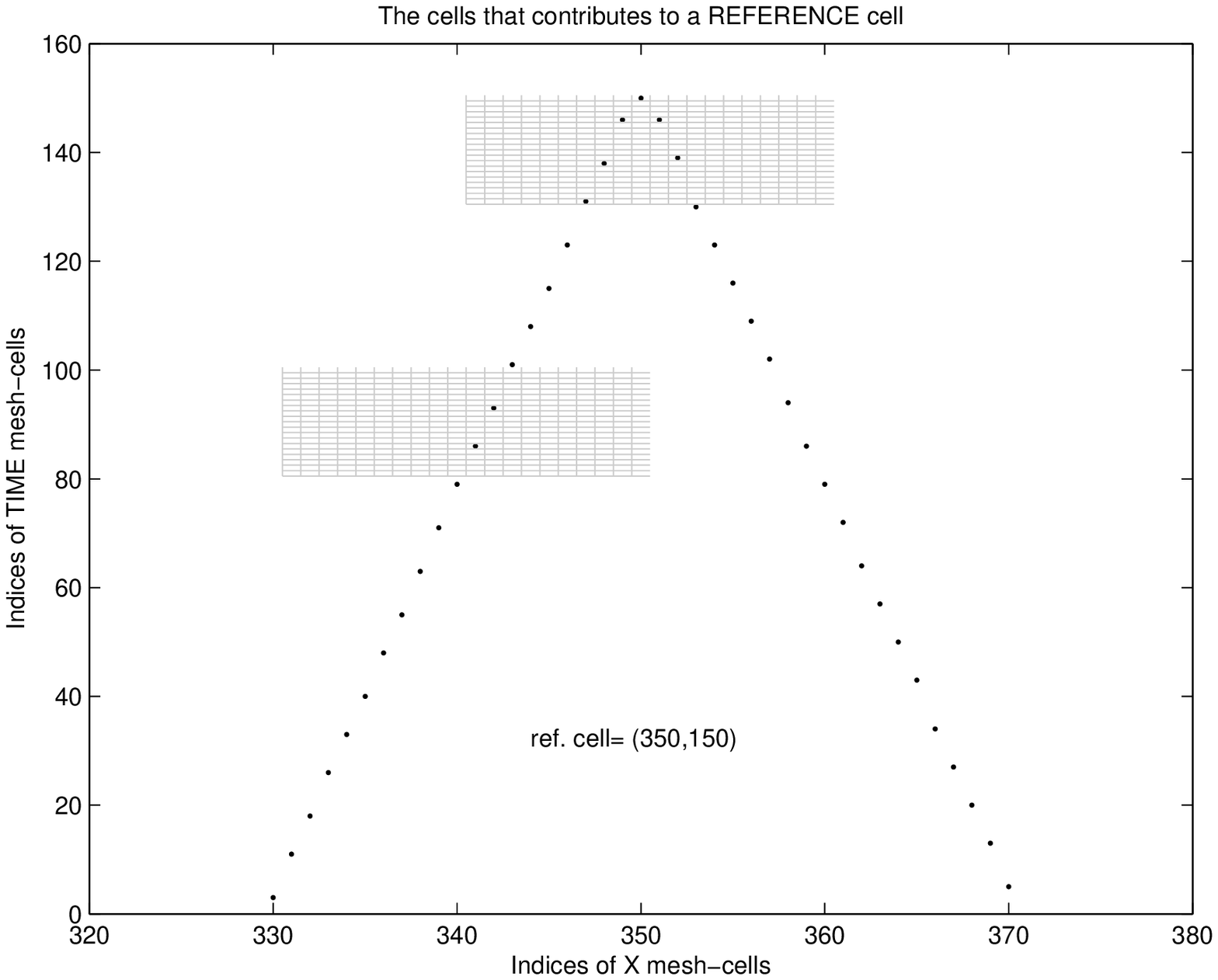}
\caption{\small{The mesh-cells that contribute to the current density that is
\textquotedblleft measured\textquotedblright\ in a reference cell $\left(
r,t\right) $, with indices of discretization $ir=350$ and $itime=150$. From
the cells that are marked by a dot the ions are moving to larger radii (for
the dots that are at the left of the axis of symmetry of the figure) and
respectively to smaller radii (for the dots that are at the right of the
axis of symmetry). Accordingly the contributions must have been generated by
ionization in a position $r-\rho $ and respectively $r+\rho $,  and at a
time $t-\rho /v_{Di}$, for them to reach the reference cell at $\left(
r,t\right) $. The real mesh is too detailed to be shown and only two patches
are shown, for illustration.
}}
\label{figure32}
\end{figure}
%

We have adopted a profile $S\left( r\right) $ which is constant in time,
Figure 2. Its support is $\left[ r_{a},r_{b}\right] $ inside the interval $%
\left[ r_{1},r_{2}\right] $. The empty radial regions on both sides are
necessary because the excursions of lengths $\Delta ^{t\pm }$ of the ions
generated at the ends of the support must be recorded in these regions. The
result (the current $J\left( r,t\right) $) depends on time since the process
that starts at $t=0$ (no-ionization) rises slowly by accumulating current
contributions, before saturation. More interesting is the spatial profile
confirming that the sign of $\partial S/\partial r$ is decisive and that
the total torque, \emph{i.e. }integrated over the plasma volume, is zero, as
expected from conservation of angular momentum and from the fact that no ion
is lost from plasma in our picture. The current is plotted in Figure 3. In
Figure 4 we show the mesh-cells that contribute to the current calculated in
a reference cell, chosen arbitrarily in $\left( r,t\right) \equiv \left(
ix=350,itime=150\right) $. The dots aligned on the a straight line (since $%
v_{Di}=$const) at left represent cells from where the new ions with positive
(parallel) initial velocity arrive in $r$ at $t$. The straight line at right
represents ions with anti-parallel initial velocity.

%
\begin{figure}[h]
\includegraphics[height=8cm]{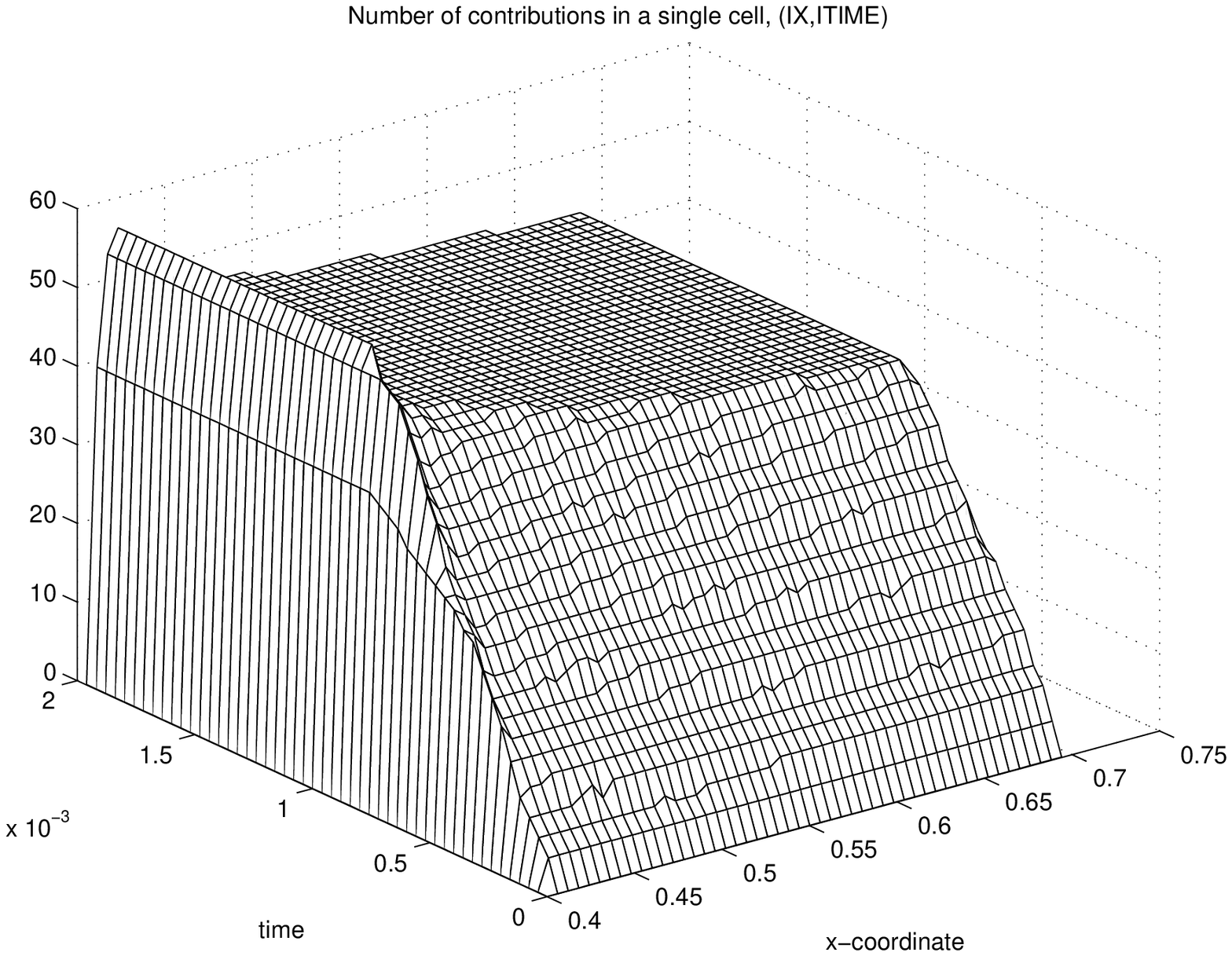}
\caption{\small{The $\left( r,t\right) $ profile of the number of contributions that are
registered in the cells of the mesh, coming from neighbor cells, as
described in Figure \ref{figure32}.
}}
\label{figure33}
\end{figure}
%

%
\begin{figure}[h]
\includegraphics[height=8cm]{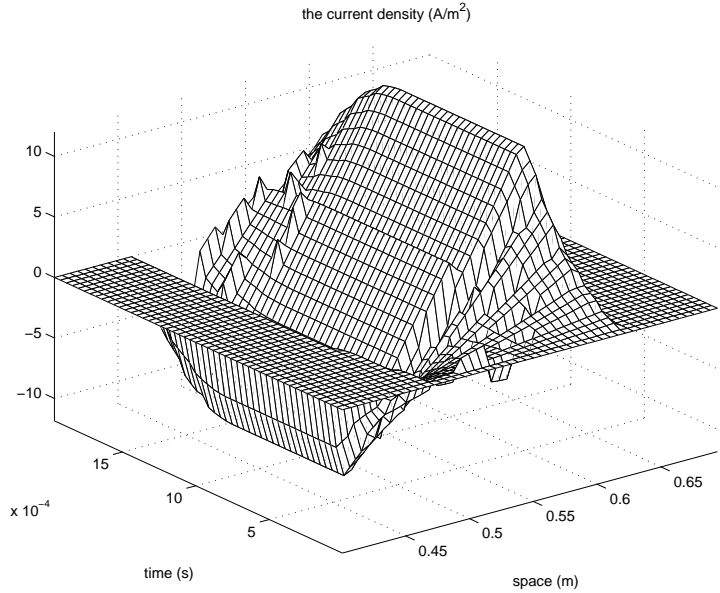}
\caption{\small{The $\left( r,t\right) $ profile of the current density $J\ \left(
A/m^{2}\right) $.
}}
\label{figure34}
\end{figure}
%

%
\begin{figure}[h]
\includegraphics[height=8cm]{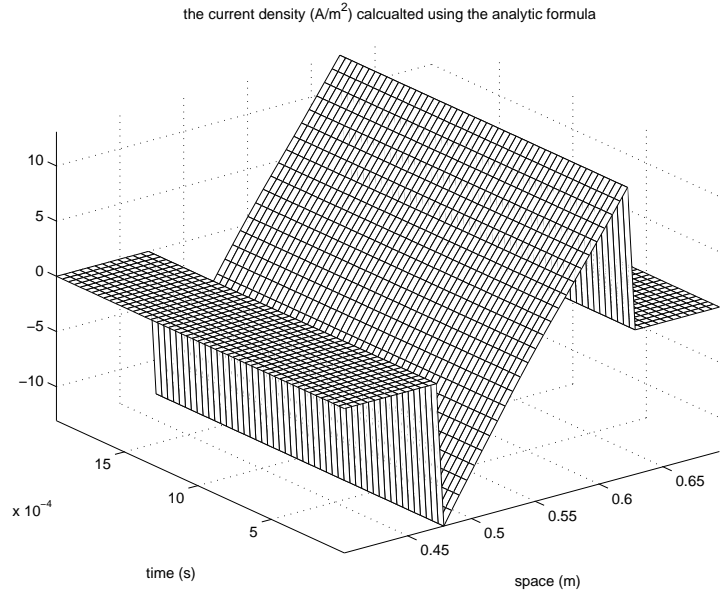}
\caption{\small{The first analytic formula that obtains the current density.
}}
\label{figure35}
\end{figure}
%

%
\begin{figure}[h]
\includegraphics[height=8cm]{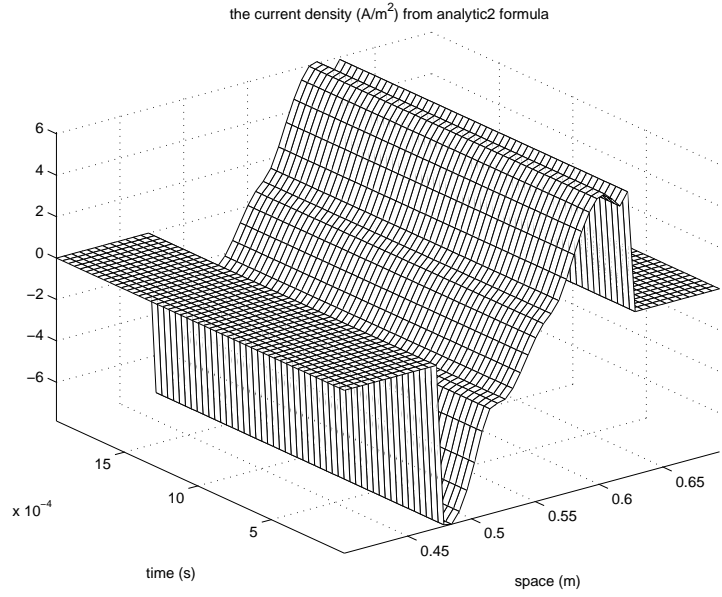}
\caption{\small{The second analytic formula for the current density.
}}
\label{figure36}
\end{figure}
%

%
\begin{figure}[h]
\includegraphics[height=8cm]{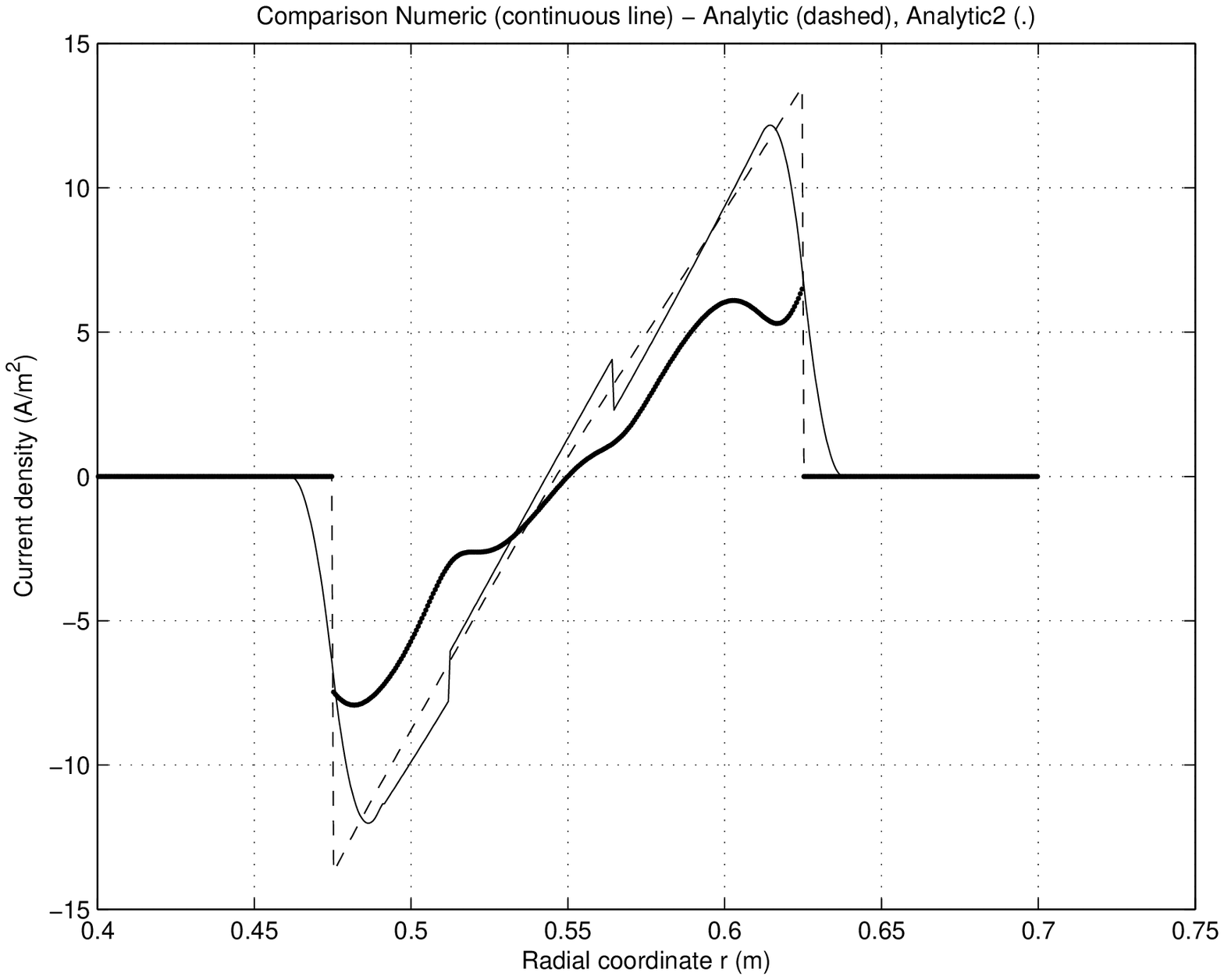}
\caption{\small{The comparison between the two analytic formulas and the result of the
numerical calculation. The section is made at time that is half the total
duration of the ionization process.
}}
\label{figure37}
\end{figure}
%

\bigskip

\section{Drift-kinetic calculation of the ionization-induced current density} \label{section4}

Similar problems are treated in the Ref. \cite{rosenbluthhintonalpha} (for
alpha particles) and the Ref. \cite{hintonrosenbluthnbi} (for NBI). For the
present case the drift-kinetic equation for the \emph{new} ions, is 
\begin{equation}
\frac{\partial f}{\partial t}+\left( v_{\parallel }\widehat{\mathbf{n}}+%
\mathbf{v}_{Di}\right) \cdot \mathbf{\nabla }f=S^{ioniz}  \label{rh1}
\end{equation}%
where the drift velocity of the guiding centre is $\mathbf{v}%
_{Di}=-v_{\parallel }\widehat{\mathbf{n}}\times \mathbf{\nabla }\left( \frac{%
v_{\parallel }}{\Omega _{ci}}\right) $ and the neoclassical notations will
be used: $\xi =v_{\parallel }/v=\left( 1-\lambda /h\right) ^{1/2}$, $\lambda
=hv_{\perp }^{2}/v^{2}$, $h=1+\varepsilon \cos \theta $. The \emph{limits}
of the trapped particle region in the variable $\lambda $ are $1-\frac{r}{R}%
<\lambda <1+\frac{r}{R}$. The velocity space variables are $v,\lambda $ and $%
\sigma \;(=$sign of $v_{\parallel }\,)$.

Since in this simple treatment we neglect the effect of collisions the
neoclassical small parameter is the ratio of the banana half-width to the
minor radius, $\delta =\Delta ^{t\pm }/a\ll 1$. In usual neoclassical
perturbative solution of the drift kinetic equation the zero order
distribution function is the Maxwellian. In the present case, the
perturbative expansion of $f$ , the solution of the drift-kinetic equation
for the new ions, must contain a term which is directly related to the
source and, since this is determined by external factors, it cannot be
ordered as powers of $\delta $. The first term is formally of order $-1$, $%
f_{-1}$. 
\begin{equation}
f=f_{-1}+f_{0}+f_{1}+...  \label{fexpnas1}
\end{equation}

The lowest order 
\begin{equation}
\widehat{\mathbf{n}}\cdot \mathbf{\nabla }f_{-1}=0  \label{fminus1onsurf}
\end{equation}%
shows that $f_{-1}$ is constant along the magnetic lines, or $%
f_{-1}=f_{-1}\left( \psi \right) $. This result is connected with an
assumption about the distribution of ionization processes in space: they
have a rate which is constant on a magnetic surface.

The zeroth order equation is 
\begin{equation}
v_{\parallel }\widehat{\mathbf{n}}\cdot \mathbf{\nabla }f_{0}=-\mathbf{v}%
_{Di}\cdot \mathbf{\nabla }\psi \frac{\partial f_{-1}}{\partial \psi }-\frac{%
\partial f_{-1}}{\partial t}+S^{ioniz}  \label{eqf01}
\end{equation}%
As in any multiple space-time scale analysis we average at this level ($0$)
to obtain a solution on the level $\left( -1\right) $. We apply the operator
of bounce averaging to eliminate the function $f_{0}$. This gives the
equation%
\begin{equation}
\frac{\partial f_{-1}}{\partial t}=\overline{S}^{ioniz}-\overline{\mathbf{v}%
_{Di}\cdot \mathbf{\nabla }\psi }\frac{\partial f_{-1}}{\partial \psi }
\label{eqav1}
\end{equation}

The operator of \emph{bounce averaging} is $\overline{A}=\frac{1}{T}\oint 
\frac{d\theta }{v_{\parallel }\widehat{\mathbf{n}}\cdot \mathbf{\nabla }%
\theta }A=\frac{1}{T}\int \frac{d\theta }{v_{\parallel }/\left( qR\right) }A$%
. The bounce time is $T=\oint \frac{d\theta }{v_{\parallel }\widehat{\mathbf{%
n}}\cdot \mathbf{\nabla }\theta }=\oint \frac{d\theta }{v_{\parallel
}/\left( qR\right) }$. The limits of integrations\textit{\ }for untrapped
ions are $\left[ -\pi \;,\;\pi \right] \;$and for trapped ions the integral
is defined 
\begin{equation}
\oint d\theta =\sum_{\sigma }\sigma \int_{-\theta _{0}}^{+\theta _{0}}d\theta
\label{rh10}
\end{equation}%
where $-\theta _{0}$ and $+\theta _{0}$ are the \textit{turning points} of
the banana. The radial projection of the guiding centre drift velocity can
be written 
\begin{equation}
\mathbf{v}_{Di}\cdot \mathbf{\nabla }\psi =Iv_{\parallel }\left( \widehat{%
\mathbf{n}}\cdot \mathbf{\nabla }\right) \left( \frac{v_{\parallel }}{\Omega
_{ci}}\right) =Iv_{\parallel }\widehat{\mathbf{n}}\cdot \mathbf{\nabla }%
\theta \frac{\partial }{\partial \theta }\left( \frac{v_{\parallel }}{\Omega
_{ci}}\right)  \label{vdnablapsi}
\end{equation}%
where $I=R^{2}\mathbf{B}\cdot \mathbf{\nabla }\varphi =RB_{T}\equiv I\left(
\psi \right) $ is a function of only the magnetic surface variable $\left(
\psi \right) $ and $\widehat{\mathbf{n}}\cdot \mathbf{\nabla }\theta
=1/\left( qR\right) $. \ At this point we assume that the new ion has
reached the asymptotic periodic motion on the banana. Then the radial
displacements average to zero 
\begin{equation}
\overline{\left( \mathbf{v}_{Di}\cdot \mathbf{\nabla }\psi \right) }=\frac{1%
}{T}\sum\limits_{\sigma }\int_{-\theta _{0}}^{+\theta _{0}}\frac{d\theta }{%
v_{\parallel }/\left( qR\right) }I\left( \psi \right) \frac{v_{\parallel }}{%
qR}\frac{\partial }{\partial \theta }\left( \frac{v_{\parallel }}{\Omega
_{ci}}\right) =0  \label{rh11}
\end{equation}

Eq.(\ref{eqav1}) becomes 
\begin{equation}
\frac{\partial f_{-1}}{\partial t}=\overline{S}^{ioniz}  \label{fminus1}
\end{equation}%
and indeed $f_{-1}$ appears as a direct result of the \textquotedblleft
external\textquotedblright\ source. The source of new ions of velocity $%
v_{0} $, with direction $\sigma _{0}$ and trapping parameter $\lambda _{0}$
is \cite{rosenbluthhintonalpha}%
\begin{equation}
\overline{S}^{ioniz}=\overset{\cdot }{n}^{ioniz}\left( \psi ,t\right) \delta
_{\sigma ,\sigma _{0}}\Theta \left( t\right) \delta \left( \lambda -\lambda
_{0}\right) \frac{\delta \left( v-v_{0}\right) }{\pi v_{0}^{2}}  \label{rh12}
\end{equation}%
The Eq.(\ref{fminus1}) simply describes the accumulation of new ions with $%
\left( v_{0},\sigma _{0},\lambda _{0}\right) $ on the surface $\psi $. The
motion of these ions toward the banana trajectories and the periodic motion
that follows must be found at higher orders. Returning to Eq.(\ref{eqf01})
we express $f_{0}$ in terms of $f_{-1}$, using (\ref{vdnablapsi})%
\begin{equation}
v_{\parallel }\widehat{\mathbf{n}}\cdot \mathbf{\nabla }f_{0}\equiv
v_{\parallel }\nabla _{\parallel }f_{0}=-v_{\parallel }\nabla _{\parallel
}\left( \frac{Iv_{\parallel }}{\Omega _{ci}}\right) \frac{\partial
f_{-1}\left( \psi \right) }{\partial \psi }  \label{rh13}
\end{equation}%
with the solution%
\begin{equation}
f_{0}\left( \psi ,\theta ,t\right) =-I\left( \frac{v_{\parallel }}{\Omega
_{ci}}\right) \frac{\partial f_{-1}\left( \psi \right) }{\partial \psi }%
+g\left( \psi ,\theta ,t\right)  \label{fzero}
\end{equation}%
where a constant of integration of the operator $\nabla _{\parallel }$ is
introduced. We make few remarks. First, the time dependence of $f_{0}$
inherited from $f_{-1}$ will be essential for the presence in the theory of
the first, transitory and unique, part of the trajectory. Further, the first
term can be approximated, using for circular geometry $\frac{I}{B_{T}}\frac{%
\partial }{\partial \psi }\simeq \frac{1}{B_{\theta }}\frac{\partial }{%
\partial r}$, 
\begin{equation}
-I\left( \frac{v_{\parallel }}{\Omega _{ci}}\right) \frac{\partial
f_{-1}\left( \psi \right) }{\partial \psi }\approx -\frac{B}{B_{\theta }}%
\frac{v_{\parallel }}{\Omega _{ci}}\frac{\partial f_{-1}\left( r\right) }{%
\partial r}=-\frac{v_{\parallel }}{\Omega _{\theta ci}}\frac{\partial }{%
\partial r}f_{-1}\left( r\right)  \label{rh14}
\end{equation}%
and we see that the difference between the distribution function $f_{0}$ and
that of the previous level $\left( f_{-1}\right) $ consists of a radial
shift of the space argument, of the order of the poloidal Larmor radius $%
v_{\parallel }/\Omega _{\theta ci}\sim \rho _{\theta }$. This is the same
relationship as between the first order neoclassical distribution function
relative to the Maxwellian equilibrium distribution \cite{hintonhazeltinermp}%
. In the particular case of trapped particles, the correction needs also to
reflect the approximative relation between the thermal speed and the
parallel velocity of ions \cite{rhh1972}, and we have $v_{\parallel }/\Omega
_{\theta ci}\approx \left( v_{th}/\Omega _{\theta ci}\right) \left(
r/R\right) ^{1/2}=\rho _{\theta }\varepsilon ^{1/2}$. Finally we note that $%
g $ is constant on the magnetic lines, \emph{i.e.} on surfaces, $\widehat{%
\mathbf{n}}\cdot \mathbf{\nabla }g=0$. A distribution function for \emph{%
bananas} can never be constant on the magnetic surfaces because the
trajectory stops somewhere. Therefore $g$ must only be added to (\ref{fzero}%
) if we consider circulating particles. For our purpose it is not retained.

\bigskip

The next step is the equation for the \emph{first order} $f_{1}$, which is
derived from the equation written at \emph{zero}-order%
\begin{equation}
\frac{\partial f_{0}}{\partial t}+v_{\parallel }\widehat{\mathbf{n}}\cdot 
\mathbf{\nabla }f_{1}+\mathbf{v}_{Di}\cdot \mathbf{\nabla }f_{0}=0
\label{order1}
\end{equation}%
This involves the variation of the first order correction function $f_{1}$
along the magnetic lines $v_{\parallel }\nabla _{\parallel }f_{1}+...$. 
\begin{equation}
\mathbf{v}_{Di}\cdot \mathbf{\nabla }f_{0}=\mathbf{v}_{Di}\cdot \mathbf{%
\nabla }\theta \frac{\partial f_{0}}{\partial \theta }+\mathbf{v}_{Di}\cdot 
\mathbf{\nabla }\psi \frac{\partial f_{0}}{\partial \psi }  \label{rh15}
\end{equation}%
where $\mathbf{v}_{Di}=-v_{\parallel }\widehat{\mathbf{n}}\times \mathbf{%
\nabla }\left( \frac{v_{\parallel }}{\Omega _{ci}}\right) $.%
\begin{equation}
\mathbf{v}_{Di}\cdot \mathbf{\nabla }\theta =v_{\parallel }\frac{1}{r}%
RB_{\theta }\frac{\partial }{\partial \psi }\left( \frac{v_{\parallel }}{%
\Omega _{ci}}\right)  \label{rh18}
\end{equation}%
and similarly%
\begin{equation}
\mathbf{v}_{Di}\cdot \mathbf{\nabla }\psi =-\frac{v_{\parallel }}{r}%
RB_{\theta }\frac{\partial }{\partial \theta }\left( \frac{v_{\parallel }}{%
\Omega _{ci}}\right)  \label{rh20}
\end{equation}%
Returning to the initial expression%
\begin{equation}
\mathbf{v}_{Di}\cdot \mathbf{\nabla }f_{0}=I\frac{v_{\parallel }}{qR}\left[ 
\frac{\partial }{\partial \psi }\left( \frac{v_{\parallel }}{\Omega _{ci}}%
\right) \frac{\partial f_{0}}{\partial \theta }-\frac{\partial }{\partial
\theta }\left( \frac{v_{\parallel }}{\Omega _{ci}}\right) \frac{\partial
f_{0}}{\partial \psi }\right]  \label{rh21}
\end{equation}%
The \emph{bounce average} is%
\begin{equation}
\overline{\left( \mathbf{v}_{Di}\cdot \mathbf{\nabla }f_{0}\right) }=\frac{1%
}{T}I\int_{-\theta _{0}}^{+\theta _{0}}\frac{d\theta }{v_{\parallel }/\left(
qR\right) }\frac{v_{\parallel }}{qR}\left[ \frac{\partial }{\partial \psi }%
\left( \frac{v_{\parallel }}{\Omega _{ci}}\right) \frac{\partial f_{0}}{%
\partial \theta }-\frac{\partial }{\partial \theta }\left( \frac{%
v_{\parallel }}{\Omega _{ci}}\right) \frac{\partial f_{0}}{\partial \psi }%
\right]  \label{rh23}
\end{equation}%
Here we replace $f_{0}$ with its expression in terms of $f_{-1}$;%
\begin{equation}
f_{0}\left( \psi ,\theta ,t\right) =-I\left( \frac{v_{\parallel }}{\Omega
_{ci}}\right) \frac{\partial f_{-1}\left( \psi \right) }{\partial \psi }
\label{rh24}
\end{equation}%
\begin{eqnarray}
&&\overline{\left( \mathbf{v}_{Di}\cdot \mathbf{\nabla }f_{0}\right) }
\label{rh241} \\
&=&-\frac{I^{2}}{T}\int_{-\theta _{0}}^{+\theta _{0}}d\theta \left\{ \frac{%
\partial }{\partial \psi }\left( \frac{v_{\parallel }}{\Omega _{ci}}\right) 
\frac{\partial }{\partial \theta }\left( \frac{v_{\parallel }}{\Omega _{ci}}%
\right) \frac{\partial f_{-1}\left( \psi \right) }{\partial \psi }+\frac{%
\partial }{\partial \psi }\left( \frac{v_{\parallel }}{\Omega _{ci}}\right)
\left( \frac{v_{\parallel }}{\Omega _{ci}}\right) \frac{\partial
^{2}f_{-1}\left( \psi \right) }{\partial \theta \partial \psi }\right. 
\notag \\
&&\left. -\frac{\partial }{\partial \theta }\left( \frac{v_{\parallel }}{%
\Omega _{ci}}\right) \frac{\partial }{\partial \psi }\left( \frac{%
v_{\parallel }}{\Omega _{ci}}\right) \frac{\partial f_{-1}\left( \psi
\right) }{\partial \psi }-\frac{\partial }{\partial \theta }\left( \frac{%
v_{\parallel }}{\Omega _{ci}}\right) \left( \frac{v_{\parallel }}{\Omega
_{ci}}\right) \frac{\partial ^{2}f_{-1}}{\partial \psi ^{2}}\right\}  \notag
\end{eqnarray}%
The first and third terms cancel. In addition we know from (\ref%
{fminus1onsurf}) that $f_{-1}$ is constant on the magnetic surfaces. \emph{%
i.e.} the second term is zero. It remains%
\begin{equation}
\overline{\left( \mathbf{v}_{Di}\cdot \mathbf{\nabla }f_{0}\right) }=\frac{%
I^{2}}{T}\frac{\partial ^{2}f_{-1}}{\partial \psi ^{2}}\int_{-\theta
_{0}}^{+\theta _{0}}d\theta \frac{\partial }{\partial \theta }\left[ \frac{1%
}{2}\left( \frac{v_{\parallel }}{\Omega _{ci}}\right) ^{2}\right] =0
\label{vdinablaf0}
\end{equation}

\bigskip

We can now calculate the radial current, using the distribution functions in
orders $-1$, $0$, $1$. This is obtained from the radial projection of the
drift velocity, Eq.(\ref{vdnablapsi})%
\begin{equation}
\mathbf{v}_{Di}\cdot \mathbf{\nabla }\psi =I\frac{v_{\parallel }}{qR}\frac{%
\partial }{\partial \theta }\left( \frac{v_{\parallel }}{\Omega _{ci}}\right)
\label{neo101}
\end{equation}%
The current is projected on the radial direction and the result is averaged
over the magnetic surface, with the operator $\left\langle A\right\rangle
=w^{-1}\int Ard\theta /B_{\theta }$, $w=\int rd\theta /B_{\theta }$, 
\begin{equation}
\left\langle \mathbf{j\cdot \nabla }\psi \right\rangle =\left\vert
e\right\vert \left\langle \int d^{3}v\left( \mathbf{v}_{Di}\cdot \mathbf{%
\nabla }\psi \right) f\right\rangle =-\left\vert e\right\vert I\left\langle
\int d^{3}v\left( \frac{v_{\parallel }}{\Omega _{ci}}\right) v_{\parallel }%
\widehat{\mathbf{n}}\cdot \mathbf{\nabla }f\right\rangle  \label{neo102}
\end{equation}%
An integration by parts over $\theta $ has been done. Two terms are absent:
(1) the order $-1$ distribution function $f_{-1}$ does not contribute due to
(\ref{fminus1onsurf}); and (2) the order $0$ does not contribute, due to (%
\ref{vdinablaf0}). The first order to have a contribution to this current
(averaged over surface) is $f_{1}$. From the equation (\ref{order1}) we take
the term%
\begin{equation}
v_{\parallel }\widehat{\mathbf{n}}\cdot \mathbf{\nabla }f_{1}=-\frac{%
\partial f_{0}}{\partial t}-\mathbf{v}_{Di}\cdot \mathbf{\nabla }f_{0}
\label{neo103}
\end{equation}%
and Eq.(\ref{neo102}) becomes%
\begin{equation}
\left\langle \mathbf{j\cdot \nabla }\psi \right\rangle =-\left\vert
e\right\vert I\left\langle \int d^{3}v\left( \frac{v_{\parallel }}{\Omega
_{ci}}\right) \left( -\frac{\partial f_{0}}{\partial t}-\mathbf{v}_{Di}\cdot 
\mathbf{\nabla }f_{0}\right) \right\rangle  \label{neo104}
\end{equation}%
The surface average operator applied on the second term in the bracket
vanishes%
\begin{equation}
\left\langle \int d^{3}v\left( \frac{v_{\parallel }}{\Omega _{ci}}\right)
\left( \mathbf{v}_{Di}\cdot \mathbf{\nabla }f_{0}\right) \right\rangle =0
\label{neo105}
\end{equation}%
This is shown by a calculation analogous to that of Eq.(\ref{rh15}), using (%
\ref{rh18}) and (\ref{rh20}) followed by the substitution of (\ref{rh24}).
The current is%
\begin{equation}
\left\langle \mathbf{j}\cdot \mathbf{\nabla }\psi \right\rangle =\left\vert
e\right\vert I\left\langle \int d^{3}v\left( \frac{v_{\parallel }}{\Omega
_{ci}}\right) \frac{\partial f_{0}}{\partial t}\right\rangle  \label{neo106}
\end{equation}%
We use the zero-order function $f_{0}$ from Eq.(\ref{rh24}) and take the
time derivative,%
\begin{equation}
\left\langle \mathbf{j\cdot \nabla }\psi \right\rangle =-\left\vert
e\right\vert I^{2}\left\langle \int d^{3}v\left( \frac{v_{\parallel }}{%
\Omega _{ci}}\right) ^{2}\frac{\partial ^{2}f_{-1}}{\partial \psi \partial t}%
\right\rangle  \label{neo107}
\end{equation}%
where, in (\ref{rh12}) we keep the isotropic velocity space integration $%
d^{3}v=4\pi v^{2}dv$ and introduce the factorization (\ref{sxt})%
\begin{equation}
\frac{\partial f_{-1}}{\partial t}=\overline{S}^{ioniz}=\overset{\cdot }{n}%
_{0}^{ioniz}S\left( r,t\right) \frac{1}{4\pi v^{2}}\delta \left(
v-v_{0}\right)  \label{neo108}
\end{equation}%
For circular surfaces%
\begin{equation}
\left\langle j_{r}\right\rangle \approx -\left\vert e\right\vert \frac{%
B_{T}^{2}}{B_{\theta }^{2}}\left\langle \int d^{3}v\left( \frac{v_{\parallel
}}{\Omega _{ci}}\right) ^{2}\frac{\partial ^{2}f_{-1}}{\partial r\partial t}%
\right\rangle  \label{neo109}
\end{equation}%
We note again the presence of the poloidal gyroradius, corrected for trapped
particles ($v_{\parallel }=\left( r/R\right) ^{1/2}v_{th}$),%
\begin{equation}
\frac{B_{T}^{2}}{B_{\theta }^{2}}\left( \frac{v_{\parallel }}{\Omega _{ci}}%
\right) ^{2}\approx \left( \frac{v_{\parallel }}{\Omega _{\theta ci}}\right)
^{2}=\rho _{\theta }^{2}\varepsilon =\rho _{i}^{2}q^{2}\varepsilon ^{-1}
\label{neo110}
\end{equation}%
This length $\left( \rho _{\theta }\varepsilon ^{1/2}\right) $ corresponds
to the radial excursion of the new ion, as defined in our previous approach,
Eq.(\ref{deltatpm})%
\begin{equation}
\rho _{\theta }\varepsilon ^{1/2}=\rho _{i}q\varepsilon ^{-1/2}\approx
\Delta ^{t\pm }  \label{neo110a}
\end{equation}%
After replacing the expression of $\partial f_{-1}/\partial t$ Eq.(\ref%
{neo108}) we have%
\begin{eqnarray}
\left\langle j_{r}\right\rangle &=&-\left\vert e\right\vert \left\langle
\int d^{3}v\rho _{\theta }^{2}\varepsilon \frac{\partial }{\partial r}%
\overset{\cdot }{n}_{0}^{ioniz}S\left( r,t\right) \frac{1}{4\pi v^{2}}\delta
\left( v-v_{0}\right) \right\rangle  \label{neo111} \\
&=&-\left\vert e\right\vert \overset{\cdot }{n}_{0}^{ioniz}\frac{\partial
S\left( r,t\right) }{\partial r}\left\langle \rho _{\theta }^{2}\varepsilon
\right\rangle =-\left\vert e\right\vert \overset{\cdot }{n}_{0}^{ioniz}\frac{%
\partial S\left( r,t\right) }{\partial r}\left\langle \left( \Delta ^{t\pm
}\right) ^{2}\right\rangle  \notag
\end{eqnarray}%
It is understood that $\rho _{\theta }$ and further $\Delta ^{t\pm }=\rho
_{i}q\varepsilon ^{-1/2}$ are calculated at the velocity $v_{0}$. As before
we replace the $\lambda $-integration with multiplication with $\sqrt{%
\varepsilon }$, fraction of trapped particles%
\begin{equation}
j_{r}=-\gamma \left\vert e\right\vert \overset{\cdot }{n}_{0}^{ioniz}\frac{%
\partial S\left( r,t\right) }{\partial r}\rho _{i}^{2}q^{2}\varepsilon
^{-1/2}  \label{neo112}
\end{equation}

The constant $\gamma $ is a purely neoclassical constant and is calculated,
for more general conditions, in \cite{rosenbluthhintonalpha}. It includes
the exact integration over the trapping parameter $\lambda $, which is
contained in $v_{\parallel }$ at $v_{0}$ fixed. The result is $\gamma \sim
0.38$.

We note that the analytic structure of our result Eq.(\ref{jtapp}) and of
its rederivation in the neoclassical drift-kinetic theory, Eq.(\ref{neo112})
are the same as the expression obtained in the treatment of Rosenbluth and
Hinton for the current induced by the $\alpha $ particles \cite%
{rosenbluthhintonalpha}. The coefficient in our approximate treatment Eq.(%
\ref{jtapp}) is $\gamma \sim 1/2$. Figure 5 represents the space dependence
of the current $J\left(r,t_{0}\right) $ obtained from the analysis of the
physical picture, Eq.(\ref{jrtr}), from the numerical model and respectively
from Eq.(\ref{jtapp}) which is also the result of the drift-kinetic
approach. The time $t_{0}$ is chosen at half the total time interval, to
avoid the transient after the onset of ionization.

We note that the conclusion of the mentioned paper, that the rotation
induced by the creation of \emph{alpha} particles is insignificant is a
consequence of the very small nuclear reaction rate. The equivalent
parameter, in the present problem, is the rate of generation of new ions,
which is three orders of magnitude higher in the case of pellets.

The two treatments (the simple arguments related with the fluxes of ions
and, respectively, the drift-kinetic equation) lead to the same result but
there is an apparent difference between them. In the first treatment the
separation of the trajectory of a new ion in a transitory part, where
effective current exist, and a periodic part with no effective radial
current is the key element that identifies the source of current, torque,
rotation. In the drift-kinetic approach this separation is not obvious. One
would expect an \textquotedblleft initial value problem\textquotedblright\
where the distribution function would result as integral over the history of
the ion's motion. This is not visible. The Heaviside function of the source
is not helpful either: it marks the beginning of the ionization process but
after that every moment of time is a source of new ions and this is not
represented. We understand however that the separation is implicitly done
through the velocity space integration. The current is defined as $j\sim
\left\vert e\right\vert \left\langle v_{Di}\int d^{3}v\overline{f}\left(
r,v\right) \right\rangle $. The late phase of the ion orbit is periodic and
the integral mixes to zero the two-way travels on banana, leaving only the first, non-periodic, part.

\section{Discussion and conclusion}

The gas puff, the pellets, the impurity seeding and in general any inflow of
neutrals into plasma produce a substantial radial current and implicitly a
torque that can be higher than the magnetic pumping damping. It can be shown
that it can also be higher than the turbulent Reynolds stress and the
Stringer mechanism. We have derived a simple analytical expression which is
confirmed by numerical simulation. Furthermore, we have re-derived it within
the neoclassical drift-kinetic approach. All three methods have close
quantitative results, as shown in Figure 5. We mention, qualitatively, few
possible consequences.

PEP regimes \cite{pepjetreview} seem to be connected with the
ionization-induced rotation that improves the local confinement by creating
effective barriers through the sheared poloidal flow \cite{itbalcatorfiore}.
The duration of the PEP and the density peaking are compatible with
ionization-induced rotation.

The ionization-induced radial current leads to density peaking, in at least
three different ways. If the gradient of the rate of ionization of a pellet
is negative ($\partial S/\partial r<0$, higher ionization rate close to the
plasma center, as for pellets launched from high-field side) then the
current Eq.(\ref{jtapp}) is directed toward the edge. The bulk ions must
move toward the magnetic axis to compensate this current. Schematically, we
consider a new ion that moves a distance $\Delta $ toward the edge and then
\textquotedblleft stops\textquotedblright\ (actually it moves periodically
on banana). An ion of the background must move in opposite direction the
same distance $\Delta $. But in that moment another new ion is created at a
distance $\Delta $ from this position, closer to the center and starts
moving toward the edge. Then the background ion must continue its
displacement toward the center to compensate this new current. While any new
ion move a distance $\Delta $ then stops, the background ion must continue
to move to compensate the small currents. Quantitatively the two fluxes are
balanced but ions from edge can travel very far toward the center. Impurity
accumulation in the center can also be produced in this process.

Second, the rotation produced at ionization is necessarily sheared, \emph{i.e.%
} $v_{\theta }=v_{\theta }\left( r\right) $, for two reasons. In the regions
of positive and respectively negative radial derivative of the rate of
ionization $\left( \partial S/\partial r\gtrless 0\right) $ the rotation has
opposite direction (Figures 3 and 5). In addition, the background ions must
have a local rotation that is opposite to that of the new ions. The rate of
extraction of the free energy from density gradients is reduced and the
turbulence will have shorter radial correlation length. The rate of
transport decreases and the peaking of the density in the center is enhanced
by the smaller density diffusion.

Third, the shear of the poloidal velocity is actually vorticity $\omega
=\partial v_{\theta }/\partial r$ and when this occurs the Ertel's theorem $%
\frac{d}{dt}\left( \frac{\omega +\Omega _{ci}}{n}\right) =0$ imposes a
redistribution of density.

A change of the density at the edge, by impurity seeding \cite%
{impurityseedingd3d} or by other strong ionization event, must now also be
regarded as an electric process, due to the charge separation and the radial
current of the new ions. It implies that very fast plasma responses should
be expected \cite{florinmadipop}. This may explain observed fast propagation
of perturbations, sometimes called \textquotedblleft
non-local\textquotedblright . Fast increase of the radial electric field is
able to determine, as a neoclassical effect, the reversal of the toroidal
rotation \cite{ricereversal}, \cite{ricetorrot}, \cite{florinmadinf12}.

\bigskip

In conclusion, we have presented arguments that the neoclassical
displacements of the new ions generated at ionization (of gas puff, pellet,
impurity seeding) produce a radial current that can be substantial. The
current is generated from the first part, transitory, unique for any
ionization event, of the trajectory: between the ionization and the moment
where the new ion reaches the stationary periodic motion, trapped or
circulating. The torque resulting from ionization can be substantial and it
can generate internal transport barriers. Our perspective on some particular
regimes may need reconsideration: Pellet Enhanced Performance, density
peaking, density pinch, regimes with density higher than the Greenwald
limit, fast propagation of edge effects, influence of the density on the
transition to \emph{H}-mode, connection between density and rotation,
reversal of toroidal rotation, etc. A more detailed investigation of the
ionization-induced rotation is requested, for each of these cases.

\bigskip

\begin{appendices}
\section{Appendix. Comparison of the currents carried by trapped and
respectively circulating ions} \label{app:appendixa}

\renewcommand{\theequation}{A.\arabic{equation}} \setcounter{equation}{0}

The equation of the closed orbit (poloidal projection of the orbit of a
circulating ion) is%
\begin{equation}
\left( x+\alpha r_{0}\right) ^{2}+y^{2}=r_{0}^{2}  \label{a1}
\end{equation}%
a circle of radius $r_{0}$ that is displaced from the magnetic axis with the
amount $x_{0}=\alpha r_{0}$ where $\alpha =\left( B/B_{T}\right)
v_{Di}/v_{\parallel }=$const. \cite{morozovsolovev}. We have $\alpha \ll 1$
(since $v_{Di}\ll v_{\parallel }\sim v_{th,i}$)$\ $for circulating ions.
Therefore the radial displacement of the \textquotedblleft
center\textquotedblright\ of the orbit of a new ion that is circulating,
relative to the center of the magnetic surface where it is created, is
small. The displacement $x_{0}$ of circulating ions that are created closer
to the edge (at higher $r_{0}$) are larger than that created closer to the
magnetic axis. For two ions born in the same point on the equatorial plane
and with velocities parallel respectively anti-parallel to $\mathbf{B}$ the
difference between these displacements is linear in $r$,%
\begin{eqnarray}
\left\vert x_{0}^{\left( +\right) }-x_{0}^{\left( -\right) }\right\vert
&=&\left\vert \left( x_{0c}+\alpha \frac{\partial r_{0}}{\partial r}%
x_{0}^{\left( +\right) }\right) -\left( x_{0c}-\alpha \frac{\partial r_{0}}{%
\partial r}x_{0}^{\left( -\right) }\right) \right\vert  \label{a2} \\
&=&\alpha \frac{\partial r_{0}}{\partial r}\left( x_{0}^{\left( +\right)
}+x_{0}^{\left( -\right) }\right) =2\alpha ^{2}r_{0}  \notag
\end{eqnarray}%
This is indeed very small, due to $\alpha ^{2}$ factor. The closed orbit for
parallel initial velocity, $\left( +\right) $, is fully contained inside the
magnetic surface, which means that the displacement $x_{0}^{\left( +\right)
} $ is positive. The closed orbit for the anti-parallel initial velocity, $%
\left( -\right) $, fully encloses the magnetic surface, which means that the
displacement $x_{0}^{\left( -\right) }$ is negative. In absolute value $%
x_{0}^{\left( +\right) }$ is greater than $x_{0}^{\left( -\right) }$. Note
that we here use the geometric \textquotedblleft center of the closed
orbit\textquotedblright\ and \emph{not} the asymptotic value of the average $%
\overline{x}\left( t\rightarrow \infty \right) $. The latter are closer the
main axis of symmetry, for both sign of the initial velocities.

The current carried by the new ions for the short time until they access the
stationary periodic motion relies on the difference between the
displacements $n^{c}\left\vert x_{0}^{\left( +\right) }-x_{0}^{\left(
-\right) }\right\vert =n^{c}2\alpha ^{2}r$ is directed towards the main axis
of the torus. The ratio between the width of the banana and the displacement
of the center for a circulating particle originating from the same point is
approximately \cite{morozovsolovev}%
\begin{equation}
\frac{\Delta ^{\pm }}{x_{0}^{\left( \pm \right) }}=4\sqrt{\frac{R_{0}}{r}}
\label{Deltaonx0}
\end{equation}

The second term in Eq.(\ref{jrtr}) is in general greater than the first. We
then compare the current from circulating ions with only this first term. We
find that the circulating ions' contribution is smaller than this term,
which justifies their neglect adopted in the main text. We have to compare
the charge displacements, including the densities of trapped $\left(
n^{t}\right) $ and circulating $\left( n^{c}\right) $ particles. We expand,
taking as reference position the point $r$ where the two ions (with parallel
and anti-parallel $\mathbf{v}_{0}$) are born, $\Delta ^{\left( +\right)
}-\Delta ^{\left( -\right) }=\frac{\partial \Delta }{\partial r}\left[
\Delta ^{\left( +\right) }+\Delta ^{\left( -\right) }\right] $. In the right
hand side, and in all expressions where the difference between $\Delta ^{\pm
}$ is not involved, we can approximate $\Delta =\Delta ^{\pm }\left(
r\right) \approx \rho _{i}q\left( r\right) \varepsilon ^{-1/2}/2$. 
\begin{equation}
\frac{n^{t}\left\vert \Delta ^{\left( +\right) }-\Delta ^{\left( -\right)
}\right\vert }{n^{c}\left\vert x_{0}^{\left( +\right) }-x_{0}^{\left(
-\right) }\right\vert }\approx \frac{\frac{1}{2}n^{t}\frac{\left( \rho
_{i}q\right) ^{2}}{\varepsilon }\frac{\partial }{\partial r}\ln \left(
qr^{-1/2}\right) }{2n^{c}\alpha ^{2}r}  \label{a4}
\end{equation}%
We use $\partial \varepsilon /\partial r\approx 1/R$ and $\left( \rho
_{i}q\right) =2\Delta \sqrt{\varepsilon }$ obtaining the ratio%
\begin{equation}
\frac{1}{\alpha ^{2}r^{2}}R\left( \rho _{i}q\right) ^{2}=r\left( \frac{%
\Delta }{x_{0}}\right) ^{2}  \label{a5}
\end{equation}%
Then%
\begin{equation}
\frac{n_{i}^{t}\left\vert \Delta ^{\left( +\right) }-\Delta ^{\left(
-\right) }\right\vert }{n_{i}^{c}\left\vert x_{0}^{\left( +\right)
}-x_{0}^{\left( -\right) }\right\vert }=\frac{n_{i}^{t}}{n_{i}^{c}}r\left( 
\frac{\Delta }{x_{0}}\right) ^{2}\frac{1}{4}\frac{\partial }{\partial r}\ln
\left( qr^{-1/2}\right)  \label{a6}
\end{equation}

The ratio of the densities of trapped and circulating particles is%
\begin{equation}
\frac{n_{i}^{t}}{n_{i}^{c}}\approx \frac{\sqrt{\varepsilon }}{1-\sqrt{%
\varepsilon }}  \label{a7}
\end{equation}%
and employing Eq.(\ref{Deltaonx0})

Then%
\begin{equation}
\frac{n^{t}\left\vert \Delta ^{\left( +\right) }-\Delta ^{\left( -\right)
}\right\vert }{n^{c}\left\vert x_{0}^{\left( +\right) }-x_{0}^{\left(
-\right) }\right\vert }\approx 4R_{0}\frac{\sqrt{\varepsilon }}{1-\sqrt{%
\varepsilon }}\frac{\partial }{\partial r}\ln \left( qr^{-1/2}\right)
\label{a8}
\end{equation}

In the region of interest $\left[ r_{1},r_{2}\right] $ it is sufficient to
approximate $q\sim 1+\beta \left( r/a\right) ^{2}$ with $\beta $ a constant.
Retaining the dominant term, $\frac{\partial }{\partial r}\ln \left(
qr^{-1/2}\right) \sim \frac{3}{2}\frac{1}{r}$ we have the estimation%
\begin{equation}
\frac{n^{t}\left\vert \Delta ^{\left( +\right) }-\Delta ^{\left( -\right)
}\right\vert }{n^{c}\left\vert x_{0}^{\left( +\right) }-x_{0}^{\left(
-\right) }\right\vert }\approx \frac{6}{\sqrt{\varepsilon }-\varepsilon }\gg
1  \label{a9}
\end{equation}%
This justifies the neglect of the current from the circulating ions in Eqs.(%
\ref{jrttr}) and (\ref{jrtr}).

\section{Appendix. The momentum of the plasma rotation induced by ionization} \label{app:appendixb}

\renewcommand{\theequation}{B.\arabic{equation}} \setcounter{equation}{0}

As mentioned in Section \ref{section3} in the present case the total torque
is zero and there is no problem of conservation of the angular momentum. For
more general $S\left( x,t\right) $ the source of angular momentum and energy
related to the \textquotedblleft spontaneous rotation\textquotedblright\
requires a discussion, which we here attempt in general terms. 

Assume a \emph{slab-like} geometry with the plasma immersed in a static
magnetic field $\mathbf{B=}B\widehat{\mathbf{e}}_{z}$. From external sources
it is applied a transversal electric field $\mathbf{E}=E\widehat{\mathbf{e}}%
_{x}$ field.  There is a motion of the plasma in the direction $\widehat{%
\mathbf{e}}_{y}$ with the speed $\mathbf{v}_{E}=\mathbf{E}\times \mathbf{B}%
/B^{2}$, which apparently violates the conservation of the momentum along
the $y$ direction: before applying $\mathbf{E}$ there was no momentum along $%
y$ but after that we find plasma moving along $y$ with all its particles. 

There is, of course, no \textquotedblleft spontaneous generation of
momentum\textquotedblright . The momentum that would ensure the conservation
and which seems to be missing is actually taken over by the external fields $%
\mathbf{B}$ and $\mathbf{E}$. They are acting as an intermediate medium
transferring the momentum of the guiding center $m_{i}\mathbf{v}_{E}$ to the
external structure (in general coils and condensers) that maintains these
fields.

We consider charges of density $\rho \left( \mathbf{r},t\right) $ and
currents of density $\mathbf{j}\left( \mathbf{r},t\right) $ in a limited
volume $V$ bounded by the surface $\Sigma $. From the Maxwell equation one
derives the local balance%
\begin{equation}
\mathbf{\nabla \cdot }\mathcal{S}=\rho \mathbf{E}+\mathbf{j\times B}%
+\varepsilon _{0}\frac{\partial }{\partial t}\left( \mathbf{E\times B}%
\right)   \label{v10}
\end{equation}%
with $\mathcal{S}=\mathcal{S}^{\left( e\right) }+\mathcal{S}^{\left(
m\right) }$, where the components of the order-two tensor $\mathcal{S}%
^{\left( e\right) }$ are \cite{stratton}
\begin{equation}
\left( \mathcal{S}^{\left( e\right) }\right) _{ij}=\varepsilon
_{0}E_{i}E_{j}-\varepsilon _{0}\frac{1}{2}\delta _{ij}E^{2}  \label{v11}
\end{equation}%
and respectively 
\begin{equation}
\left( \mathcal{S}^{\left( m\right) }\right) _{ij}=\frac{1}{\mu _{0}}%
B_{i}B_{j}-\frac{1}{\mu _{0}}\frac{1}{2}\delta _{ij}B^{2}  \label{v13}
\end{equation}%
For a point-like particle of charge $\left\vert e\right\vert $ with
trajectory $\mathbf{r}_{0}\left( t\right) $, 
\begin{eqnarray}
\rho \left( \mathbf{r},t\right)  &=&\left\vert e\right\vert \delta ^{3}\left[
\mathbf{r-r}_{0}\left( t\right) \right] \ \ \ \left( C/m^{3}\right) 
\label{v14} \\
\mathbf{j}\left( \mathbf{r},t\right)  &=&\left\vert e\right\vert \mathbf{v}%
\delta ^{3}\left[ \mathbf{r-r}_{0}\left( t\right) \right] \ \ \left(
A/m^{2}\right)   \notag
\end{eqnarray}%
Eq.(\ref{v10}) is integrated over the volume $V$ 
\begin{equation}
\int_{\Sigma }\mathcal{S\cdot }\widehat{\mathbf{n}}\ da=\mathbf{F}^{\left(
e\right) }+\mathbf{F}^{\left( m\right) }+\varepsilon _{0}\frac{\partial }{%
\partial t}\int_{V}dv\left( \mathbf{E\times B}\right)   \label{v15}
\end{equation}%
where one notes that the quantity $\varepsilon _{0}\mathbf{E\times B}$ is a
volume density of momentum and the last term%
\begin{equation}
\mathbf{P}^{em}\equiv \varepsilon _{0}\int_{V}dv\left( \mathbf{E\times B}%
\right)   \label{v16}
\end{equation}%
represents the amount of \emph{momentum of the electromagnetic field} inside
the volume $V$. This underlies the role of the tensor $\mathcal{S}$: the
quantity $\mathcal{S\cdot }\widehat{\mathbf{n}}$ is the flux of momentum
through the element of area $da$ of the surface $\Sigma $. The total
mechanical force acting on the particles and currents can be expressed as
the time derivative of the \emph{mechanical momentum} inside $V$ 
\begin{equation}
\mathbf{F}^{\left( e\right) }+\mathbf{F}^{\left( m\right) }=\frac{d}{dt}%
\mathbf{P}^{mech}  \label{v17}
\end{equation}%
Then the equation becomes%
\begin{equation}
\int_{\Sigma }\mathcal{S\cdot }\widehat{\mathbf{n}}\ da=\frac{d}{dt}\left( 
\mathbf{P}^{mech}+\mathbf{P}^{em}\right)   \label{v18}
\end{equation}%
Anything that changes inside $V$ (field, motion) must be balanced by a
reaction from the world exterior to $V$. Now we imagine that $\Sigma $ is
very large, enclosing the toroidal coils, etc. such that the fields have
vanished on the boundary. Then%
\begin{equation}
\frac{d}{dt}\mathbf{P}^{mech}=-\frac{d}{dt}\mathbf{P}^{em}  \label{v19}
\end{equation}%
which expresses in the most clear way the idea of this discussion: any
modification of the mechanical momentum of the charged particle must be
balanced by an opposite modification of the electromagnetic field. For
externally applied $\left( \mathbf{B,E}\right) $ fields, the momentum (and
angular momentum) is sustained by a reaction against the sources of the
fields, coils and condensers.

When the electric field is generated by charge separation inside plasma, the
momentum and the energy of the guiding centers will ncessarly involve the
momentum and energy inside plasma, besides those external to it. After
ionization, the new ions take energy by interacting with the background
plasma and it is with this energy that they move to settle on final periodic
orbits. Their motion produces the layer of unbalanced ion charge at the edge
of the ionization region and the resulting electric ($E^{I}$) field is fully
dependent on the energy of the new ions. Then the momentum and energy of the
plasma moving with $\mathbf{v}_{E}=\mathbf{E}^{I}\mathbf{\times B}/B^{2}$
have their origin in the energy that the new ions could get from the
background plasma. The rotation $\mathbf{v}_{E}$ is a backreaction, like an
inertia. The plasma responds to $E^{I}$ by polarization drift of the
background ions. Moving on a distance $\delta x^{L}=\left( \Omega
_{ci}B\right) ^{-1}E^{I}$ in the field $E^{I}$ the variation in energy is $%
\delta W=n^{bg}\left\vert e\right\vert E^{I}\times \left( \Omega
_{ci}B\right) ^{-1}E^{I}=\delta \left( n^{bg}\frac{m_{i}v_{E}^{2}}{2}\right) 
$ which is the variation in energy of the rotation of the plasma. We can see that  the plasma rotation $%
v_{E}$ is significant only if 
the background plasma feeds the new ions with sufficient energy for them to
reach trapping orbits with substantial $\Delta ^{t\pm }$ \emph{i.e.} such that the
charge separation produces a large $\mathbf{E}^{I}$.

\section{Appendix. Plasma response to the charge separation induced by the
displacement of the new ions} \label{app:appendixc}

\renewcommand{\theequation}{C.\arabic{equation}} \setcounter{equation}{0}

After ionization the new ions move to take their neoclassical periodic
orbit. Between the point of creation and the \textquotedblleft
center\textquotedblright\ of the periodic motion they carry a transitory,
short, finite current. As in the main text we consider the ionization to
take place in a volume limited between the radii $r_{1}$ (left side, closer
to the center) $r_{2}$ (right side, closer to the last closed magnetic
surface). The plasma is considered homogeneous and the ionization generates
ions that move to the right a distance $\Delta ^{t}$ with velocity $v_{Di}$
while the new electrons can be considered imobile. Then most of the volume
between $r_{1}$ and $r_{2}$ is neutral but at the right end $\sim r_{2}$ of
the ionization interval it results a layer (denoted $I$) of positive charge,
of width $\Delta ^{t}$. This is the source of electric field, resulted from
ionization, $E^{I}$, directed from $r_{2}$ towards $r_{1}$. The charge layer 
$I$ and its field $E^{I}$ are built up on a time scale $\delta t=\Delta
^{t}/v_{Di}$ in which $\partial E^{I}/\partial t>0$. The background plasma
responds by modifying the Larmor gyration orbit from the usual circle to a
cycloid (actually the new orbit is a \emph{prolate trochoid}). The
deformation of the gyration directly indicates the expected $\mathbf{v}_{E}=%
\mathbf{E\times B/}B^{2}$ motion and also the asymmetry of the charge
distribution along the new orbit. The asymmetry creates a new layer (denoted 
$L$) of positive (ion) charge, at the left end, $\sim r_{1}$ and an electric
field $E^{L}$ opposite to $E^{I}$. This field is sufficient to almost cancel 
$E^{I}$ inside plasma, leaving in the interior a small $E^{int}=E^{I}-E^{L}$%
, directed to the left, like $E^{I}$. The asymmetry of the modified Larmor
orbit is a manifestation of the polarization drift induced by the variation
in time of the electric field $E^{I}$ (implicitely $E^{ind}$), a
displacement of the background ions in the direction to which points $E^{I}$ (\emph{i.e.}
to the left). Since the ionization continues to accumulate new ions in the
layer $I$, hence $\partial E^{I}/\partial t>0$ , the ion's drift of
polarization $v_{Di}^{\left( pol\right) }$ fills the layer $L$ at the left
end, whose electric field $E^{L}$ (from $r_{1}$ toward $r_{2}$) continues to
quasi-compensate $E^{I}$ inside plasma.

Qualitatively, this picture conforms to the concept of \emph{return current}%
, which is universally invoked as the plasma response to any mechanism that
is able to produce rotation of only some component of the density: NBI,
ICRH, \emph{alpha} particle, etc., to which we add: ionization. In the
following we examine the density of charge and respectively the current
density arising from ionization, deformation of the Larmor orbit and finally
the polarization drift. The ionization-induced charge separation and current
are regarded as \textquotedblleft external\textquotedblright\ factors since
they are requested by the geometry of the field.

We first include a justification of the neglect of the volume-charge accumulation
that can be associated with the strong vorticity.

\subsection{Charge and current related to the vorticity}

At the edge of the tokamak in the $H$-mode regime there is a layer of strong
poloidal rotation, with radial extension of about a banana width calculated
for the poloidal magnetic field. The variation of the velocity magnitude is
very fast in this layer, or, equivalently, the layer is a concentration of
vorticity $\mathbf{\omega }=\mathbf{\nabla \times v}$ or $\omega \sim \frac{%
\partial v_{\theta }}{\partial r}$. Taking as usual $\mathbf{v=v}_{E}=\frac{-%
\mathbf{\nabla }\phi \times \widehat{\mathbf{e}}_{z}}{B}$ we have $\mathbf{%
\omega }=\mathbf{\nabla \times v=-}\frac{\mathbf{\nabla }^{2}\phi }{B}%
\widehat{\mathbf{e}}_{z}$ or%
\begin{equation}
\omega =-\mathbf{\nabla }^{2}\phi /B  \label{w10}
\end{equation}%
directed along the magnetic field line. The Laplacian of the electric
potential $\Delta \phi $ is the electric charge density and we have the well
known situation that a vorticity is formally equivalent to a density of
electric charge. If this charge is quantitatively important, it must be
taken it into account together with the currents%
\begin{equation}
\frac{\partial \rho ^{V}}{\partial t}+\mathbf{\nabla \cdot J}=0  \label{w11}
\end{equation}%
where the \textquotedblleft charge\textquotedblright\ $\mathbf{\nabla }%
^{2}\phi =-\rho ^{V}/\varepsilon _{0}$\ is the \emph{vorticity}%
\begin{equation}
\rho ^{V}=-\varepsilon _{0}\omega B  \label{w12}
\end{equation}

We can estimate the magnitude of the charge density $\rho ^{V}$. If the
poloidal velocity has a spatial variation from $v_{\theta }=0$ at the edge
of the rotation layer and reaches amplitude of $\sim 10\ \left( km/s\right) $
on a radial extension of $10^{-2}$ $\left( m\right) $ then $\omega \sim
10^{7}\ \left( s^{-1}\right) $ and this means%
\begin{equation}
\rho ^{V}\approx 3\times 10^{-4}\ \left( C/m^{3}\right)  \label{w13}
\end{equation}

If the formation of this vorticity layer takes place on an interval
controlled by the drift of the ions then%
\begin{equation}
\delta t\sim \frac{\delta r}{v_{Di}}=\frac{10^{-2}\left( m\right) }{30\left(
m/s\right) }=3\times 10^{-4}\left( s\right)   \label{w14}
\end{equation}%
and the time variation of the charge is 
\begin{equation}
\frac{\partial \rho ^{V}}{\partial t}\sim \frac{\rho ^{V}}{\delta t}=\frac{%
3\times 10^{-4}\left( C/m^{3}\right) }{3\times 10^{-4}\left( s\right) }%
=1\left( \frac{A}{m^{3}}\right)   \label{w15}
\end{equation}%
This must be compared with $\mathbf{\nabla \cdot J}^{I}$. Taking the value
estimated in the text $J^{I}\sim 10\ \left( A/m^{2}\right) $ and a spatial
variation on the same radial extension $\delta r\sim 10^{-2}\left( m\right) $
we have $\left\Vert \mathbf{\nabla \cdot J}^{I}\right\Vert \sim 10^{3}\left(
A/m^{3}\right) $. This is much higher than the time derivative of the
vorticity-charge, so we can neglect this latter component of the physical
picture. We must remember however that $J^{I}\sim 10\ \left( A/m^{2}\right) $
is obtained for pellets while in other cases (\emph{e.g.} neutrals
penetrating from the edge) can be orders of magnitude smaller. In addition the interaction between ions and neutrals in this region is complex \cite{fulop1}. Then we have
to check the possibility to neglect the vorticity-charge.

\bigskip

\subsection{The charge accumulation and the current induced by ionization}

The rate of increase of the density of charge $\rho ^{I}$ by influx of the
new ions in the region of unbalanced charge at the right end $\sim r_{2}$ of
the segment of ionization (the charge layer $I$) is $d\rho ^{I}/dt\sim
\left\vert e\right\vert \dot{n}_{ioniz}\left[ \Theta \left(
r-r_{2}\right) \Theta \left( r_{2}+\Delta ^{t}-r\right) \right] \ (C/m^{3}/s)
$, where $\Theta $ is the Heaviside function. The width of the layer $I$ is $%
\Delta ^{t}$ and the time scale to fill with newly born ions is $\delta
t\sim \Delta ^{t}/v_{Di}$. Using the Gauss law%
\begin{equation}
\frac{\partial E^{I}}{\partial x}=\frac{1}{\varepsilon _{0}}\rho ^{I}
\label{w16}
\end{equation}%
we take the time derivation and integrate over $x$ (actually $x\equiv r$ and
we use $x$ instead of $r$ to underline the $1D$ geometry assumed here), 
\begin{equation}
\frac{\partial E^{I}}{\partial t}=\Delta ^{t}\frac{1}{\varepsilon _{0}}%
\left\vert e\right\vert \dot{n}_{ioniz}\ \left( \frac{V}{ms}%
\right)   \label{w17}
\end{equation}%
According to the source of ionization (neutrals penetrating from the edge,
pellets, etc.) $\left\Vert \partial E^{I}/\partial t\right\Vert $ can vary
over an interval of three orders of magnitude $\sim 10^{9}...10^{12}\ \left( 
\frac{V}{ms}\right) $. The static magnitude of $E^{I}$ can be obtained from
the surface charge density $\sigma ^{I}=\rho ^{I}\Delta ^{t}$, as $%
E^{I}=\sigma ^{I}/\varepsilon _{0}$.

The current density induced by ionization is calculated in the main text

\begin{equation}
J^{I}\left( r\right) \approx -\frac{1}{2}\left\vert e\right\vert %
\dot {n}_{0}^{ioniz}\left( \frac{\partial S}{\partial r}\right) \rho
_{i}^{2}q^{2}\varepsilon ^{-1/2}  \label{w18}
\end{equation}%
However this calculation was adapted to a particular class of cases and it
must be reconsidered for other cases. We just mention the order of magnitude $%
J^{I}\left( x_{C},t\right) \approx 10\ \left[ A/m^{2}\right] $.

\bigskip

\subsection{The charge accumulation and the current produced by the ion's
drift of polarization}

An ion in a constant magnetic field $\mathbf{B}=\widehat{\mathbf{e}}_{z}B$
performs the Larmor gyration in the transversal plane $xOy$, on a circle of
radius $\rho _{i}$ with frequency $\Omega _{ci}=\left\vert e\right\vert
B/m_{i}$. When a constant electric field $\mathbf{E}=\widehat{\mathbf{e}}%
_{x}E$ is added, the circle is deformed into a curve of the cycloid type.
Integrating the equation 
\begin{equation}
\frac{d^{2}\mathbf{r}}{dt^{2}}=\frac{\left\vert e\right\vert E}{m_{i}}%
\widehat{\mathbf{e}}_{x}+\frac{\left\vert e\right\vert B}{m_{i}}\frac{d%
\mathbf{r}}{dt}\times \widehat{\mathbf{e}}_{z}  \label{w19}
\end{equation}%
we obtain%
\begin{equation}
x\left( t\right) =\frac{1}{\Omega _{ci}^{2}}\frac{\left\vert e\right\vert E}{%
m_{i}}+\frac{v_{y0}}{\Omega _{ci}}+x_{0}-\frac{1}{\Omega _{ci}}\left( \frac{1%
}{\Omega _{ci}}\frac{\left\vert e\right\vert E}{m_{i}}+v_{y0}\right) \cos
\left( \Omega _{ci}t\right) +\frac{v_{x0}}{\Omega _{ci}}\sin \left( \Omega
_{ci}t\right)  \label{w20}
\end{equation}%
\begin{equation}
y\left( t\right) =y_{0}-\frac{v_{x0}}{\Omega _{ci}}-\frac{1}{\Omega _{ci}}%
\frac{\left\vert e\right\vert E}{m_{i}}t+\frac{1}{\Omega _{ci}}\left( \frac{1%
}{\Omega _{ci}}\frac{\left\vert e\right\vert E}{m_{i}}+v_{y0}\right) \sin
\left( \Omega _{ci}t\right) +\frac{v_{x0}}{\Omega _{ci}}\cos \left( \Omega
_{ci}t\right)  \label{w21}
\end{equation}%

\begin{figure}[h]
\includegraphics[height=10cm]{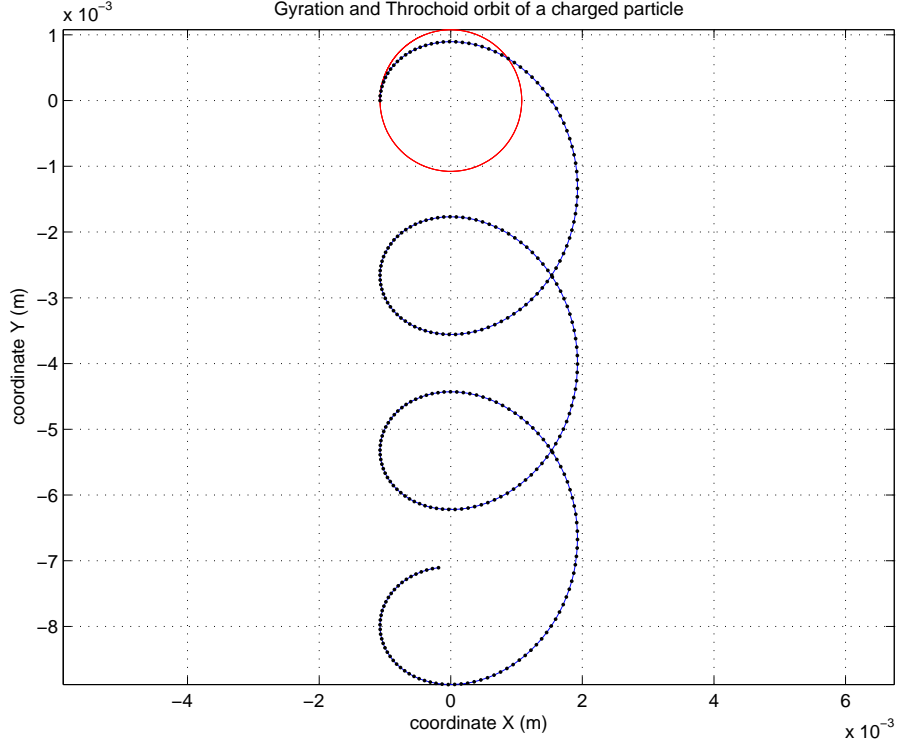}
\caption{The deformation of the pure gyration orbit (red) into a trochoid (blue) under the effect of an electric field. A very high value of the electric field (350 kV/m) was used in order to make more visible the deformation. The plasma center is at right.}
\label{figure6}
\end{figure}
%

This curve (see Figure \ref{figure6} ) is a \emph{prolate trochoid}. The electric $E\times B$ motion is
along the negative $y$ axis. We adopt initial conditions ($x_{0}=-\rho _{i}$%
, $y_{0}=0$, $v_{x0}=0$, $v_{y0}=v_{th,i}$) that are identical for the
static $E\neq 0$ as well as for pure gyration ($E\equiv 0$). In this way we
can see how the \emph{trochoid} is different of the Larmor circle. The orbit
has, broadly, two unequal lobes. This asymmetry makes that the
\textquotedblleft center\textquotedblright\ of the positions of the particle
to be shifted relative to the one of the pure Larmor gyration. For ions the
shift is in the direction of the electric field $\mathbf{E}^{I}$. The ions
are now more frequently present to the left of the symmetry axis of the
previously symmetric (circle) orbit. There is an effective concentration of
ions at the end of the interval on $r$ to which points the electric field
produced by the \textquotedblleft external\textquotedblright\ $\mathbf{E}%
^{I} $ (generated by ionization). For electrons there is a shift to the
opposite direction but much smaller and will be neglected.

Therefore for a given $\mathbf{E}^{I}$ there is an excess of ion charge at
the left ($\sim r_{1}$) end of the ionization domain. A new layer (called $L$%
) of positive charge is generated at the left end, opposite to the
ionization-induced layer $I$. The width $\delta x^{L}$ is the amount of
deformation relative to the pure Larmor gyration orbit, \emph{i.e.} the
distance between the center of the \emph{prolate trochoid} and the center of
the Larmor circle, when both trajectories start from the same initial
conditions, but with $E\neq 0$ respectively $E=0$. \ To find it, we
calculate the time evolution of the averages%
\begin{equation}
\overline{x}\left( t\right) =\frac{1}{t}\int_{0}^{t}x\left( t^{\prime
}\right) dt^{\prime }\ \ ,\ \ \ \overline{y}\left( t\right) =\frac{1}{t}%
\int_{0}^{t}y\left( t^{\prime }\right) dt^{\prime }  \label{w23}
\end{equation}

For large $t$%
\begin{equation}
\overline{x}\left( t\rightarrow \infty \right) =\frac{1}{\Omega _{ci}^{2}}%
\frac{\left\vert e\right\vert E}{m_{i}}+\frac{v_{y0}}{\Omega _{ci}}+x_{0}=%
\frac{1}{\Omega _{ci}B}E  \label{w24}
\end{equation}%
and%
\begin{equation}
\overline{y}\left( t\rightarrow \infty \right) =y_{0}-\frac{v_{x0}}{\Omega
_{ci}}+\frac{1}{2}\left( -\frac{1}{\Omega _{ci}}\frac{\left\vert
e\right\vert E}{m_{i}}\right) t=-\frac{E}{2B}t  \label{w25}
\end{equation}

The \emph{center} of the new orbit has a shift $\delta x^{L}=\overline{x}%
\left( t\rightarrow \infty \right) =\frac{1}{\Omega _{ci}B}E$. The electric
field that occurs in the above equation is the \emph{internal} field $%
E^{int} $, \emph{i.e.} the ionization-induced field $E^{I}$ from which we
substract the field generated by the new layer $L$, $E^{int}=E^{I}-E^{L}$.
Using the shift $\delta x^{L}=\frac{1}{\Omega _{ci}B}E^{int}$ the surface
charge density in the layer $L$ is 
\begin{equation}
\sigma ^{L}=\left\vert e\right\vert n^{bg}\frac{1}{\Omega _{ci}B}E^{int}\ \
\ \left( \frac{C}{m^{2}}\right)  \label{w26}
\end{equation}%
The electric field produced by the deformation of the Larmor gyration is%
\begin{equation}
E^{L}=\frac{\sigma ^{L}}{\varepsilon _{0}}=\frac{1}{\varepsilon _{0}}%
\left\vert e\right\vert n^{bg}\frac{1}{\Omega _{ci}B}E^{int}=\frac{c^{2}}{%
v_{A}^{2}}E^{int}  \label{w27}
\end{equation}%
from where we find $E^{int}=E^{I}-E^{L}=E^{I}-\frac{c^{2}}{v_{A}^{2}}E^{int}$
, or%
\begin{equation}
E^{int}=\frac{E^{I}}{1+c^{2}/v_{A}^{2}}  \label{w28}
\end{equation}

If the \textquotedblleft external\textquotedblright , ionization-induced,
electric field $E^{I}$ continues to increase, there is \emph{increase} in
time of the deformation of the trochoid%
\begin{equation}
\frac{d}{dt}\overline{x}\left( t\rightarrow \infty \right) =\frac{1}{\Omega
_{ci}B}\frac{dE^{int}}{dt}  \label{w29}
\end{equation}%
which is precisely the \emph{polarization drift} of the ions%
\begin{equation}
v_{Di}^{\left( pol\right) }=\frac{1}{\Omega _{ci}B}\frac{dE^{int}}{dt}
\label{w30}
\end{equation}%
\emph{i.e.} the drift of polarization simply consists of the time variation
of the deformation $v_{Di}^{\left( pol\right) }=\frac{d}{dt}\overline{x}%
\left( t\rightarrow \infty \right) $.

The velocity of the polarization drift of the background ions (of density $n^{bg}$)%
\begin{equation}
v_{Di}^{\left( pol\right) }=\frac{1}{\Omega _{ci}B}\frac{1}{1+c^{2}/v_{A}^{2}%
}\frac{dE^{I}}{dt}\approx \frac{\varepsilon _{0}}{\left\vert e\right\vert
n^{bg}}\frac{dE^{I}}{dt}  \label{w31}
\end{equation}%
is in general much smaller than the first order drift $v_{E}$ and than the
neoclassical drift $v_{Di}$. The estimated magnitude varies between $%
v_{Di}^{\left( pol\right) }\sim 0.02...10\ \left( m/s\right) $ according to $%
\dot{n}_{ioniz}$ is determined by slow gas input or pellets. We
would be tempted to expect a slower response of the background ions. The
build-up of the charge layer induced by ionization is $\delta t=\Delta
^{t}/v_{Di}$ while the build up of the charge layer induced by polarization
drift is $\delta t^{\left( pol\right) }=\delta x^{L}/v_{Di}^{\left(
pol\right) }$. However these two time scales are identical. Using Eqs.(\ref%
{w29}-\ref{w30}) and (\ref{w28})%
\begin{equation}
\delta t^{\left( pol\right) }=\frac{\delta x^{L}}{v_{Di}^{\left( pol\right) }%
}=\left( \frac{d}{dt}\ln E^{I}\right) ^{-1}  \label{w312}
\end{equation}%
and inserting Eq.(\ref{w17}) and $\delta t=\Delta ^{t}/v_{Di}$ we find%
\begin{equation}
\delta t^{\left( pol\right) }=\delta t  \label{w313}
\end{equation}

Using again Eq.(\ref{w17}) we obtain%
\begin{equation}
v_{Di}^{\left( pol\right) }=\frac{1}{\Omega _{ci}B}\frac{1}{1+c^{2}/v_{A}^{2}%
}\Delta ^{t}\frac{1}{\varepsilon _{0}}\left\vert e\right\vert \dot%
{n}_{ioniz}=\Delta ^{t}\frac{\dot{n}_{ioniz}}{n^{bg}}
\label{eqcu}
\end{equation}%
This can be translated in the language of currents. By definition%
\begin{equation}
J^{\left( pol\right) }=\left\vert e\right\vert n^{bg}v_{Di}^{\left(
pol\right) }  \label{w35}
\end{equation}%
and 
\begin{equation}
J^{I}=\left\vert e\right\vert \Delta ^{t}\dot{n}_{ioniz}
\label{w36}
\end{equation}%
and Eq.(\ref{eqcu}) shows that the polarization current $J^{\left(
pol\right) }$ is equal and opposite to the \textquotedblleft
externally\textquotedblright\ imposed current $J^{I}$.%
\begin{equation}
J^{I}=J^{\left( pol\right) }  \label{w37}
\end{equation}%

The plasma response $J^{\left( pol\right) }$ is the return current,
involving the background ions. 

\end{appendices}

\bigskip

\bigskip 

\begin{thebibliography}{10}

\bibitem{improvedcorefuelling}
L.~R. Baylor, T.~C. Jernigan, S.~K. Combs, W.~A. Houlberg, M.~Murakami,
  P.~Gohil, K.~H. Burrell, C.~M. Greenfield, R.~J. Groebner, C.-L. Hsieh,
  R.~J.~La Haye, P.~B. Parks, G.~M. Staebler, DIII-D Team, G.~L. Schmidt, D.~R.
  Ernst, E.~J. Synakowski, and M.~Porkolab.
\newblock Improved core fueling with high field side pellet injection in the
  diii-d tokamak.
\newblock {\em Phys. Plasmas}, 7:1878--1885, 2000.

\bibitem{baylorjetd3d}
L.R. Baylor, T.C. Jernigan, P.B. Parks, G.~Antar, N.H. Brooks, S.K. Combs, D.T.
  Fehling, C.R. Foust, W.A. Houlberg, and G.L. Schmidt.
\newblock Comparison of deuterium pellet injection from different locations on
  the diii-d tokamak.
\newblock {\em Nuclear Fusion}, 47(11):1598, 2007.

\bibitem{baylorjettftr}
L.R. Baylor, G.L. Schmidt, W.A. Houlberg, S.L. Milora, C.W. Gowers, W.P.
  Bailey, M.~Gadeberg, P.~Kupschus, J.A. Tagle, D.K. Owens, D.K. Mansfield, and
  H.K. Park.
\newblock Pellet fuelling deposition measurements on jet and tftr.
\newblock {\em Nuclear Fusion}, 32(12):2177, 1992.

\bibitem{corepoloidaltftr}
R~E Bell, F~M Levinton, S~H Batha, E~J Synakowski, and M~C Zarnstorff.
\newblock Core poloidal rotation and internal transport barrier formation in
  tftr.
\newblock {\em Plasma Physics and Controlled Fusion}, 40(5):609, 1998.

\bibitem{berkgaleev}
H.A. Berk and A.A. Galeev.
\newblock Velocity space instabilities in a toroidal geometry.
\newblock {\em Phys. Fluids}, 10:441--450, 1967.

\bibitem{itbalcatorfiore}
C.~L. Fiore, D.~R. Ernst, J.~E. Rice, K.~Zhurovich, N.~Basse, P.~T. Bonoli,
  M.~J. Greenwald, E.~S. Marmar, and S.~J. Wukitch.
\newblock Internal transport barriers in alcator c-mod.
\newblock {\em Fusion Science and Technology}, 51:303--316, 2007.

\bibitem{fonghahm}
B.H. Fong and T.S. Hahm.
\newblock Bounce averaged kinetic equations and neoclassical polarization
  density.
\newblock {\em Physics of Plasmas}, 6:189--199, 1999.

\bibitem{fulop1}
T.~Fulop, Peter~J. Catto, and P.~Helander.
\newblock Neutral diffusion and anomalous effects on collisional ion flow shear
  in tokamaks.
\newblock {\em Physics of Plasmas (1994-present)}, 5(11):3969--3973, 1998.

\bibitem{galeevsagdeev}
A.A. Galeev and R.Z. Sagdeev.
\newblock Theory of neoclassical diffusion.
\newblock In M.A. Leontovich, editor, {\em Reviews of Plasma Physics},
  volume~7, pages 257--343. Consultants Bureau, New York, 1979.

\bibitem{hassamkulsrud}
Adil~B. Hassam and Russell~M. Kulsrud.
\newblock Time evolution of mass flows in a collisional tokamak.
\newblock {\em Physics of Fluids (1958-1988)}, 21(12):2271--2279, 1978.

\bibitem{hintonhazeltinermp}
F.L. Hinton and R.D. Hazeltine.
\newblock Theory of plasma transport in toroidal confinement systems.
\newblock {\em Rev. Mod. Phys.}, 48:239--308, 1976.

\bibitem{hintonrosenbluthnbi}
F.L. Hinton and M.N. Rosenbluth.
\newblock The mechanism for the toroidal momentum input to tokamak plasma from
  neutral beams.
\newblock {\em Phys. Letters}, A259:267--275, 1999.

\bibitem{pelletjet1}
W.A. Houlberg, S.E. Attenberger, L.R. Baylor, M.~Gadeberg, T.C. Jernigan,
  P.~Kuschus, S.L. Milora, G.L. Schmidt, D.W. Swain, and M.L. Watkins.
\newblock Pellet penetration experiments on jet.
\newblock {\em Nucl. Fusion}, 32:1951--1965, 1992.

\bibitem{Hugon}
M.~Hugon, B.Ph. van Milligen, P.~Smeulders, L.C. Appel, D.V. Bartlett,
  D.~Boucher, A.W. Edwards, L.-G. Eriksson, C.W. Gowers, T.C. Hender,
  G.~Huysmans, J.J. Jacquinot, P.~Kupschus, L.~Porte, P.H. Rebut, D.F.H. Start,
  F.~Tibone, B.J.D. Tubbing, M.L. Watkins, and W.~Zwingmann.
\newblock Shear reversal and mhd activity during pellet enhanced performance
  pulses in jet.
\newblock {\em Nuclear Fusion}, 32(1):33, 1992.

\bibitem{impurityseedingd3d}
G.L. Jackson, M.~Murakami, G.R. McKee, D.R. Baker, J.A. Boedo, R.J.~La Haye,
  C.J. Lasnier, A.W. Leonard, A.M. Messiaen, J.~Ongena, G.M. Staebler,
  B.~Unterberg, M.R. Wade, J.G. Watkins, and W.P. West.
\newblock Effects of impurity seeding in diii-d radiating mantle discharges.
\newblock {\em Nuclear Fusion}, 42(1):28, 2002.

\bibitem{jetrefuelling}
P~T Lang, B~Alper, L~R Baylor, M~Beurskens, J~G Cordey, R~Dux, R~Felton,
  L~Garzotti, G~Haas, L~D Horton, S~Jachmich, T~T~C Jones, A~Lorenz, P~J Lomas,
  M~Maraschek, H~W Müller, J~Ongena, J~Rapp, K~F Renk, M~Reich, R~Sartori,
  G~Schmidt, M~Stamp, W~Suttrop, E~Villedieu, D~Wilson, and EFDA-JET
  workprogramme collaborators.
\newblock High density operation at jet by pellet refuelling.
\newblock {\em Plasma Physics and Controlled Fusion}, 44(9):1919, 2002.

\bibitem{globaltoresupra}
Wandong Liu and M.~Talvard.
\newblock Rapid global response of the electron temperature during pellet
  injection on tore supra.
\newblock {\em Nuclear Fusion}, 34(3):337, 1994.

\bibitem{morozovsolovev}
A.I. Morozov and L.S. Solovev.
\newblock Motion of charged particles in electro-magnetic fields.
\newblock In M.A. Leontovich, editor, {\em Reviews of Plasma Physics},
  volume~2, pages 201--297. Consultants Bureau, New York, 1966.

\bibitem{nycanderyankovtrapped}
J.~Nycander and V.V. Yankov.
\newblock H-mode in tokamaks attributed to absence of trapped ions in
  poloidally rotating plasma.
\newblock {\em Pis’ma Zh. Eksp. Teor. Fiz.}, 63(6):427--430, 1996.

\bibitem{ricetorrot}
J.~E. Rice, W.~D. Lee, E.~S. Marmar, N.~P. Basse, P.~T. Bonoli, M.~J.
  Greenwald, A.~E. Hubbard, J.~W. Hughes, I.~H. Hutchinson, A.~Ince-Cushman,
  J.~H. Irby, Y.~Lin, D.~Mossessian, J.~A. Snipes, S.~M. Wolfe, S.~J. Wukitch,
  and K.~Zhurovich.
\newblock Toroidal rotation and momentum transport in alcator c-mod plasmas
  with no momentum input.
\newblock {\em Physics of Plasmas (1994-present)}, 11(5):2427--2432, 2004.

\bibitem{ricereversal}
J.E. Rice, B.P. Duval, M.L. Reinke, Y.A. Podpaly, A.~Bortolon, R.M. Churchill,
  I.~Cziegler, P.H. Diamond, A.~Dominguez, P.C. Ennever, C.L. Fiore, R.S.
  Granetz, M.J. Greenwald, A.E. Hubbard, J.W. Hughes, J.H. Irby, Y.~Ma, E.S.
  Marmar, R.M. McDermott, M.~Porkolab, N.~Tsujii, and S.M. Wolfe.
\newblock Observations of core toroidal rotation reversals in alcator c-mod
  ohmic l-mode plasmas.
\newblock {\em Nuclear Fusion}, 51(8):083005, 2011.

\bibitem{rhh1972}
M.~N. Rosenbluth, R.~D. Hazeltine, and F.~L. Hinton.
\newblock Plasma transport in toroidal confinement systems.
\newblock {\em Physics of Fluids (1958-1988)}, 15(1):116--140, 1972.

\bibitem{rosenbluthhintonalpha}
M.N. Rosenbluth and F.L. Hinton.
\newblock Plasma rotation driven by alpha particles in a tokamak reactor.
\newblock {\em Nucl. Fusion}, 36:55--67, 1996.

\bibitem{pepjetreview}
P.~Smeulders, L.C. Appel, B.~Balet, T.C. Hender, L.~Lauro-Taroni, D.~Stork,
  B.~Wolle, S.~Ali-Arshad, B.~Alper, H.J.~De Blank, M.~Bures, B.~De Esch,
  R.~Giannella, R.~Konig, P.~Kupschus, K.~Lawson, F.B. Marcus, M.~Mattioli,
  H.W. Morsi, D.P. O'Brien, J.~O'Rourke, G.J. Sadler, G.L. Schmidt, P.M.
  Stubberfield, and W.~Zwingmann.
\newblock Survey of pellet enhanced performance in jet discharges.
\newblock {\em Nuclear Fusion}, 35(2):225, 1995.

\bibitem{florinmadipop}
F.~Spineanu and M.~Vlad.
\newblock Fluctuation of the ambipolar equilibrium in magnetic perturbations.
\newblock {\em Physics of Plasmas}, 9(12):5125--5128, 2002.

\bibitem{florinmadinf12}
F.~Spineanu and M.~Vlad.
\newblock A model for the reversal of the toroidal rotation in tokamak.
\newblock {\em Nuclear Fusion}, 52:114019, 2012.

\bibitem{florinmadieps13}
F.~Spineanu and M.~Vlad.
\newblock The role of the rotation in the correlated transient change of the
  density and confinement.
\newblock $40^{th}$ EPS Conference on Plasma Physics, 2013.
\newblock Helsinki, Finland, 1-5 July 2013. Paper P1.178.

\bibitem{stratton}
Julius~Adams Stratton.
\newblock {\em Electromagnetic Theory}.
\newblock McGraw-Hill Book Company, 1941.

\bibitem{startrotationtournianski}
M.R. Tournianski, P.G. Carolan, N.J. Conway, G.F. Counsell, A.R. Field, and
  M.J. Walsh.
\newblock Poloidal rotation and associated edge behaviour in start plasmas.
\newblock {\em Nuclear Fusion}, 41(1):77, 2001.

\bibitem{valovicmast}
M.~Valovic, L.~Garzotti, C.~Gurl, R.~Akers, J.~Harrison, C.~Michael, G.~Naylor,
  R.~Scannell, and the MAST~team.
\newblock H-mode access by pellet fuelling in the mast tokamak.
\newblock {\em Nuclear Fusion}, 52(11):114022, 2012.

\bibitem{wongburrell}
S.K. Wong and K.H. Burrell.
\newblock Transport theory of tokamak plasmas with large toroidal rotation.
\newblock {\em Phys. Fluids}, 25:1863--1870, 1982.

\end{thebibliography}

\providecommand{\noopsort}[1]{}\providecommand{\singleletter}[1]{#1}%

\end{document}